\documentclass{article}

\usepackage{setspace}
\RequirePackage[OT1]{fontenc}
\usepackage{verbatim}
\usepackage{multirow}
\usepackage{makecell}
\usepackage{tabularx}
\usepackage{tabu}
\usepackage{float}
\usepackage{stfloats}

\RequirePackage{amsthm,amsmath,amsfonts,amssymb,mathtools}
\usepackage[utf8]{inputenc} % allow utf-8 input
\usepackage[T1]{fontenc}    % use 8-bit T1 fonts
\usepackage{url}            % simple URL typesetting
\usepackage{booktabs}       % professional-quality tabless
\usepackage{nicefrac}       % compact symbols for 1/2, etc.
\usepackage{microtype}      % microtypography
\usepackage{natbib}
\usepackage{doi}
\usepackage{xcolor}

\usepackage{cancel}
\usepackage{graphicx}
\usepackage{enumitem}
\usepackage{tabularx}
\newcommand\round[1]{\left[#1\right]}
\usepackage{comment}
\newtheorem{definition}{Definition}
\newtheorem{proposition}{Proposition}
\newtheorem{theorem}{Theorem}
\newtheorem{property}{Property}

\usepackage{geometry}
\onecolumn 

\DeclareMathOperator{\sign}{sign}

\usepackage{hyperref} %aggiungere link
    \hypersetup{
    	colorlinks=true,
    	linkcolor=blue,
    	filecolor=blue,      
    	urlcolor=blue,
    	citecolor=blue
    }

\title{Alternative Approaches for Estimating Highest-Density Regions}

\author{Nina Deliu$^{1,2}$ and Brunero Liseo$^{1}$\\
\normalsize $^{1}$MEMOTEF Department, Sapienza University of Rome, Italy\\ \normalsize$^{2}$MRC -- Biostatistics Unit, University of Cambridge, UK}

\date{}

\begin{document}
\maketitle

\begin{abstract}
Among the variety of statistical intervals, highest-density regions (HDRs) stand out for their ability to effectively summarize a distribution or sample, unveiling its distinctive and salient features. An HDR represents the minimum size set that satisfies a certain probability coverage, and current methods for their computation require knowledge or estimation of the underlying probability distribution or density $f$. In this work, we illustrate a broader framework for computing HDRs, which generalizes the classical density quantile method introduced in the seminal paper of \cite{hyndman_computing_1996}. The framework is based on \textit{neighbourhood} measures, i.e., measures that preserve the order induced in the sample by $f$, and include the density $f$ as a special case. We explore a number of suitable distance-based measures, such as the $k$-nearest neighborhood distance, and some probabilistic variants based on \textit{copula models}. An extensive comparison is provided, showing the advantages of the copula-based strategy, especially in those scenarios that exhibit complex structures (e.g., multimodalities or particular dependencies). Finally, we discuss the practical implications of our findings for estimating HDRs in real-world applications. 

\end{abstract}

\section{Introduction}
\begin{figure}[ht]
    \centering
    \includegraphics[scale=0.6]{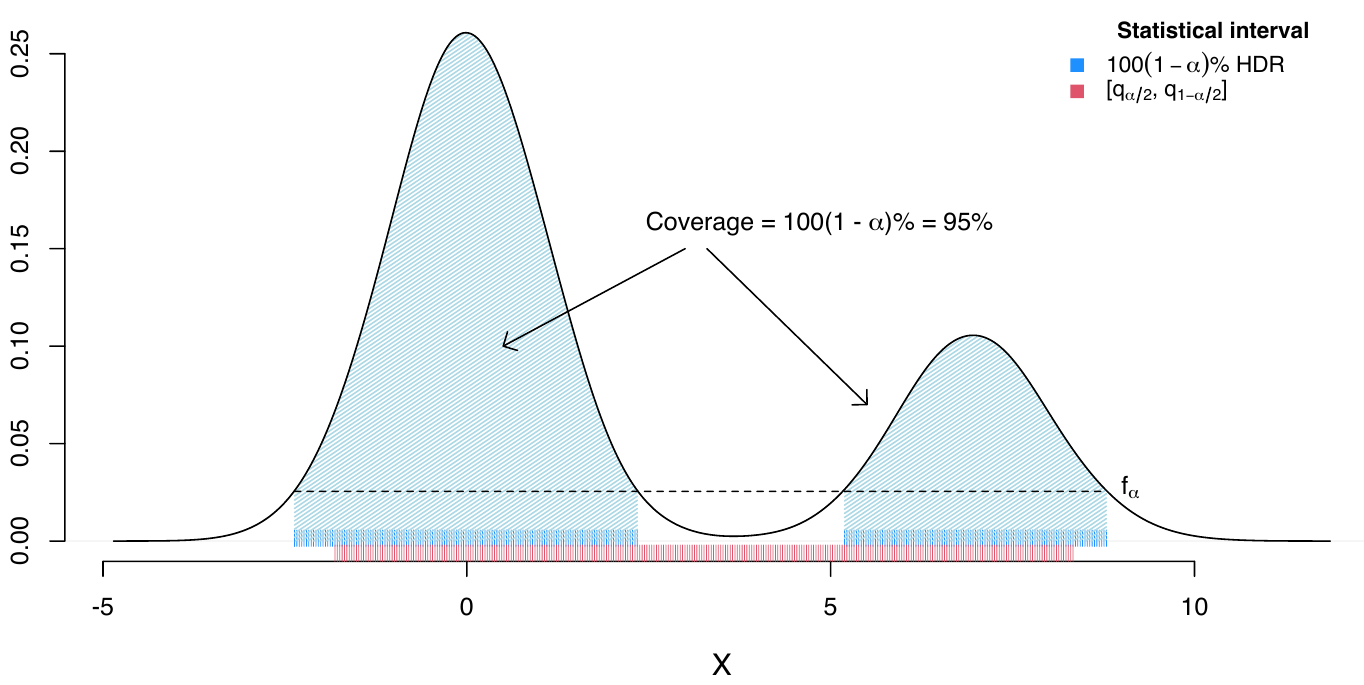}
    \caption{Comparison between two $100(1-\alpha)\%$ probability intervals for a normal mixture density: an HDR and an equal-tailed interval. The coverage parameter $\alpha$ is set to $0.05$.}
    \label{fig: HDR_symmQ}
\end{figure}

A ubiquitous problem in statistics is to derive statistical intervals or regions--especially in the multivariate setting--for population parameters or other unknown quantities. Their role is to provide a way to quantify and describe the uncertainty about a quantity of interest, or simply a way to summarize the information contained in a distribution. Statistical regions may address different problems. For example, a \textit{confidence interval} (CI) describes the uncertainty related to an estimate for an unknown \textit{parameter}, while a \textit{prediction interval} provides bounds for one or more future \textit{observations}. Alternatively, a \textit{tolerance interval} is the interval expected to contain a specified proportion of the sampled population. In a Bayesian setting, \textit{highest posterior density} or \textit{credible} regions provide, in a natural way, set estimates for a specific parameter~\citep{box_bayesian_1992, turkkan_computation_1993}. % In that context, they are based on a posterior distribution, but the underlying idea remains unchanged. 
We refer to~\cite{meeker_statistical_2017} and~\cite{krishnamoorthy_statistical_2009} for an overview.
Furthermore, even when the focus is on a specific type of interval, e.g., a two-sided 95\% prediction interval, questions on how this should be defined may still arise. 
Should we use the interval symmetric about the mean or the interval of shortest length, among others? Although each of these intervals has 95\% coverage, they may all be different. Consider, for example, the case of nonsymmetric and/or multimodal distributions, with an illustration given in Figure~\ref{fig: HDR_symmQ}. An additional layer of complexity arises in multivariate settings, where there are no unique agreed definitions, and generalizations include the concept of simultaneous CIs~\citep{guilbaud_simultaneous_2008}, multivariate CIs~\citep{korpela_multivariate_2017}, or different definitions for multivariate quantiles~\citep{figalli_continuity_2018,cai_multivariate_2010,coblenz_nonparametric_2018}, among others. 

The multivariate setting will be the target of this work, with a focus on bivariate distributions. In particular, our interest is devoted to statistical \textit{regions} for summarizing probability distributions in the form of \textit{highest-density regions}~\citep[HDRs;][]{hyndman_computing_1996}. Statistical regions other than HDRs are beyond the scope of the present work, and we refer to~\cite{meeker_statistical_2017} and ~\cite{krishnamoorthy_statistical_2009} for a comprehensive survey on the broader topic. As the name suggests, an HDR specifies the set of points of highest density: the density for points inside the region must be higher than that for points outside it. More specifically, considering a $d$-dimensional continuous variable of interest $X \in \mathbb{R}^d, d \geq 1$, with probability density function $f$, the problem is to estimate minimum volume sets of the form $C(f_\alpha) = \{ x\!: f(x) \geq f_\alpha\}$, such that $P(X \in C(f_\alpha)) \geq 1-\alpha$, where $1-\alpha$, with $\alpha \in (0,1)$, represents a prespecified coverage probability. Although they share substantial similarities with \textit{multivariate quantiles}, estimating an HDR differs from estimating level sets in that one is interested in specifying a probability content rather than the level directly. This complicates the problem, and we refer to~\cite{doss_bandwidth_2018} for more details.

The scope of an HDR can be wide and diverse; the following are possible applications.
\begin{description}
    \item[Forecasting] To obtain a prediction or forecast region for a set of observable variables in order to 
    inform the most likely future realizations and convey in a simple way the accuracy of a forecast~\citep[for illustrative examples, see e.g.,][]{hyndman_computing_1996,kim_improved_2011}.
    \item[Anomaly detection] To detect abnormal observations from a sample: if a data point does not belong to a region of ``normal'' data (the HDR), then it is regarded anomalous~\citep[see e.g.,][and references therein]{steinwart_classification_2005}.
    \item[Unsupervised or semi-supervised classification] To identify areas or clusters with a relatively high concentration of a given phenomenon; see, e.g., the work of~\cite{saavedra-nieves_nonparametric_2022}, aimed at finding areas of high incidence of coronavirus.
\end{description}

This work will primarily be driven by the problem of anomaly detection, characterizing a broad spectrum of applied domains, going from astrophysics to diagnostics and sport analytics. More specifically, our interest is to develop a valid and efficient framework in support of the worldwide doping detection mission headed by the World Anti-Doping Agency \citep{wada_world_2021}. In practice, WADA's current analytical implementation is based on identifying reference ranges that discriminate well between normal and abnormal values for predefined biomarkers of interest \citep{sottas_bayesian_2007}. This is done following a univariate approach, with reference ranges, in the form of equal-tailed intervals, derived for each biomarker separately. Clearly, addressing this problem over increased dimensions presents significant challenges, including the presence of data with complex dependence structures, in addition to distributional multimodalities or skewness. 

Due to their flexibility ``to convey both multimodality and asymmetry'', HDRs are argued to be a more effective summary of the distribution~\citep{hyndman_highest_1995}. In the case of unimodal symmetric distributions, such as the normal distribution, an HDR coincides with the usual probability region symmetric about the mean, spanning the $\alpha/2$ and $1-\alpha/2$ quantiles. However, in the case
of a multimodal distribution, it may consist of several disjoint subregions, each containing a local mode. This provides useful information that could not be traced by other probability regions such as an equal-tailed interval (see Figure~\ref{fig: HDR_symmQ} for an illustrative example). 

As the name suggests, estimating an HDR is based on knowing the density function $f$ of the variable of interest $X$. However, as typically occurs in practice, this quantity is unknown and the estimation of an HDR requires estimating $f$ first. The seminal paper of \cite{hyndman_computing_1996} discusses the \textit{density-quantile approach} for computing HDRs in such settings. Although for unidimensional problems this task can be achieved very accurately using methods such as the kernel density estimator~\citep[KDE;][]{parzen_estimation_1962} or the local likelihood approach~\citep{hjort_locally_1996}, it may be inefficient for multidimensional problems~\citep{liu_sparse_2007}. In fact, over increased dimensions, KDE suffers from the difficulty of finding optimal kernel functions and the corresponding bandwidths (i.e.,  the smoothing parameters). In particular, bandwidth selection in KDE is recognized as the most crucial and difficult step~\citep[see e.g., Chapter 2 in][]{wand_kernel_1994}, with no definite and widely accepted solution. Furthermore, high-dimensional data also pose challenges from an algorithmic/computational perspective when deriving the associated HDR. 

In this work, we illustrate a broad framework for estimating HDRs, which generalizes the current density-quantile approach implementable on the basis of a consistent estimator of the (multivariate) density~\citep{hyndman_computing_1996}. The proposed framework is based on \textit{neighbourhood} measures~\citep{munoz_estimation_2006}, that is, measures that preserve the order induced in the sample by the density function. Notably, it includes the widely-used density estimation procedure as a special case. 
We then elaborate on and evaluate a number of suitable probabilistic- and distance-based measures, including a variation of the measure adopted by the \textit{k-nearest neighbors} algorithm. In particular, motivated by the ubiquitous role of \textit{copula} modeling~\citep{nelsen_introduction_2006} in modern statistics, among probabilistic-based measures, we explore the use of copulae in an HDR estimation context. Interestingly, copulae introduce more flexibility to deal with multivariate random vectors, by separately estimating the marginals and their dependence structure, i.e., the copula model. In addition, by placing a strong focus on the dependence model, a copula approach has the advantage of better capturing data specificities, especially when these exhibit complex relationships such as asymmetric and/or tail dependencies. 

The remainder of this manuscript is organized as follows. In Section~\ref{sec: problem}, we introduce the problem of interest and review existing methods for HDR estimation. The general neighborhood-quantile framework is described in Section~\ref{sec: neigh_quantile}. In Section~\ref{sec: measures}, we discuss and propose alternative measures in the context of HDRs, including some variants based on copulae. Section \ref{sec: results} provides a comprehensive comparison among the introduced measures and the standard kernel density estimator. Empirical studies are focused on bivariate scenarios that vary according to the complexity of the data (marginal and dependence structure, multimodality, etc) and the sample size. A distribution on a compact support, namely, the Dirichlet distribution, is also considered to support practical applications with compositional data. An application to the MAGIC data set, which classifies high-energy Gamma particles in the atmosphere~\citep{bock_methods_2004}, is illustrated in Section~\ref{sec: magic_data}. We conclude in Section~\ref{sec: conclusions} by summarizing the main findings and discussing their implications for estimating HDRs in real-world problems.

\section{A general framework for HDR estimation} \label{sec: problem}
Let $\{X_1,\dots, X_n\}$ be a sample of $n$ independent and identically distributed (iid) replications of a random variable $X$ defined on $\mathbb{R}^d$, with $d\geq1$. In our problem, we assume to have access to a sample $s_n = \{x_1,\dots,x_n \} \in S_n$ of their actual realizations, with $S_n$ the sample space. We then wish to use $s_n$ for estimating an HDR, that is, a statistical region containing those sample values of relatively high density (see Definition 1, attributed to~\citealp{hyndman_computing_1996}). We denote by $X_{ij}\ (x_{ij})$ the $j$-th component of $X_i\ (x_i)$, for $j=1,\dots,d$ and $i=1,\dots,n$. We restrict our discussion to continuous random variables $X$ and, unless otherwise stated, we denote by $f$ their probability density function (PDF) and by $F$ their cumulative density function (CDF).
 
\begin{definition}[Highest-density region;~\citealp{hyndman_computing_1996}]
    Denote by $f$ the PDF of a continuous, possibly multivariate, random variable $X \in \mathbb{R}^d$; the $100(1-\alpha)\%$ HDR is defined as the subset $C(f_\alpha)$ of the sample space of $X$ such that:
\begin{align*}
    C(f_\alpha) = \{ x: f(x) \geq f_\alpha\},
\end{align*}
where $f_\alpha$ is the largest constant such that $P(X \in C(f_\alpha)) \geq 1-\alpha$, with $\alpha \in (0,1)$. 
\end{definition}

One of the most distinctive properties of HDRs is that, among all regions of probability coverage $100(1-\alpha)\%$, the HDR has the smallest possible volume. The notion of ``smallest'' is to be understood with respect to some measure such as the usual Lebesgue measure; in the continuous one-dimensional case that would lead to the shortest-length set, while in two dimensions that would be the smallest-area set. It also follows from the definition that the boundary of an HDR consists of those values of the sample space with equal density. Hence a plot of a bivariate HDR has as the boundary a contour plot. 

\subsection{Density quantile approach} The study of HDRs has been largely enhanced by Hyndman, who proposed the \textit{density quantile approach} (outlined in Proposition~\ref{prop1}) to estimate multivariate HDRs~\citep{hyndman_computing_1996}. Today, this still represents the typical strategy and involves estimating the density. 
%Let $X$ be the continuous random variable of interest with density $f$ for which we want to estimate an HDR. Let $Y = f(X)$ be the random variable obtained by transforming $X$ by $f$ (bounded and continuous in $x$). Define $f_\alpha$ as the $\alpha$ quantile of $Y$, that is, $f_\alpha$ is such that $\mathbb{P}(f(X) > f_\alpha) = 1 - \alpha$. Clearly, if $s_m \doteq \{x_1,\dots,x_m\}$ is a set of independent observations from the distribution of $X$, then $\{f(x_1),\dots,f(x_m)\}$ represents a set of independent observations from the distribution of $Y$. Furthermore, the following holds. 
\begin{proposition}[\citealp{hyndman_computing_1996}] \label{prop1}
    Let $\{f(x_1),\dots,f(x_m)\}$ be a sample of independent observations of size $m$ of the random variable $Y = f(X)$, with $f$ a bounded and continuous function in $x$. Consider the ordered sample $\{f_{(1)},\dots,f_{(m)}\}$ with $f_{(j)}$ the $j$-th largest among the $f(x_{i})$'s so that $f_{(j)}$ is the $(j/m)$ sample quantile of $Y$. Then, given a constant $\alpha \in [0,1]$, and denoted with $\lfloor j \rfloor$ the greatest integer less than or equal to $j$, in probability,
    \begin{align*}
         \hat{f}_\alpha &\doteq f_{(\lfloor \alpha m \rfloor)} \to f_\alpha \quad \text{as}\ m \to \infty,\\
         C_m(\hat{f}_\alpha) &\doteq \{x\!:f(x) > \hat{f}_\alpha\} \to C(f_\alpha) \quad \text{as}\ m \to \infty.
    \end{align*}
\end{proposition}

Basically, the HDR is derived based on the sample quantile of the density $f$. However, the density function itself is often unknown and one has to estimate it based on a set of available iid observations $s_n \doteq \{x_1,\dots,x_n\}$. In this case, the $100(1-\alpha)\%$ HDR can be estimated as
\begin{align}\label{eq: est_C_dq}
    \hat{C}_n(\hat{f}_\alpha) &\doteq \{x\!:f_n(x) > f_{(\lfloor \alpha n \rfloor)}\},
\end{align}
with $f_n$ being a possibly consistent estimator of $f$. Note that for small $n$, it may not be possible to get a reasonable density estimate. Also, with few observations and no prior knowledge on the underlying density function, there seems to be little point in attempting to summarize the density.

\subsection{One-class neighbor machines (OCNM) approach} An alternative approach to HDR estimation, inspired by the theory of support vector machines~\citep{scholkopf_estimating_2001} has been introduced in the machine learning literature by~\cite{munoz_estimation_2006}. The procedure is outlined in Proposition~\ref{prop2}, and involves the notion of \textit{neighborhood measures} (see Definition~\ref{def: neigh}, attributed to~\citealp{munoz_estimation_2006}).

\begin{definition}[Neighbourhood measure;~\citealp{munoz_estimation_2006}] \label{def: neigh}
    Let $X$ be a random variable with density function $f$ defined on $\mathbb{R}^d$. Denoted by $S_n$ the set of random iid samples $s_n = \{x_1,\dots,x_n\}$ of size $n$ (drawn from $f$), the real-valued function $g\!:\mathbb{R}^d \times S_n \to \mathbb{R}$ is a neighbourhood measure if one of the following holds:
    \begin{align*}
        \text{(a)}\ f(x) < f(y)& \Longrightarrow \lim_{n\to \infty}\mathbb{P}\left(g(x,s_n) > g(y,s_n)\right) = 1\quad x, y \in s_n, \forall s_n \in S_n, \\
        \text{(b)}\ f(x) < f(y)& \Longrightarrow \lim_{n\to \infty}\mathbb{P}\left(g(x,s_n) < g(y,s_n)\right) = 1\quad x, y \in s_n,  \forall s_n \in S_n.
    \end{align*}
    The function $g$ is called either a (a) \textit{sparsity} or a (b) \textit{concentration} measure. 
\end{definition}

\begin{proposition}[\citealp{munoz_estimation_2006}] \label{prop2}
     Consider an iid sample $s_n = \{x_1,\dots,x_n\}$ and a sparsity measure $g$. Define $\rho^* = g(x_{(\nu n)}, s_n)$, with $x_{(\nu n)}$ being the $(\nu n)$-th sample in the order induced in $s_n$ by $g$, provided that $\nu n \in \mathbb{N}$; otherwise, the least integer greater than $\nu n$, denoted by $\lceil \nu n \rceil$, is taken. Then, the binary decision function $h(x) = \sign(\rho^*-g(x, s_n))$ is such that:
     \begin{align*}
          \mathbb{P}\left(\frac{1}{n}\sum_{i=1}^n\mathbb{I}(h(x_i) \neq -1) = \nu \right) \to 1\quad \text{as}\ n \to \infty.
      \end{align*} 
      where $\mathbb{I}$ denotes the indicator function, and
     \begin{align*}
         C^{\text{OCNM}}_n \doteq \{x\!: h(x) \geq 0\} \to C(f_\nu) \doteq \{x\!: f(x) \geq f_\nu\} \quad \text{as}\ n \to \infty,
     \end{align*}
     where $C(f_\nu)$ is the minimum-volume set such that $\mathbb{P}(C(f_\nu)) \geq \nu$, with $\nu \in [0,1]$.
\end{proposition}
Notably, the OCNM algorithm relaxes the HDR estimation problem in the following sense: Instead of estimating and evaluating the density $f$, a more general and potentially simpler measure $g$ that asymptotically preserves the order induced by the density, can be considered. It is, however, important to remark that the quality of the estimation procedure heavily depends on using a neighborhood measure. If the measure used is neither a concentration nor a sparsity measure, there is no reason why the method should work.

\subsection{Neighborhood-quantile approach} \label{sec: neigh_quantile}

We now discuss a hybrid approach that can be viewed as: i) a generalization of the density quantile approach and; ii) a restatement of the OCNM method directly in terms of its solution $\rho^*$. Compared to the former, it ensures a wider applicability allowing for a more general set of functions, including the density $f$ as a special case; with respect to the latter, it offers a more direct, interpretable, and computationally efficient method.

\begin{theorem} \label{th: quantile_neigh}
Let $X$ be a continuous random variable defined on $\mathbb{R}^d$ with density function $f$, and consider a set $s_n = \{x_1,\dots,x_n\} \in S_n$ of size $n$ (drawn from $f$). Assume $g\!: \mathbb{R}^d \times S_n \to \mathbb{R}$ is a neighbourhood measure. Then an estimate of the $100(1-\alpha)\%$ HDR can be obtained as:
\begin{align*}
    C_n &= \{x\!: g(x, s_n) \leq g(x_{(\lceil(1-\alpha) n\rceil)}, s_n)\}\quad \text{if}\ g\ \text{is a sparsity measure},\\
    C_n &= \{x\!: g(x, s_n) \geq g(x_{(\lfloor \alpha n\rfloor)}, s_n)\} \quad \quad\ \  \text{if}\ g\ \text{is a concentration measure},
\end{align*}
where $g(x_{(\lfloor\alpha n\rfloor)}, s_n)$ is the $\alpha$-quantile of the sample $\{g(x_1, s_n),\dots,g(x_n, s_n) \}$, and $\lfloor j \rfloor$ denotes the greatest integer less than or equal to $j$.%, while $\lceil j \rceil$ denotes the least integer greater than or equal to $j$.
\end{theorem}

The proof is straightforward when considering the relationship between the region $C^{\text{OCNM}}_n$ as defined in Proposition~\ref{prop2} and the region $C_n$ as defined in Theorem~\ref{th: quantile_neigh}. 
In fact, without loss of generality, taking $g$ to be a sparsity measure, and noticing that $\nu$ plays the role of $1-\alpha$, one can observe that:
\begin{align*}
    C^{\text{OCNM}}_n &= \{x\!: h(x) =  \sign(\rho^*-g(x, s_n)) \geq 0\}\\
        &= \{x\!: \sign(g(x_{(\lceil \nu n\rceil)}, s_n)-g(x, s_n)) \geq 0\}\\
        %&= \{x\!: g_n(x_{\lfloor\alpha n\rfloor})-g_n(x) \geq 0\}\\
        & = \{x\!: g(x, s_n) \leq g(x_{(\lceil(1-\alpha) n\rceil)}, s_n)\}\\
        & = C_n.
\end{align*}

\textbf{Remark} If $g$ is chosen to be a concentration measure, then to ensure a $100(1-\alpha)\%$ coverage (notice that in this case $\mathbb{P}(C(f_\nu)) = 1-\nu$), the decision value $\rho^*$ induced by the concentration measure is given by $\rho^* = g(x_{(\lfloor \alpha n\rfloor)}, s_n)$. 

Noticing that the density represents a concentration measure, the resemblance with the density quantile approach in Proposition~\ref{prop1} should now be clear. In fact, from the perspective of the density quantile approach, one can view Theorem~\ref{th: quantile_neigh} as a generalization of Proposition~\ref{prop1}, where $f$ is replaced by any function $g$ that satisfies the criteria of neighborhood measures. Intuitively, provided that the density function $f$ is replaced by a function that preserves the order induced in the sample by $f$, the estimated HDR is asymptotically valid. Neighborhood measures ensure this ranking. 
To see it, without loss of generality, consider a sparsity measure $g$ and a sample $s_n = \{x_1,\dots,x_{(1-\alpha) n},\dots,x_n\}$ ordered so that
\begin{align*}
    f(x_1)<\dots<f(x_{(1-\alpha) n})<\dots<f(x_n),
\end{align*}
where, for simplicity, we suppose $(1-\alpha) n \in \mathbb{N}$ and $f(x_j) \neq f(x_j)$ for all $i \neq j$. From Definition~\ref{def: neigh} (a), for each pair $(x_i, x_j)$, with $i<j$, it holds that $\mathbb{P}\left(g(x_i, s_n) > g(x_j, s_n)\right) \to 1$. Hence, given $\epsilon \in (0,1)$, there exists $n_{ij} \in \mathbb{N}$ such that $\mathbb{P}\left(g(x_i, s_{n_{ij}}) > g(x_j, s_{n_{ij}})\right) > 1- \epsilon$. Taking $n \geq \max\{n_{ij}\}$, it is guaranteed that $\mathbb{P}\left(g(x_{1}, s_n) > \dots > g(x_{(1-\alpha) n}, s_n) > \dots > g(x_{n}, s_n)\right) > 1- \epsilon$. Therefore, as $n \to \infty$
\begin{align*}
    \mathbb{P}\left(g(x_{1}, s_n) > \dots > g(x_{(1-\alpha) n}, s_n) > \dots > g(x_{n}, s_n)\right) \to 1.
\end{align*}

\textbf{Remark} Proposition~\ref{prop1} is a special case of Theorem~\ref{th: quantile_neigh} with
$g(x, s_n)$ being either: (a) a concentration measure $g(x, s_n) \propto \hat{f}(x, s_n)$ or; (b) a sparsity measure $g(x, s_n) \propto \frac{1}{\hat{f}(x, s_n)}$, where $\hat{f}$ can be any consistent density estimator. Among the plethora of density estimators, in this work we focus on KDE~\citep{parzen_estimation_1962}, and use it as a benchmark measure for the proposed comparators in Section~\ref{sec: measures}. Specifically, given a set of iid observations $s_n \doteq \{x_1,\dots,x_n\}$ drawn from an unknown target density $f$, the KDE measure $M_0(x, s_n, h)$ at the location $x$ is defined as 
\begin{align}\label{eq: kde}
    M_0(x, s_n, h) = \frac{1}{nh^d}\sum_{i=1}^n K\left(\frac{ \lVert x-x_i \rVert}{h}\right),
\end{align} 
where $K\!: \mathbb{R}^d \to \mathbb{R}$ denotes the kernel function, satisfying $\int K(x) dx = 1$ and $K(x) \geq 0, \forall x$, and $h > 0$ the bandwidth hyperparameter. Details on the chosen kernel and bandwidth value will be given in Section~\ref{sec: results}. %It is well known that the bandwidth selection is the most crucial and difficult step in order to obtain a good estimate~\cite[see e.g., Chapter 2 in][]{wand1994kernel}, and there exists no definite and unique solution to this problem. The bandwidth of the kernel can exhibit a strong influence on the resulting estimate: analogously to the binwidth in a histogram, this parameter controls the amount of smoothing, with small values of $h$ \textit{undersmoothing} the function, and large value of $h$ \textit{oversmoothing} it, obscuring much of the underlying structure. %In this work we consider the asymptotically optimal solution proposed in~\cite{chacon2011asymptotics}.

\section{Alternative measures for estimating HDRs} \label{sec: measures}

We now propose a number of neighborhood measures that could be used to estimate HDRs. We elaborate on some measures that have been successfully employed in areas such as classification or clustering and introduce some novel ideas, including a copula-based approach. Although some of the discussed distances are popular in existing statistical domains, e.g., the $k$-nearest neighbors distance and its use in classification and regression, these have not been considered in the context of HDRs.  

\subsection{\texorpdfstring{$\boldsymbol{k}$}-NN-Euclidean distance } \label{sec: M1}

This corresponds to the sum of all Euclidean distances of a given point $x$ from its $k$-nearest neighbors. The concept of closeness (``nearest'') is defined according to the Euclidean metric or $L^2$-norm denoted by $\lVert \cdot\rVert_2$; for $k=1$, the nearest neighbor is the point $x$ itself and, in this case, the distance is zero.

\begin{definition}[$k$NN-Euclidean distance] \label{def: M1}
Given a data point $x \in \mathbb{R}^d$ of a sample set $s_n$ of size $n$, and an integer $k \in [1,n]$, we define the k-nearest neighborhood Euclidean distance of point $x$ from sample $s_n$ as:
\begin{align*}
    M_1(x, s_n, k) \doteq \sum_{i=1}^k \lVert x-x_{(i)}\rVert_2,
\end{align*}
where $x_{(i)}$ denotes the $i$-th observation in the reordered data such that $0 = \lVert x - x_{(1)} \rVert_2 \leq \dots \leq \lVert x - x_{(i)} \rVert_2 \leq \dots \lVert x - x_{(n)} \rVert_2$.
\end{definition}

$M_1(x, s_n, k)$ is entirely based on a metric distance, that is, the Euclidean metric, and represents a sparsity measure. The property follows from the convergence in probability of the $k$-nearest neighbor density estimator~\citep{silverman_density_1996}. Similar distances have been used in the literature for nonparametric classification and regression starting from the seminal work of~\cite{fix1951discriminatory} and~\cite{cover1967nearest}, which led to the well-known $k$-nearest neighbors ($k$NN) algorithm. %Despite representing different problems from HDR estimation (particularly given the supervised vs. nonsupervised setting, respectively), the intuition behind the strategy is the same. Observations that share similar features tend to be similar, thus closer (according to a user-defined distance). 
Notably, the distance to the nearest neighbors can also be seen as a local density estimate~\citep{loftsgaarden_nonparametric_1965,silverman_density_1996} and as a special case of a variable-bandwidth KDE with a uniform kernel~\citep{terrell_variable_1992}. A large $k$NN distance indicates that the density is small and vice versa. Furthermore, by ranking each data point according to the distance from its $k$-nearest neighbors, this measure can also be used as an outlier score in anomaly detection~\citep{ramaswamy_efficient_2000}.

Compared to other methods, the $k$NN approach has several advantages such as: (i) being purely nonparametric, hence able to flexibly adapt to any continuous distribution; (ii) having a reasonable time complexity; (iii) depending on a unique hyperparameter $k$, whose tuning is relatively simple. The choice of $k$ should be made based on the sample data. Generally, higher values of $k$ reduce the effect of noise; however, they underfit the model--making boundaries between classes less distinct--and are computationally more expensive, especially for large $d$. A general rule of thumb in classification is $k = \lfloor\sqrt{n} \rfloor$, with $n$ the number of samples in the dataset. Further investigation of the role of $k$ in different settings and how it relates to the missclassification error is given in the Supplementary Material B and in~\cite{meeker_statistical_2017}. %Further details related to our specific HDR problem will be given in Section~\ref{sec: results}.

%\textcolor{red}{[Work for future] Alternative to $k$NN could be another local density estimator, defined as the number of datapoints belonging to the $\epsilon$-hyperrectangular of dimension $d$ (similar to other measure considered below). It could be useful to connect this alternative measure with the DBSCAN approach for clustering. I think the neighborhood-quantile approach provides a way to unify the problem of HDR and clustering from a density-based view.}

\subsection{ \texorpdfstring{$\boldsymbol{k}$}-NN-CDF distance} \label{sec: M2}

Although the Euclidean distance and other general notions of measures (including concepts like area or volume) can certainly be useful for capturing relevant insights on the topology of a given set, they may be equipped with a probability measure to better resemble the notion of density. Let $\mathbb{P}$ be a probability measure defined on $\mathbb{R}^d$ and denote by $F$ its associated CDF and by $F_i$ the CDF of the $i$-th marginal. %that is
%\begin{align*}
%    F_i(x_i) = \lim_{x_1,\dots,x_{i-1},x_{i+1},\dots,x_d \to \infty} F(x),\quad i=1,\dots,d,
%\end{align*}
%where $\mathbb{P}$ will be omitted for ease of notation, i.e., $F_i = F_i^\mathbb{P}$ and $F = F^\mathbb{P}$.

Given two data points $x_1 = (x_{11},\dots,x_{1d}) \in \mathbb{R}^d$ and $x_2 = (x_{21},\dots,x_{2d}) \in \mathbb{R}^d$ from the sample set $s_n$, a preliminary version of the CDF distance between the two data points, denoted by $d_\mathbb{P}(x_1, x_2)$, has been defined in~\cite{venturini_statistical_2015} in the context of clustering problems as:
\begin{align} \label{eq: cdf_distance}
    d_\mathbb{P}(x_1, x_2) = \sqrt{\sum_{j=1}^d \left(F_j(x_{1j})-F_j(x_{2j})\right)^2}.
\end{align}
In the typical case of unknown marginal CDFs, their empirical counterpart $F_{n,j}(t) = \frac{1}{n}\sum_{i=1}^n \mathbb{I}(x_{ij} \leq t)$, $j=1,\dots,d$, could be considered.

\begin{figure}[ht]
    \centering
    \includegraphics[scale=0.63]{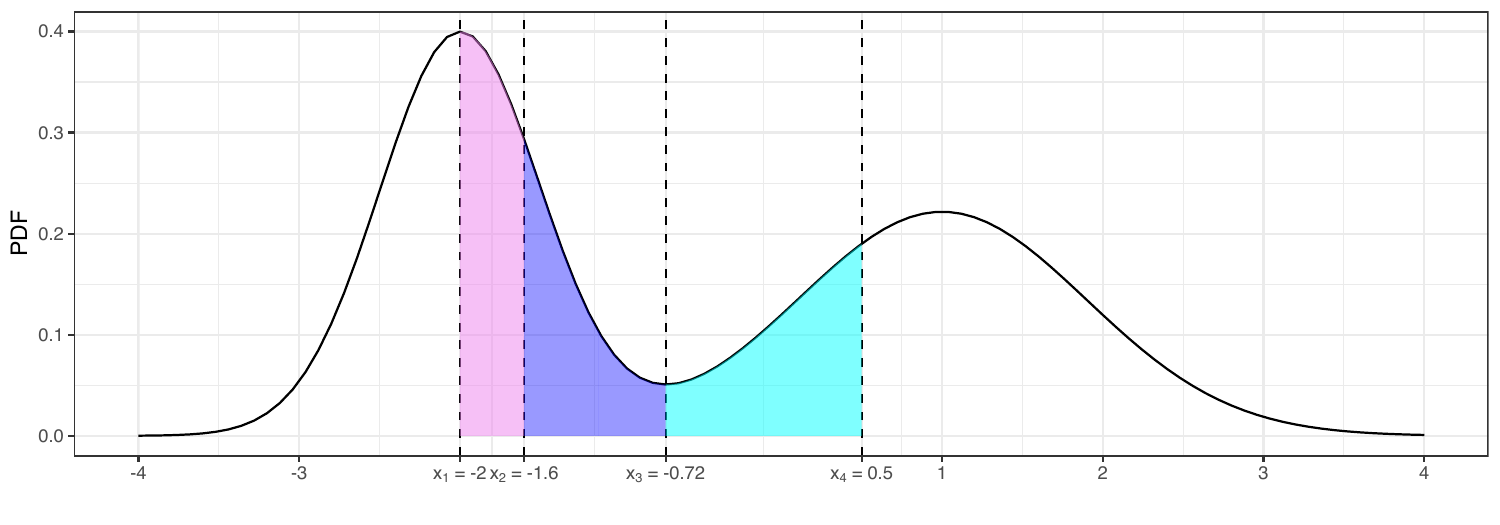}
    \caption{Illustration of the CDF-distances computed on the set of points $x_1 = -2$, $x_2 = -1.6$, $x_3 = -0.72$, $x_4 = 0.5$ based on a Gaussian mixture model.}
    \label{fig: Gaussian_mix}
\end{figure}

The distance $d_\mathbb{P}$ can be interpreted as the composition of the (nonlinear) transformation $F\!: \mathbb{R}^d \to [0,1]$ and the computation of the ordinary Euclidean distance. It fulfills all the properties of being a proper metric. %; for any $x_1, x_2, x_3 \in s_n$ we have the following: (i) $d_\mathbb{P}(x_1,x_2) \geq 0$, with $d_\mathbb{P}(x_1,x_2) = 0 \iff x_1=x_2$, (ii) $d_\mathbb{P}(x_1,x_2) = d_\mathbb{P}(x_2,x_1)$, and (iii) $d_\mathbb{P}(x_1,x_2) \leq d_\mathbb{P}(x_1,x_3) + d_\mathbb{P}(x_3,x_2)$.
Furthermore, it has the nice property that the distance between two points is proportional to the probability contained between the two points. However, as defined in Eq.~\eqref{eq: cdf_distance}, it presents some key limitations for inferring the underlying density and estimating HDRs. Consider for simplicity the univariate Gaussian mixture model illustrated in Figure~\ref{fig: Gaussian_mix} and the set of points $x_1 = -2$, $x_2 = -1.6$, $x_3 = -0.72$, $x_4 = 0.5$. When evaluating the CDF distance between $x_2$ and all other points, we have that $d_\mathbb{P}(x_2,x_1) \approx d_\mathbb{P}(x_2,x_3)$, but there is little to say about the density around $x_1$ and $x_3$, apart from realizing that they are clearly different. One could gain more information if one were to compute the CDF distance from the global mode $x_1$, assuming we have this information, noticing that the higher $d_\mathbb{P}(x_1,x_i)$, for all $i$, the lower the density of $x_i$. However, this is no longer verified for multimodal densities. In fact, we have $d_\mathbb{P}(x_1,x_4) > d_\mathbb{P}(x_1,x_3)$, but $f(x_4) > f(x_3)$, contrasting with the definition of a neighborhood measure. %Note also that by using the mode, one would need to have information on its value and any multimodal patterns. 
%So the neighborhood property is not satisfied. 

Motivated by these limitations and inspired by the idea of the $k$NN distance in Definition~\ref{def: M1}, we propose a new variant consisting of the sum of CDF-distances between a point $x$ and its $k$-nearest neighbors. Here, ``nearest'' should again be understood according to the Euclidean distance. The rationale is simple: the higher the CDF distances between a point and its neighbors, the higher one expects the density to be at that point. Since the $k$ neighbors of each data point will be at different (Euclidean) distances compared to the $k$ neighbors of the other data points, one needs to properly scale or weight the CDF distances according to this information. The proposed measure is reported in Definition~\ref{def: M2a}.

\begin{definition}[$\boldsymbol{k}$NN-CDF distance] \label{def: M2a}
Given $F_i$, the $i$-th marginal of a CDF $F$, for $i=1,\dots,d$, and an integer $k \in [1,n]$, we define the $k$-nearest neighborhood CDF distance as:
\begin{align*}
    M_2(x, s_n, k) = 
    \begin{cases} \sum\limits_{i=2}^k\frac{d_\mathbb{P}(x, x_{(i)})}{\lVert x-x_{(i)}\rVert_2} = \sum\limits_{i=2}^k\frac{\sqrt{\sum_{j=1}^d \left(F_j(x_j)-F_j(x_{(i)j})\right)^2}}{\sqrt{\sum_{j=1}^d \left(x_j-x_{(i)j}\right)^2}}&\quad k>1\\
     0 &\quad k=1,
    \end{cases}
\end{align*}
where $x = (x_1,\dots,x_d) \in \mathbb{R}^d$, and $x_{(i)} = (x_{(i)1},\dots,x_{(i)d}) $ is the $i$-th observation in the sample $s_n$ such that $\lVert x - x_{(1)} \rVert_2 \leq \dots \leq \lVert x - x_{(i)} \rVert_2 \leq \dots \lVert x - x_{(n)} \rVert_2$. %To ensure the measure is well defined, for $x = x_{(1)}$, we set $\frac{d_\mathbb{P}(x, x_{(1)})}{\lVert x-x_{(1)}\rVert_2} = 0$.
\end{definition}

\begin{property}[Semimetric]
    Consider the $M_2$ measure as defined in Definition~\ref{def: M2a}. It can be easily verified that, for any $k \in [1,n]$, if one restricts the CDF distance to $x$ and its $k$-th neighbor $x_{(k)}$, then $M_2(x, x_{(k)}, k)\!: \mathbb{R}^d \times \mathbb{R}^d \to \mathbb{R}$ is a semimetric. In fact, for any set $x, x_{(k)} \in \mathbb{R}^d$, it follows that:
\begin{itemize}
    \item If $x=x_{(k)}$, $M_2(x, x_{(k)}, 1) = 0$ by definition.
    \item If $x \neq x_{(k)}$, $M_2(x, x_{(k)}, k) > 0$. In fact, $(x_j-x_{(k)j})^2 > 0$ and $(F_j(x_j)-F_j(x_{(k)j}))^2 > 0$ for all $j$ and all $k$, when $x \neq x_{(k)}$ and $X$ is continuous. 
    \item  $M_2(x, x_{(k)}, k) = M_2(x_{(k)}, x, k)$, as $(x_j-x_{(k)j})^2 = (x_{(k)j}-x_j)^2$ and $(F_j(x_j)-F_j(x_{(k)j}))^2 = (F_j(x_{(k)j})-F_j(x_j))^2$ for all $j$ and all $k$.
\end{itemize}
The triangular inequality, stating that $M_2(x, x_{(k)}, k) \leq M_2(x, x_{(k')}, k) + M_2(x_{(k')}, x_{(k)}, k)$, for all $x, x_{(k)}, x_{(k')} \in \mathbb{R}^d$, does not hold. It is enough to consider a simple counterexample such as: $d=1$, $F$ the CDF of a standard normal distribution, and three points $x=1 < x_{(k)}=2 < x_{(k')}=3$. 
\end{property}

\subsection{\texorpdfstring{$\boldsymbol{\epsilon}$-}-neighborhood multivariate CDF distance} \label{sec: M3}

The measure $M_2(x, s_n, k)$ %and $M_2(x, s_n, \epsilon)$ 
introduced in Section~\ref{sec: M2} has several advantages; for example: (i) it takes into account the probabilistic information in the data, (ii) it is based on marginal CDFs which may be easy to estimate, and (iii) it has convenient computational times. However, using only the marginal CDFs may compromise their validity whenever the marginal components of a multivariate random variable $X \in \mathbb{R}^d$ show a significant dependence structure. In that case, the use of a multivariate CDF may be more appropriate. If the multivariate CDF is unknown, one could use its empirical counterpart. Notice, however, that this estimation part may compromise computational efficiency, especially in high dimensions and for large values of $k$. %(this will be later illustrated in Section~\ref{sec: results}). 
We therefore focus on an $\epsilon$-neighborhood version, which reduces the number of operations from $k$ to $1$, once $\epsilon$ is determined. %, each for each of the $k$ neighbors, nor requires their determination.

\begin{definition}[$\epsilon$-neighborhood multivariate CDF distance] \label{def: M3}
Given the CDF $F$ of a multivariate variable $X \in \mathbb{R}^d$, we define the $\epsilon$-neighborhood multivariate CDF distance of a point $x = (x_1,\dots,x_d)$ to be the probability of that point belonging to a hyperrectangle of dimension $d$ defined on vertexes $[x_1-\epsilon_1, x_1+\epsilon_1]\times \dots \times [x_d-\epsilon_d, x_d+\epsilon_d]$ scaled by hyperrectangle's $d$-volume:
\begin{align}\label{eq: M3}
    M_3(x, \epsilon) &= \frac{\mathbb{P}(X \in [x-\epsilon, x+\epsilon])}{\prod_{j=1}^d 2\epsilon_j} = \frac{\sum_{\nu \in \mathcal{V}} (-1)^{n(\nu)} F(\nu)}{\prod_{j=1}^d 2\epsilon_j}  \propto \sum_{\nu \in \mathcal{V}} (-1)^{n(\nu)} F(\nu),
\end{align}
where $\nu = (\nu_1,\dots,\nu_d)$, with $\nu_j \in \{x_j - \epsilon_j, x_j + \epsilon_j\}$, for $j \in 1,\dots, d$, and $n(\nu) = \sum_{j=1}^d \mathbb{I}(\nu_j = x_j - \epsilon_j)$. The sum is computed over the $2^d$ vectors of the set $\mathcal{V}$.
\end{definition}

As in the case of the density $f$ itself, when the CDF $F$ is unknown, a consistent estimator, say $F_n$, may be used leading to:
\begin{align}
    M_3(x, s_n, \epsilon) & = \frac{\sum_{\nu \in \mathcal{V}} (-1)^{n(\nu)} F_n(\nu)}{\prod_{j=1}^d 2\epsilon_j}  \propto \sum_{\nu \in \mathcal{V}} (-1)^{n(\nu)} F_n(\nu). \nonumber
\end{align}

Relationship in Eq.~\eqref{eq: M3} can be derived by recursion starting from small $d$ values. For $d=2$, for example, it is easy to verify that \begin{align*}
    M_3(x, \epsilon) &\propto \mathbb{P}(X \in [x-\epsilon, x+\epsilon]) \\
    & = \mathbb{P}(X_1 \in [x_1-\epsilon_1, x_1+\epsilon_1], X_2 \in [x_2-\epsilon_2, x_2+\epsilon_2])\\
    & = \mathbb{P}(X_1 \in [x_1-\epsilon_1, x_1+\epsilon_1], X_2 \leq x_2+\epsilon_2) - \mathbb{P}(X_1 \in [x_1-\epsilon_1, x_1+\epsilon_1], X_2 < x_2-\epsilon_2)\\
    & = \mathbb{P}(X_1 \leq x_1+\epsilon_1, X_2 \leq x_2+\epsilon_2) - \mathbb{P}(X_1 < x_1-\epsilon_1, X_2 \leq x_2+\epsilon_2)\\
    &\quad - \mathbb{P}(X_1 \leq x_1+\epsilon_1, X_2 \leq x_2-\epsilon_2) + \mathbb{P}(X_1 < x_1-\epsilon_1, X_2 < x_2-\epsilon_2)\\
    & = F(x_1+\epsilon_1, x_2+\epsilon_2) -F(x_1-\epsilon_1, x_2+\epsilon_2) - F(x_1+\epsilon_1, x_2-\epsilon_2) + F(x_1-\epsilon_1, x_2-\epsilon_2)\\
    & = \sum_{\nu \in \mathcal{V}} (-1)^{n(\nu)} F(\nu),
\end{align*}
with $\mathcal{V} = \{(x_1+\epsilon_1, x_2+\epsilon_2), (x_1-\epsilon_1, x_2+\epsilon_2), (x_1+\epsilon_1, x_2-\epsilon_2), (x_1-\epsilon_1, x_2-\epsilon_2) \}$.

We now state the following result for this measure.

\begin{property}[Density equivalence] Consider the $M_3$ measure as defined in Definition~\ref{def: M3} and assume that $F$ is differentiable at any point $x$ of its support. Then, as $\epsilon \to 0$, $M_3(x, \epsilon) \to f(x)$, for any $x \in \mathbb{R}^d$. 
\end{property}
The proof follows from basic probability and calculus theory.
%The proof is given in Appendix~\ref{app: proofs} and is based on the Lagrange theorem and the notion of symmetric derivative. 

\subsection{Copula-based measures} \label{sec: copula_measures}

A $d$-dimensional copula $C\!: [0,1]^d \to [0,1]$ is a CDF with uniform marginal distribution functions~\citep{nelsen_introduction_2006}. If we consider a random vector $X = (X_1,\dots,X_d)$ with joint CDF $F$ and marginals $F_1, \dots, F_d$, then the copula of $X$ is represented by the joint distribution of $F_1({X_1}), \dots, F_d({X_d})$ and is derived as:
\begin{align*}
    C(u_1,\dots,u_d) &= \mathbb{P}(F_1({X_1}) \leq u_1,\dots,F_d({X_d}) \leq u_d) \nonumber\\
    & = \mathbb{P}(X_1 \leq F_1^{-1}(u_1),\dots,X_d \leq F_d^{-1}(u_d)) \nonumber\\
    & = F(F_1^{-1}(u_1),\dots,F_d^{-1}(u_d)).
\end{align*}
Letting $u_j \doteq F_j(x_j)$, this yields the following well-known result due to \cite{sklar_fonctions_1959}:
\begin{align} \label{eq: cop_CDF}
    F(x_1,\dots,x_d) = C(F_1(x_1),\dots,F_d(x_d)),\quad \text{for all}\ x = (x_1,\dots,x_d) \in \mathbb{R}^d.
\end{align}
Furthermore, when the random vector $X$ is continuous with density $f$, we also have that 
\begin{align} \label{eq: cop_density}
    f(x_1,\dots,x_d) = c\Big(F_1(x_1),\dots,F_d(x_d)\Big) \times f_1(x_1)\times \cdots \times f_d(x_d),
\end{align}
where $c$ is the density of the random vector $\Big(F_1(X_1),\dots,F_d(X_d)\Big) \in [0,1]^d$. 

In summary, one can decompose every $d$-dimensional CDF $F$ or PDF $f$ into a composition of their marginal distribution functions and a $d$-copula. This allows us to redefine both $M_0$ in Eq.~\eqref{eq: kde} and $M_3$ in Definition~\ref{def: M3} in terms of their copula representation. For example, in the case of the $M_3$ measure, we may construct its copula-based alternative given in Definition~\ref{def: M4}.

\begin{definition}[$\epsilon$-neighborhood copula-based CDF distance] \label{def: M4}
Given the CDF $F$ of a random vector $X = (X_1,\dots,X_d)$, we define the $\epsilon$-neighborhood copula-based CDF distance of point $x = (x_1,\dots,x_d)$ to be 
\begin{align}\label{eq: M4}
    M^{\text{Cop}}_3(x, \epsilon) &= \frac{\sum_{\nu \in \mathcal{V}} (-1)^{n(\nu)} C(F_1(\nu_1), \dots, F_d(\nu_d))}{\prod_{j=1}^d 2\epsilon_j}  \propto \sum_{\nu \in \mathcal{V}} (-1)^{n(\nu)} C(F_1(\nu_1), \dots, F_d(\nu_d)),
\end{align}
where $\nu = (\nu_1,\dots,\nu_d)$, with $\nu_j \in \{x_j - \epsilon_j, x_j + \epsilon_j\}$, for $j \in 1,\dots, d$, and $n(\nu) = \sum_{j=1}^d \mathbb{I}(\nu_j = x_j - \epsilon_j)$.
\end{definition}
For $d=2$, it is immediate to verify that 
\begin{align*}
    M^{\text{Cop}}_3(x, \epsilon) &\propto C\left(F_1(x_1+\epsilon_1), F_2(x_2+\epsilon_2)\right) - C(F_1(x_1-\epsilon_1), F_2(x_2+\epsilon_2))\\ &\quad - C(F_1(x_1+\epsilon_1), F_2(x_2-\epsilon_2)) + C(F_1(x_1-\epsilon_1), F_2(x_2-\epsilon_2)).
\end{align*}
%with $\mathcal{V} = \{(x_1+\epsilon_1, x_2+\epsilon_2), (x_1-\epsilon_1, x_2+\epsilon_2), (x_1+\epsilon_1, x_2-\epsilon_2), (x_1-\epsilon_1, x_2-\epsilon_2) \}$.

The main advantage of this representation over the one involving the joint CDF is that the estimation of the multivariate distribution--when unknown--is performed through the estimation of the (univariate) marginals, evading thus the curse of dimensionality~\citep[see e.g.,][]{nagler_evading_2016}.  %Multivariate nonparametric density estimators are indeed known to converge very slowly to the true density when more than a few variables enter the model. 
Furthermore, copulae offer a flexible framework that captures complex dependence structures while offering direct control over the marginals. %In this work, we will also show that: (i) the copula-based approach also leads to significant computational advantages, and (ii) it is much more robust to the $\epsilon$ threshold vector that defines the neighborhood.

%Notice that, while in principle Sklar's result extends to density functions as well, provided densities exist, here, we prefer to focus on a measure that works directly with CDFs, and resemble more closely the definition of copulae. Furthermore, by focusing on the CDFs, we overcome the need of estimating both marginal densities $f_{X^j}$ and marginal CDFs $F_{X^j}$, for $j=1\dots,d$. 

\section{Empirical evaluation} \label{sec: results}

\subsection{Simulation setup}
We now evaluate and compare the proposed measures in an extensive number of simulated bivariate settings that vary according to the complexity of the data. In particular, we consider scenarios with different dependence structures induced by the copula model (e.g., tail or asymmetric dependencies), different marginal distributions (e.g., heavy-tailed or multimodal distributions), as well as different sample sizes.

\begin{enumerate}
    \item[(i)] We consider different copula models for the dependence structure among marginals, all sharing the same degree of dependence, expressed via Kendall's $\tau$, and set equal to $0.5$. The list of different copulae follows. 
    \begin{description}
        \item[$C^{\text{Gauss}}_{\theta = 0.7}$] Gaussian family (Elliptical class) with zero tail dependence and radial symmetry.
        \item[$C^{\text{t}}_{\theta = 0.7, \nu = 6}$] Student-$t$ family (Elliptical class) with radial symmetry.
        \item[$C^{\text{Frank}}_{\theta = 5.75}$] Frank family (Archimedian class) with radial symmetry.
        \item[$C^{\text{Clay}}_{\theta = 2}$] Clayton family (Archimedian class) not restricted to radial symmetry.
    \end{description}
    %We consider two copula families from the Elliptical class, the Gaussian and the Student-$t$ capula, characterised by radial symmetry with zero and nonzero tail dependence, respectively, and the Frank and Clayton families from the Archimedian class, with the latter not restricted to radial symmetry:
    
    \item[(ii)] Each of the above copulae is then combined with different simulation schemes for the marginals, according to the following details. Fixing $\sigma^2 = 2$ and $w_1 = 1 - w_2 = 0.5$, and taking $\mu_{11} = 0, \mu_{12} = 9, \mu_{21} = 1,  \mu_{22} = 8$, we consider the following.
    \begin{description}
        \item[Unimodal--heavy tails] Student-$t$ model $X_i \sim t_{\nu = 2},\quad i=1,2$.
        \item[Unimodal] Gaussian model $X_i \sim \mathcal{N}(\mu_{i1}, \sigma^2),\quad i=1,2$.
        \item[Bimodal] Gaussian $X_1 \sim \mathcal{N}(\mu_{11}, \sigma_1^2)$ \& Gaussian mixture $ X_2 \sim \sum_{k=1}^2w_k\mathcal{N}(\mu_{2k} + 5\mathbb{I}(k = 2), \sigma^2)$.
        \item[Quadrimodal] Gaussian mixture $X_i \sim \sum_{k=1}^2w_k\mathcal{N}(\mu_{ik}, \sigma^2), \quad i = 1,2$.
    \end{description}
    %We combine a Gaussian and a Gaussian mixture model to induce bimodality and two Gaussian mixture marginals to induce multimodality (4 modes). We consider the Student-$t$ distribution with $\nu = 2$ degrees of freedom as an instance of heavy-tailed distribution. 
    \item[(iii)] We also consider a distribution on a compact supports, that is, the Dirichlet distribution defined on the $(K-1)$-simplex, with $K=3$ and with parameters $\mathbf{\alpha} = (1,1,2)$. In this case, the dependence structure is uniquely determined and its closed-form copula expression is given in Section~\ref{sec: Dirichlet}. 
    \item[(iv)] We finally cover different sample sizes, with $n \in \{50, 100, 500, 1000\}$. 
\end{enumerate}

\begin{figure}[!ht]
    \centering
    \includegraphics[scale=0.6]{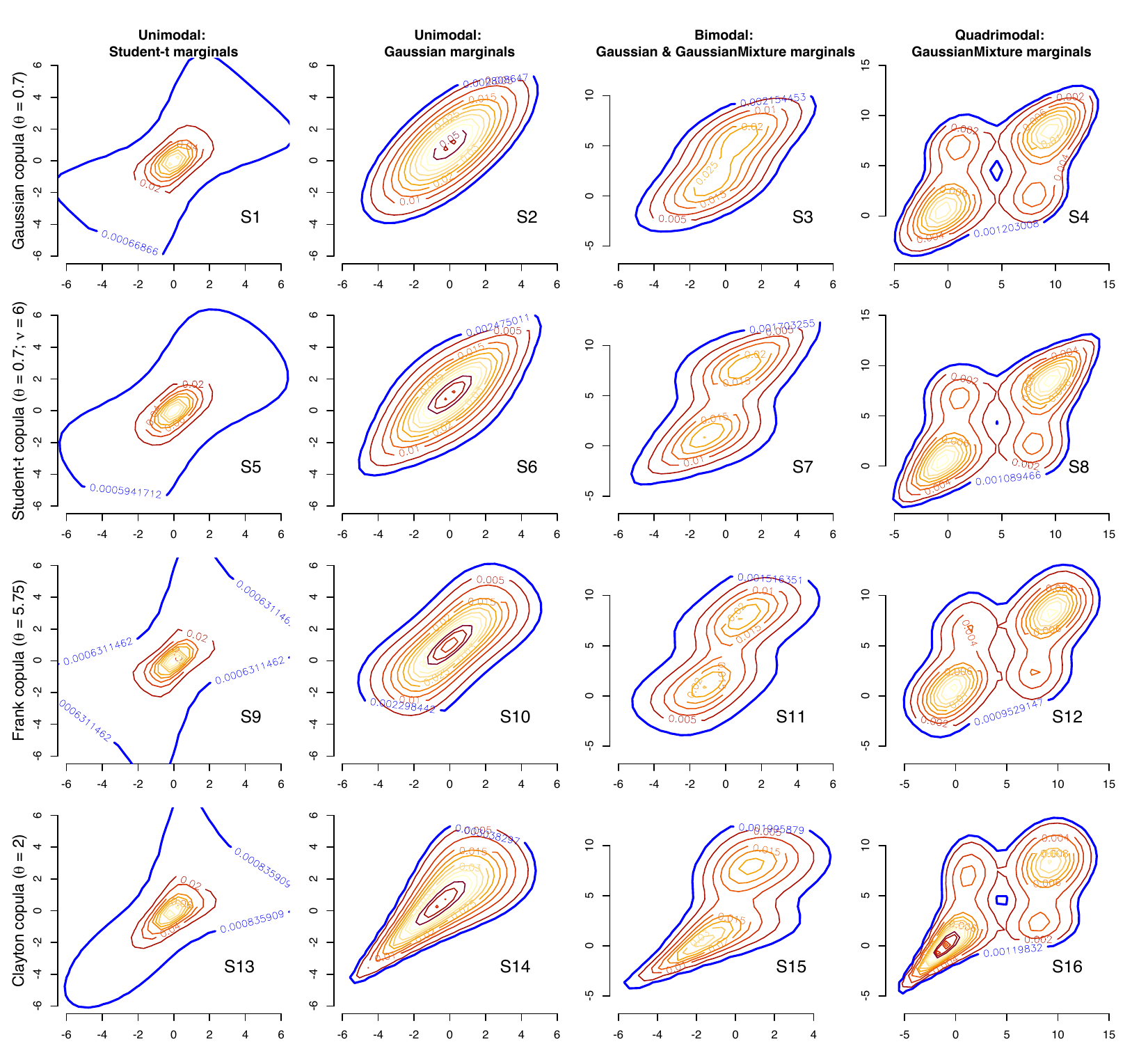}
    \caption{Evaluated scenarios with their contour plots at different levels, included the one delimiting the $95\%$ HDR (blue color).}
    \label{fig: scenarios}
\end{figure}

In total, 17 different scenarios are considered, named S1 to S17. Their graphical representation, in the form of contour plots, is provided in Figure~\ref{fig: scenarios} (S1-S16) and Figure~\ref{fig: dirichlet} (S17). 
By exploring such an extended number of simulation setups, we hope to cover a broad spectrum of potential scenarios that may occur in practice. In addition to offering a comprehensive understanding of the performances of the different measures, it may provide guidance to applied scientists who may need to choose the most ideal measure for their specific application. For example, the problem of doping detection when relying on the two primary biomarkers of the hematological module could well be related to scenario S15 (Gaussian and Gaussian mixture marginals and a Clayton copula; see~\cite{deliu2024multivariate} for an illustrative example). 

\subsection{Evaluated measures}
In all settings, we employ the proposed neighborhood-quantile method (see Section~\ref{sec: neigh_quantile}) on data samples of different sizes generated according to the aforementioned scenarios. 
%. We assume that the distribution that generated the data is unknown and derive the HDR 
For each scenario, we evaluate the following eight methods, with full details on their hyperparameter tuning reported in the Supplementary Material B.
\begin{description}
    \item[$M_0$:KDE] Direct estimation of the bivariate density. We use KDE, with Gaussian kernel and bandwidth selection based on the asympotically optimal solution proposed in~\cite{chacon_asymptotics_2011}, where its adequacy is shown in general settings, including Gaussian mixture models.

    \item[$M^{\text{NPCop}}_0$:DE] Nonparametric indirect density estimation with copula. We use standard KDE with the same optimal bandwidth of~\cite{chacon_asymptotics_2011} for the univariate marginals, and KDE with the transformation local likelihood estimator %of~\cite{geenens2017probit} 
    and nearest-neighbor bandwidth for the copula density. We refer to~\cite{nagler_evading_2016} for details.

    \item[$M^{\text{PCop}}_0$:DE] Parametric indirect density estimation with copula. We adopt a fully parametric approach (with maximum likelihood fitting) to estimate both marginals and the copula. For the copula model, we select the best model using the AIC criterion; no misspecification is introduced for the marginals.

    \item[$M_1$:$k$NN-Eucl] Cumulative Euclidean distances from the $k$NNs. We sum the Euclidean distances between each point and its $k$ neighbors defined according to the Euclidean metric. The choice of the hyperparameter $k$ is based on an extensive cross-validation procedure, leading to a general rule of thumb aligned with the existing literature: $k = \round{\sqrt{n/2}}$, with $\round{x}$ the integer closest to $x$.

    \item[$M_2$:$k$NN-CDF] Cumulative CDF distances from the $k$NNs. We sum the CDF distances between each point and its $k$ neighbors, which are defined according to the Euclidean metric. The measure follows a univariate approach as detailed in Section~\ref{sec: M2}. The choice of the hyperparameter $k$ is based on an extensive cross-validation procedure, suggesting a uniform choice across different scenarios and sample sizes: $k = 30$.

    \item[$M_3$:$\epsilon$-CDF] This represents the $\epsilon$-neighborhood multivariate CDF distance introduced in Section~\ref{sec: M3}. The empirical CDF is used as the CDF estimator. The optimal choice of the hyperparameter $\epsilon$ is based on an extensive cross-validation procedure, leading to the following heuristic for S1-S16 as a result of an exponential decay model fit with respect to the sample size: $\epsilon = \exp\left(2.13-0.3\log{n}\right)$.

    \item[$M^{\text{NPCop}}_3$:$\epsilon$-CDF] Fully nonparametric indirect estimation of the $\epsilon$-CDF measure with copula. We use the empirical CDF for estimating the univariate marginals, and KDE with the transformation local likelihood estimator and nearest-neighbor bandwidth for the copula density. The optimal choice of the hyperparameter $\epsilon$ follows the same strategy as the previous measure, leading to the following heuristic for S1-S16: $\epsilon = \exp\left(1.74-0.26\log{n}\right)$.

    \item[$M^{\text{PCop}}_3$:$\epsilon$-CDF] Parametric indirect CDF estimation with copula. We adopt a fully parametric approach (with maximum likelihood fitting) to estimate both marginals and the copula. For the copula model, we select the best model using the AIC criterion; no misspecification is introduced for the marginals. The optimal choice of the hyperparameter $\epsilon$ follows the heuristic $\epsilon = \exp\left(1.60 - 0.41\log{n} \right)$ for S1-S16.
    %\item[M8:CDF-PCop] 
\end{description}
Exception made for the measures depending on $\epsilon$ (with considerations deferred to Section~\ref{sec: Dirichlet}), the same hyperparameter choices are adopted in S17.

\subsection{Performance metrics}

The measurement of the performance of an HDR estimator can be closely related to the specific problem of interest. In a context where the interest is in detecting abnormal values, for example, estimating an HDR would allow to understand which points fall outside the normal-points region. This motivates certain metrics of common use in one-class classification problems, which we also employ in this work. However, we emphasize that the derivation of an HDR can in principle have a wider scope compared to a classification goal. Indeed, it would define the region of highest-density points (e.g., normal values), regardless of whether these points have been observed or not. Thus, it would not only allow us to classify an \textit{observed} point as normal or abnormal, but it would also provide the entire region of normal values, useful, for example, in a prediction setting \textit{prior to observing} a point. 

Let FP, TP, FN, and TN be the number of \textit{false positive}, \textit{true positive}, \textit{false negative}, and \textit{true negative} points, respectively, where \textit{positive} refers to those points that should be outside the true $(1-\alpha)\%$ HDR and \textit{negative} the others: 
\begin{align*}
    TN = \sum_{i \in s_n}\mathbb{I}(x_i \in C(f_\alpha)),\quad\quad
    FN = \sum_{i \in s_n}\mathbb{I}(x_i \in \hat{C}_n(\hat{f}_\alpha) \mid x_i \notin C(f_\alpha)),\\
    TP = \sum_{i \in s_n}\mathbb{I}(x_i \notin C(f_\alpha)),\quad\quad
    FP = \sum_{i \in s_n}\mathbb{I}(x_i \notin \hat{C}_n(\hat{f}_\alpha) \mid x_i \in C(f_\alpha)).
\end{align*}
Well-established measures of inefficiency are false negative/positive rates (FNR/FPR), and the total error rate (ERR), that is, the one-complement of accuracy:
$$\text{FNR}= \frac{FN}{FN +TP}, \quad \text{FPR} = \frac{FP}{FP+ TN}, 
\quad \text{ERR} = \frac{FN + FP}{FN + FP + TN + TP}= 1-\text{Accuracy}.$$
To account for the potentially high imbalance between positives and negatives, we also evaluate the two-sided F1 score, and the Matthews correlation coefficient~\citep[MCC;][]{matthews_comparison_1975}
alternatively known in statistics as the $\phi$-coefficient~\citep[see pg. 282 in][]{cramer_mathematical_1946}:
\begin{align*}
    \text{F1}&= \frac{2TP}{2TP+FP+FN} + \frac{2TN}{2TN+FP+FN},\\ \text{MCC}&= \frac{TP \times TN - FP \times FN}{\sqrt{(TP + FP) \times (TP + FN) \times (TN + FP) \times (TN + FN)}}.
\end{align*}
In particular, MCC has been shown to produce good scores only if the classification is adequate in all four elements of interest (true positives, false negatives, true negatives, and false positives), overcoming the overoptimistic inflated results, especially on imbalanced datasets, of other popular classification measures~\citep{chicco_advantages_2020}. 
All evaluations are based on $\alpha = 0.05$, that is, a coverage probability of 95\%, or, alternatively stated, $5\%$ and $95\%$ of positives and negatives, respectively. 

\subsection{Simulation results}
Comparative performance results are reported in Table~\ref{tab: performance_results_main} in terms of their mean (standard deviation) across a number of $1000$ independent Monte Carlo (MC) replicates. We primarily focus on discussing four scenarios that are representative of the different copula-induced dependencies and multimodalities, specifically those on the diagonal of Figure~\ref{fig: scenarios} (S1, S6, S11, S16), and the sample size $n = 500$; all the other scenarios and sample sizes are deferred to the Supplementary Material C. The Dirichlet case is covered in Section~\ref{sec: Dirichlet}. 

In general, copula-based approaches result in the most performing measures, with classification errors (ERR, FPR, FNR) uniformly smaller than those of the other measures and at no cost in terms of variability. Their advantage is particularly interesting in more complex scenarios, such as the quadrimodal case, where the parametric copula approach (both \textbf{$M^{\text{PCop}}_0$:DE} and \textbf{$M^{\text{PCop}}_3$:$\epsilon$-CDF}) practically halves the total error rate (ERR) of a standard \textbf{$M_0$:KDE}. Although the difference may be considered relatively negligible when looking at the ERR (maximum difference of $0.015$ in S16) and the FPR (maximum distance of $0.008$ in S16), it plays a significant role for the FNR, with an error difference of $0.157$. In practice, the probability of a \textit{true positive} being missed by the ``test'' or measure decreases from around $28\%$ (\textbf{$M_0$:KDE}; S16) to around $13\%$ (\textbf{$M^{\text{PCop}}_0$:DE}, \textbf{$M^{\text{PCop}}_3$:$\epsilon$-CDF}; S16). This aspect is well-captured by the alternative performance metrics of F1 and MCC, which offer a more reliable global measure of efficiency compared to the ERR or the Accuracy. As shown in Table~\ref{tab: performance_results_main}--S16, when comparing \textbf{$M_0$:KDE} to \textbf{$M^{\text{PCop}}_0$:DE} or \textbf{$M^{\text{PCop}}_3$:$\epsilon$-CDF}, the MCC, for example, increases from $0.693$ to more than $0.85$. Note that MCC varies from $-1$ (worst value) to $1$ (best value). 

When comparing the non-copula based measures, no substantial difference is noticed between \textbf{$M_0$:KDE}, \textbf{$M_1$:$k$NN-Eucl} and \textbf{$M_3$:$\epsilon$-CDF}, with a slightly improved performance of the second. In particular, \textbf{$M_1$:$k$NN-Eucl} provides the only exception to the uniform advantage of copula-based approaches. This occurs in S1 for $n = 50$, where $M_1$:$k$NN-Eucl shows an enhancement, although negligible (results are given in the Supplementary material C; Table 2 and Figure 31). Interestingly, S1 represents a scenario with heavy-tailed marginals, i.e., Student-$t$ distribution with $\nu = 2$, which translates into a substantially wider $95\%$ HDR (see Figure~\ref{fig: scenarios}). Relatively good results are also achieved in the other heavy-tailed cases (S5, S9, and S13), and only in small samples (in particular $n=50$; see Figure 31 in the Supplementary material). Outside these scenarios, compared to the classical \textbf{$M_0$:KDE}, \textbf{$M_1$:$k$NN-Eucl} has typically superior performances in more complex cases, in particular, in all quadrimodal distributions (S4, S8, S12, S16), and for smaller sample sizes ($n = 50$ and $n = 100$). As the sample size increases ($n > 100$), \textbf{$M_0$:KDE} shows some improvements, especially in simpler settings with Gaussian marginals (result are given in the Supplementary material C).

The worst behavior is shown by \textbf{$M_2$:$k$NN-CDF} (with FNR being as high as $72\%$ in S16). Note that this was expected as \textbf{$M_2$:$k$NN-CDF} follows a univariate rationale, motivating therefore the multivariate \textbf{$M_3$:$\epsilon$-CDF} proposal (see Section~\ref{eq: cdf_distance}). 
%In terms of sample size, clearly, the smaller $n$ the worse the performance of the different measures. However, 
\begin{table*}[!ht]
\setlength\tabcolsep{0pt} % make LaTeX figure out intercolumn space
\scriptsize
\caption{Performance results of the compared methods in selected scenarios for sample size $n=500$. Data are summarized in terms of mean (standard deviation) across $1000$ independent Monte Carlo (MC) replicates.} 
\label{tab: performance_results_main}
\begin{tabular*}{\textwidth}
{@{\extracolsep\fill}lc@{}c@{}c@{}cc@{}c@{}c@{}c@{}}\toprule
&\multicolumn{4}{@{}@{}@{}c@{}}{\textbf{Measures w/o copula}}
& \multicolumn{4}{@{}@{}@{}c@{}}{\textbf{Measures w copula}} \\
\cmidrule{2-5}\cmidrule{6-9}
\textbf{Metrics} & \textbf{$M_0$:KDE}  & \textbf{$M_1$:$k$NN-Eucl} & \textbf{$M_2$:$k$NN-CDF} & \textbf{$M_3$:$\epsilon$-CDF}
& \multicolumn{1}{@{}c@{}}{\textbf{$M^{\text{NPCop}}_0$:DE}}  & \textbf{$M^{\text{PCop}}_0$:DE} & \textbf{$M^{\text{NPCop}}_3$:$\epsilon$-CDF} & \textbf{$M^{\text{PCop}}_3$:$\epsilon$-CDF}\\ \\
\multicolumn{9}{@{}c@{}}{ Scenario S1 (Unimodal): Gaussian copula -- Student-$t$ marginals }\\
\toprule
ERR      & 0.015 (0.006) & \textbf{0.013 (0.006)} & 0.018 (0.007) & 0.017 (0.006) & 0.015 (0.007) & \textbf{0.009 (0.006)} & 0.014 (0.006) & \textbf{0.009 (0.006)} \\  
FPR      & 0.008 (0.005) & \textbf{0.007 (0.006)} & 0.009 (0.007) & 0.010 (0.006) & 0.008 (0.006) & \textbf{0.004 (0.005)} & 0.007 (0.006) & \textbf{0.005 (0.006)} \\  
FNR      & 0.135 (0.092) & \textbf{0.117 (0.079)} & 0.173 (0.075) & 0.132 (0.089) & 0.139 (0.088) & \textbf{0.073 (0.087)} & 0.125 (0.086) & \textbf{0.075 (0.086)} \\  
Accuracy & 0.985 (0.006) & \textbf{0.987 (0.006)} & 0.982 (0.007) & 0.983 (0.006) & 0.985 (0.007) & \textbf{0.991 (0.006)} & 0.986 (0.006) & \textbf{0.991 (0.006)} \\  
F1       & 1.842 (0.061) & \textbf{1.863 (0.063)} & 1.807 (0.080) & 1.826 (0.066) & 1.840 (0.075) & \textbf{1.908 (0.061)} & 1.854 (0.062) & \textbf{1.906 (0.061)} \\  
MCC      & 0.846 (0.058) & \textbf{0.867 (0.059)} & 0.811 (0.077) & 0.830 (0.063) & 0.844 (0.072) & \textbf{0.912 (0.055)} & 0.858 (0.058) & \textbf{0.910 (0.055)} \\ 
\bottomrule
\\
\multicolumn{9}{@{}c@{}}{Scenario S6 (Unimodal): Student-$t$ copula -- Gaussian marginals}\\
\toprule
ERR      & 0.015 (0.006) & 0.017 (0.006) & 0.060 (0.009) & 0.019 (0.006) & \textbf{0.014 (0.006)} & \textbf{0.011 (0.005)} & 0.017 (0.006) & \textbf{0.011 (0.005)} \\  
FPR      & 0.008 (0.005) & 0.009 (0.005) & 0.032 (0.006) & 0.010 (0.006) & \textbf{0.007 (0.005)} & \textbf{0.006 (0.005)} & 0.009 (0.005) & \textbf{0.006 (0.005)} \\  
FNR      & 0.140 (0.089) & 0.161 (0.094) & 0.601 (0.090) & 0.165 (0.092) & \textbf{0.128 (0.090)} & \textbf{0.096 (0.085)} & 0.157 (0.093) & \textbf{0.095 (0.085)} \\  
Accuracy & 0.985 (0.006) & 0.983 (0.006) & 0.940 (0.009) & 0.981 (0.006) & \textbf{0.986 (0.006)} & \textbf{0.989 (0.005)} & 0.983 (0.006) & \textbf{0.989 (0.005)} \\  
F1       & 1.838 (0.062) & 1.816 (0.065) & 1.363 (0.096) & 1.804 (0.068) & \textbf{1.851 (0.059)} & \textbf{1.884 (0.058)} & 1.820 (0.061) & \textbf{1.886 (0.058)} \\  
MCC      & 0.843 (0.059) & 0.820 (0.062) & 0.365 (0.096) & 0.808 (0.066) & \textbf{0.855 (0.055)} & \textbf{0.888 (0.053)} & 0.825 (0.059) & \textbf{0.890 (0.053)} \\ 
\bottomrule
\\
\multicolumn{9}{@{}c@{}}{Scenario S11 (Bimodal): Frank copula -- Gaussian \& Gaussian mixture marginals}\\
\toprule
ERR      & 0.019 (0.007) & 0.016 (0.006) & 0.061 (0.009) & 0.018 (0.006) & 0.015 (0.006) & \textbf{0.011 (0.006)} & \textbf{0.015 (0.006)} & \textbf{0.011 (0.006)} \\  
FPR      & 0.010 (0.005) & 0.008 (0.005) & 0.032 (0.006) & 0.011 (0.006) & 0.008 (0.005) & \textbf{0.006 (0.005)} & \textbf{0.008 (0.005)} & \textbf{0.006 (0.005)} \\  
FNR      & 0.175 (0.095) & 0.144 (0.093) & 0.608 (0.092) & 0.150 (0.093) & 0.139 (0.093) & \textbf{0.096 (0.089)} & \textbf{0.136 (0.093)} & \textbf{0.096 (0.089)} \\  
Accuracy & 0.981 (0.007) & 0.984 (0.006) & 0.939 (0.009) & 0.982 (0.006) & 0.985 (0.006) & \textbf{0.989 (0.006)} & \textbf{0.985 (0.006)} & \textbf{0.989 (0.006)} \\  
F1       & 1.800 (0.068) & 1.832 (0.064) & 1.355 (0.097) & 1.810 (0.065) & 1.837 (0.061) & \textbf{1.882 (0.059)} & \textbf{1.841 (0.061)} & \textbf{1.883 (0.059)} \\  
MCC      & 0.804 (0.066) & 0.836 (0.061) & 0.357 (0.097) & 0.815 (0.063) & 0.842 (0.058) & \textbf{0.887 (0.055)} & \textbf{0.845 (0.058)} & \textbf{0.887 (0.054)} \\
\bottomrule
\\
\multicolumn{9}{@{}c@{}}{Scenario S16 (Quadrimodal): Clayton copula -- Gaussian mixture marginals }\\
\toprule
ERR      & 0.029 (0.007) & 0.027 (0.008) & 0.072 (0.010) & 0.032 (0.008) & \textbf{0.021 (0.007)} & \textbf{0.014 (0.006)} & 0.024 (0.007) & \textbf{0.014 (0.006)} \\  
FPR      & 0.015 (0.006) & 0.014 (0.006) & 0.038 (0.005) & 0.019 (0.007) & \textbf{0.011 (0.006)} & \textbf{0.007 (0.006)} & 0.013 (0.006) & \textbf{0.008 (0.006)} \\  
FNR      & 0.283 (0.092) & 0.256 (0.093) & 0.720 (0.088) & 0.276 (0.095) & \textbf{0.197 (0.089)} & \textbf{0.126 (0.089)} & 0.226 (0.088) & \textbf{0.129 (0.089)} \\  
Accuracy & 0.971 (0.007) & 0.973 (0.008) & 0.928 (0.010) & 0.968 (0.008) & \textbf{0.979 (0.007)} & \textbf{0.986 (0.006)} & 0.976 (0.007) & \textbf{0.986 (0.006)} \\  
F1       & 1.689 (0.080) & 1.716 (0.081) & 1.238 (0.089) & 1.670 (0.083) & \textbf{1.777 (0.074)} & \textbf{1.850 (0.067)} & 1.748 (0.077) & \textbf{1.848 (0.069)} \\  
MCC      & 0.693 (0.079) & 0.720 (0.080) & 0.239 (0.090) & 0.675 (0.082) & \textbf{0.782 (0.072)} & \textbf{0.855 (0.064)} & 0.752 (0.075) & \textbf{0.852 (0.066)} \\ 
\bottomrule
\end{tabular*}
\end{table*}

\subsubsection{Simplex scenario: Dirichlet} \label{sec: Dirichlet}
A particular interest is dedicated to the simplex scenario, which plays an important role as the sample space of compositional data. Compositional data quantitatively describe parts of some whole and consist of vectors of positive components subject to a unit-sum constraint~\citep{aitchison_statistical_1982}. Measurements involving proportions, probabilities, or percentages can all be thought of as compositional data. These commonly arise in many disciplines; for example, in demography, cause-specific mortality rates can be studied by considering them as compositions, where no single rate is free to vary separately from the rest of the rate composition~\citep[see e.g.,][]{stefanucci_analysing_2022}. This induces a particular and unique dependence structure (represented, e.g., by the correspondent copula), which is fully entangled in the whole system. 

\begin{figure}[!ht]
    \centering
    \includegraphics[scale=0.55]{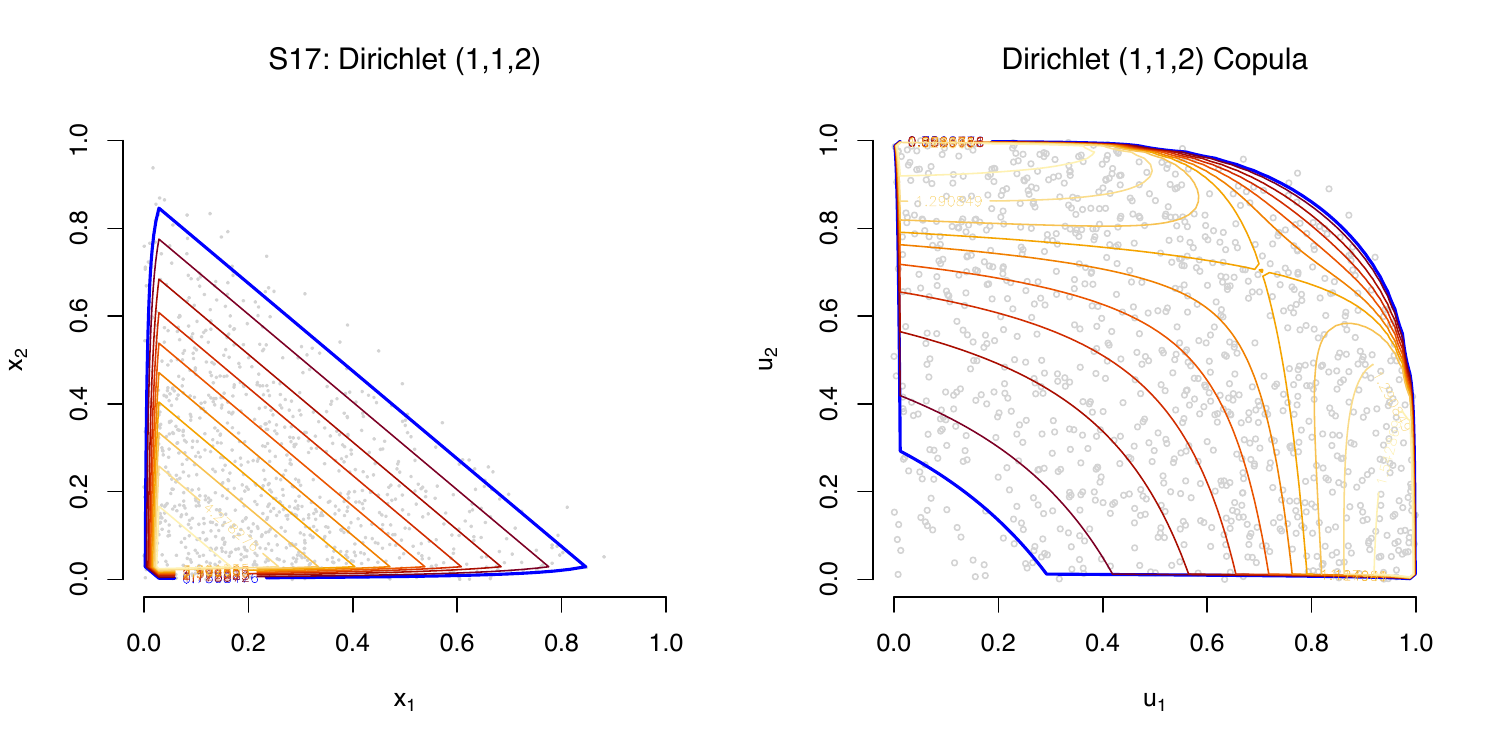}
    \caption{Contours of a Dirichlet(1,1,2) density (left) and of its induced copula density (right). The $95\%$ HDR is represented by the blue contour.}
    \label{fig: dirichlet}
\end{figure}

The Dirichlet distribution represents a natural candidate for analyzing compositional data, since its support is the simplex. Let $X \sim \mathcal{D}(\alpha_1, \alpha_2, \alpha_3)$ denote a Dirichlet random vector defined on the $2$-dimensional simplex, where $\alpha_j > 0$, for $j=1,2,3$; we refer to Chapter XI in \cite{devroye_non-uniform_1986} for a detailed exposition. Up to a normalizing constant, the density of $X$ is given by
%Briefly, given a parameter vector $\boldsymbol{\alpha} = (\alpha_1,\alpha_2,\alpha_3)$, with $\alpha_j > 0,\ j=1,2,3$, the random vector $X$ is said to follow a Dirichlet distribution if its joint density has a representation given by: 
\begin{align*}
    f(x_1, x_2; \alpha_1, \alpha_2, \alpha_3) \propto x_1^{\alpha_1-1}x_2^{\alpha_2-1}(1-x_1-x_2)^{\alpha_3-1},\quad x_1, x_2 \in [0,1]; x_1 + x_2 \leq 1.
\end{align*}
In the $2$-dimensional simplex, when $\alpha_1 = 1, \alpha_2 = 1, \alpha_3 = a$, the associated copula density has the following expression up to a normalizing constant: %and regardless of the value of the positive scalar $a$:
\begin{align*}
    c\left(u_1, u_2; \alpha_1 = 1, \alpha_2 = 1, \alpha_3 = a\right) \propto \frac{\left[ (1-u_1)^\frac{1}{a+1} + (1-u_2)^\frac{1}{a+1} - 1\right]^{a-1}}{\left[(1-u_1)(1-u_2)\right]^\frac{a}{a+1}},
\end{align*}
with $u_1, u_2$ such that $u_1, u_2 \in [0,1]$ and $(1-u_1)^\frac{1}{a+1} + (1-u_2)^\frac{1}{a+1} \geq 1$. The analytical derivation is given in Supplementary Material A, while its graphical representation--in terms of a set of random draws and different level sets--is illustrated in Figure~\ref{fig: dirichlet}.

To evaluate the proposed measures in this specific scenario, one first needs to perform an accurate tuning of their hyperparameters. As shown in the Supplementary Material B, the optimal choice of $k$--which represents the optimal number of neighbors to take into account for computing both \textbf{$M_1$:$k$NN-Eucl} and \textbf{$M_2$:$k$NN-CDF}--remains the same as for the other scenarios: $k = \round{\sqrt{n/2}}$.  However, for measures depending on $\epsilon$--which defines the length, area, or volume of the neighborhood--the optimal choice depends on the support of the underlying variable. Being defined on a simplex, in a Dirichlet scenario, the optimal values of $\epsilon$ have a smaller magnitude compared to noncompact or less restrictive scenarios such as S1-S16. In this case, the following choices are made for $\epsilon$, guided by simulation studies reported in the Supplementary Material B:
\begin{description}
    \item[$M_3$:$\epsilon$-CDF] Performances are robust to sample size, with the empirical optimal $\epsilon = 0.10$.
    \item[$M^{\text{NPCop}}_3$:$\epsilon$-CDF] Performances depend on the sample size, according to a heuristic given by: $\epsilon = \exp\left(-1.22-0.23\log{n} \right)$. This relationship is obtained by fitting a nonlinear regression (exponential decay model), with the empirical optimal $\epsilon$ and the sample size $n$ as dependent and independent variables, respectively.
    \item[$M^{\text{PCop}}_3$:$\epsilon$-CDF] Performances are robust to sample size, with the empirical optimal $\epsilon = 0.02$.
\end{description}

\begin{figure}[!ht]
    \centering
    \includegraphics[scale=0.63]{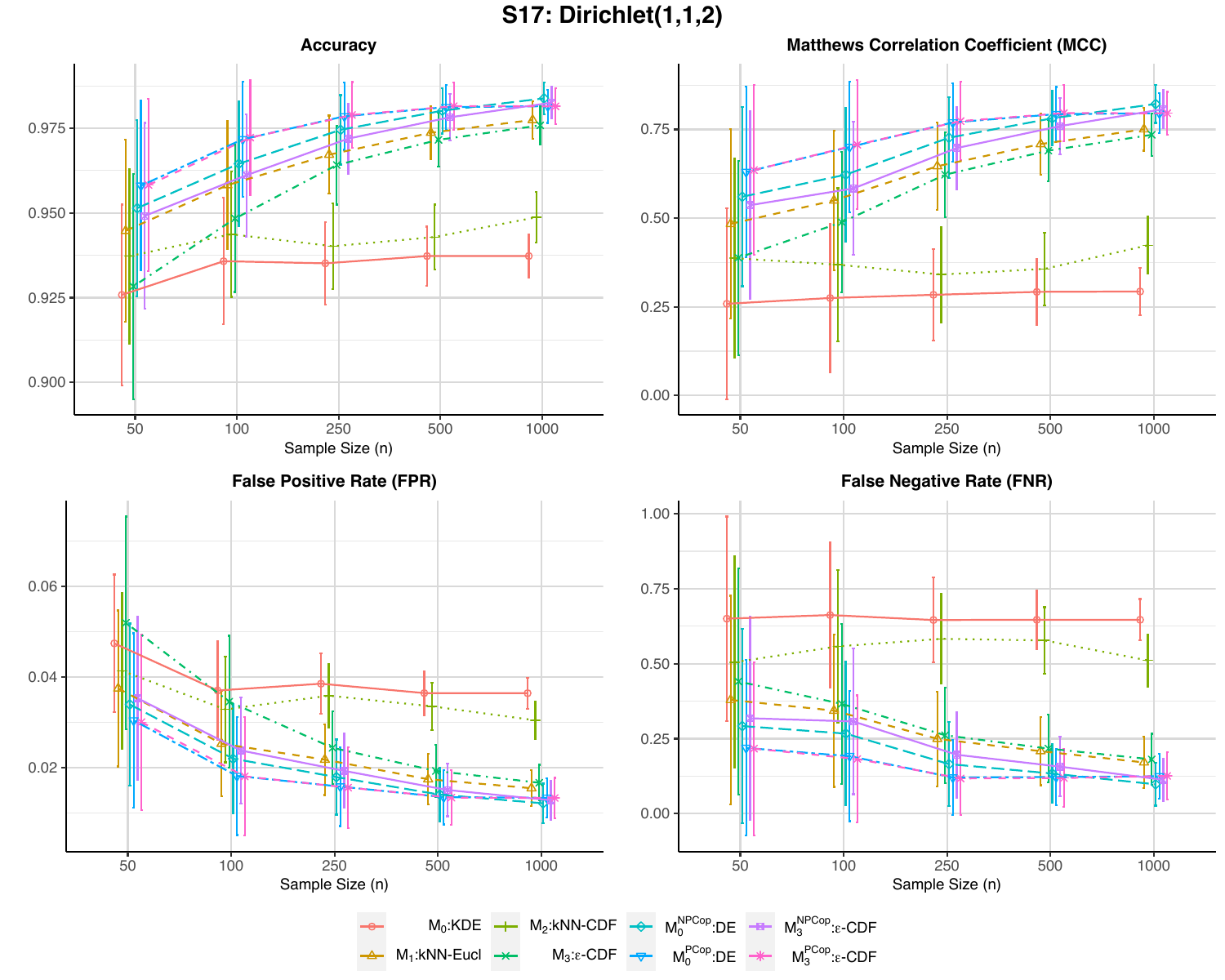}
    \caption{Performance results of the compared methods in the Dirichlet scenario for varying sample sizes. Data are summarized in terms of mean and error bounds across $1000$ independent Monte Carlo (MC) replicates.}
    \label{fig: dirichlet_errors}
\end{figure}

In terms of results, as suggested in Table~\ref{tab: performance_results_main}, the more complex the scenario, the more difficult one should expect it to be to identify the highest-density points (or \textit{true negatives}) versus the \textit{true positives}. In this particular case, as shown in Figure~\ref{fig: dirichlet_errors}, the discrepancy between the different measures is remarkable, with persistently high FNR values, even when the sample size increases, for \textbf{$M_0$:KDE} and \textbf{$M_2$:$k$NN-CDF}. All other measures improve with the sample size, achieving results comparable to the other scenarios.  

\subsection{MAGIC data} \label{sec: magic_data}
We now apply the proposed measures to derive an HDR for the joint distribution of two selected variables from the MAGIC dataset~\citep{bock_methods_2004}. These data simulate the registration of high-energy gamma particles in a ground-based atmospheric Cherenkov gamma telescope, and have been studied in classification problems~\citep{dvorak_softening_2007}, as well as to analyze the dependence structure of some of the characterizing variables~\citep[see e.g.,][]{nagler_evading_2016,grazian_approximate_2022}.

\begin{figure}[!ht]
    \centering
    \includegraphics[scale=0.66]{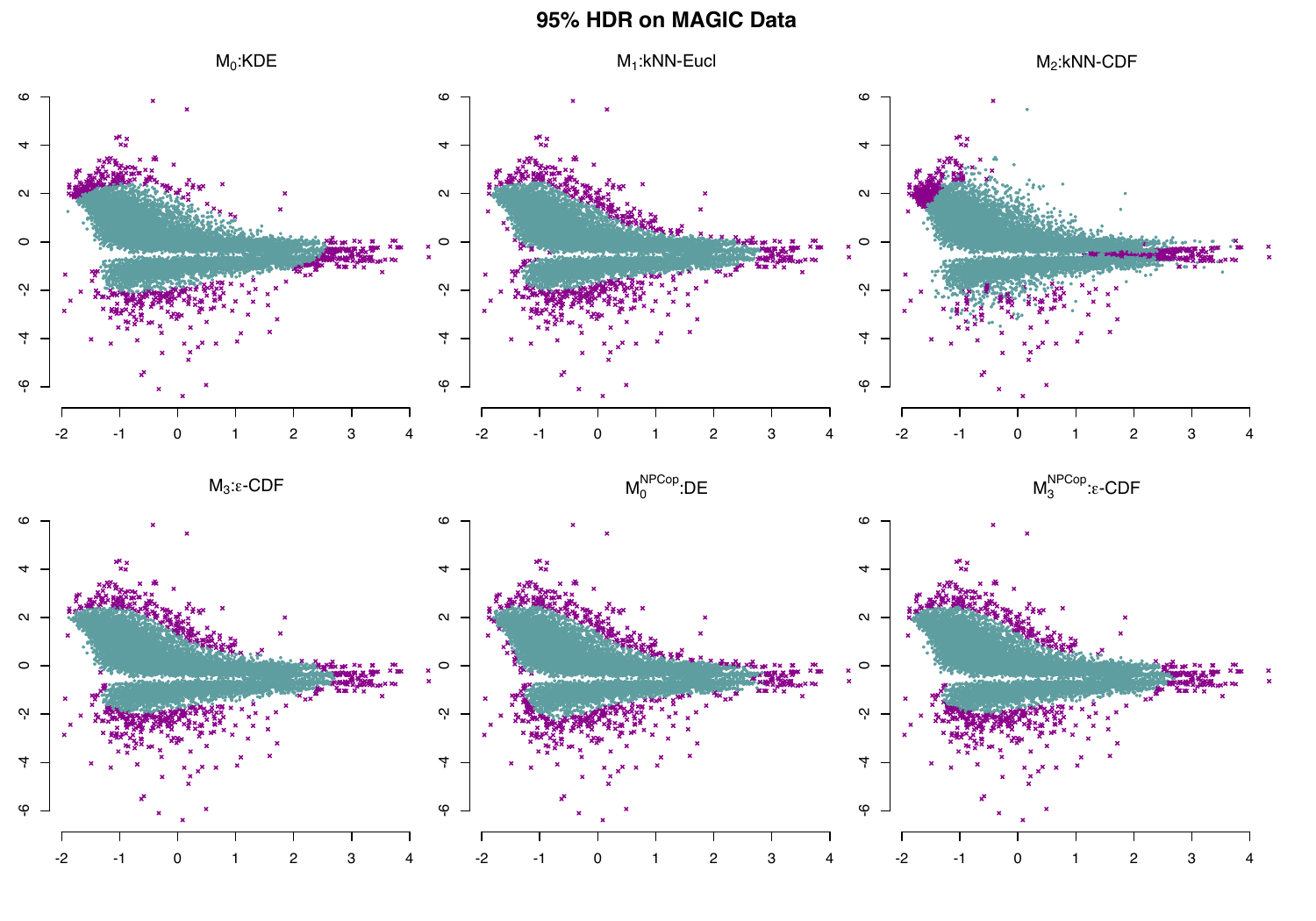}
    \caption{Estimated 95\% HDR of two scaled variables (``fConc1'' on x-axis and ``fM3Long'' on y-axis) of the MAGIC dataset. Only nonparametric measures are evaluated. Cadet-blue points define the estimated 95\% HDR; in contrast, purple points are those lying outside the HDR.}
    \label{fig: magic}
\end{figure}

In this evaluation, we focus on gamma-ray observations (overall $n=12,332$) and consider the two variables ``fConc1'' and ``fM3Long'', after scaling them. We refer to~\cite{bock_methods_2004} for a full description of the dataset. In this case (as deduced from the complex structure of the data; see Figure~\ref{fig: magic}), the parametric approach is inappropriate for both the estimation of the marginal distribution and, more importantly, the copula model. Thus the 95\% HDR is estimated using the nonparametric measures only. Although in the absence of the underlying truth it is not possible to perform a reliable evaluation, it seems that the two nonparametric copula-based approaches (\textbf{$M^{\text{NPCop}}_0$:DE} and \textbf{$M^{\text{NPCop}}_3$:$\epsilon$-CDF}), as well as the distance-based \textbf{$M_1$:$k$NN-Eucl} and \textbf{$M_3$:$\epsilon$-CDF}, more sensibly exclude tail data points (which may be expected to have a lower density) from the HDR. 

\begin{figure}[!ht]
    \centering
    \includegraphics[scale=0.7]{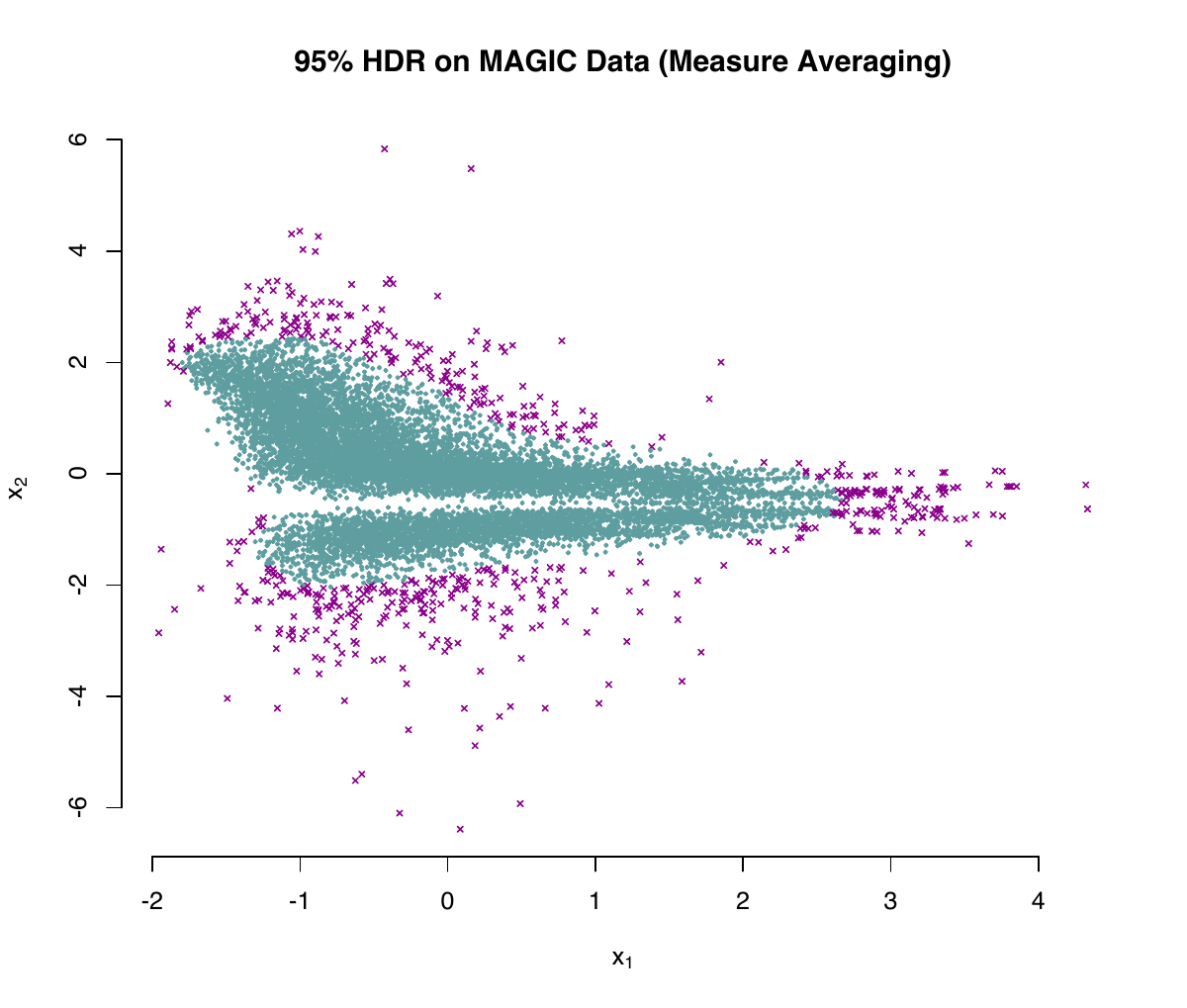}
    \caption{Estimated 95\% HDR based on a ``measure averaging'' approach of two scaled variables (``fConc1'' on x-axis and ``fM3Long'' on y-axis) of the MAGIC dataset. Cadet-blue points define the estimated 95\% HDR; in contrast, purple points are those lying outside the HDR.}
    \label{fig: magic_MA}
\end{figure}

To deal with estimation and data uncertainty, we also explore the potential of a ``measure averaging'' in a similar, but simplified, manner to \textit{model averaging}~\citep[see e.g.,][]{hoeting_bayesian_1999,hjort_frequentist_2003}. By forming a consensus between the different available measures, one would expect the resulting averaged HDR to be more robust than the individual estimates. Specifically, in Figure~\ref{fig: magic_MA}, we illustrate a 95\% HDR formed by averaging across the different nonparametric measures: the estimated HDR is defined by the set of points identified as highest-density points by more than half of the evaluated metrics. Notably, when comparing the resulting HDR in Figure~\ref{fig: magic_MA} and those obtained by the individual measures (Figure~\ref{fig: magic}), we identify \textbf{$M_3$:$\epsilon$-CDF} and \textbf{$M^{\text{NPCop}}_3$:$\epsilon$-CDF} as the regions better resembling the average (both with < 1\% classification difference). Therefore, these may result in a possible superior ability to detect anomalous values in scenarios where a parametric assumption is unrealistic. 

\section{Discussion and conclusions}  \label{sec: conclusions}
In this work, we have discussed some generalizations of the standard density-based approach for estimating highest-density regions, using neighborhood measures. Various measures, with distinct properties, have been introduced and compared to the classical kernel density estimator, across different scenarios and sample sizes. Furthermore, the use of copula models for representing a multivariate distribution through a combination of its marginals and their dependence structure has been investigated. Our results suggest that such a generalized approach may provide great advantages to HDR estimation when considering several alternative measures to KDE, especially those based on copulae. In fact, compared to traditional KDE, copula-based HDR resulted in greater accuracy and lower FPR and FNR in a number of simulation scenarios and possibly in real data (as suggested by the MAGIC data application). Such performances are particularly important when the interest is in balancing different types of errors, minimizing both \textit{false positives} and \textit{false negatives}, and maximizing therefore the probability of detecting atypical or anomalous values. In a doping detection problem, for example, this would directly translate into an enhanced ability of identifying doping abuse. 

It is important to emphasize that among the various copula-based measures, the parametric ones generally outperformed the non-parametric measures. This outcome was expected because all the considered scenarios, despite their complexity, were generated from parametric models and were evaluated under the assumption of no model misspecification for the marginals. In real-world applications, one would first need to select the most appropriate family using a criterion of fit, such as the Akaike Information Criterion (AIC) or the Bayesian Information Criterion (BIC). Notably, an important advantage of using copulae is that marginal model selection is implemented separately for each marginal, restricting the analysis to simpler univariate cases. Concerning the copula model, we account for potential misspecification, with model selection performed using the AIC (note that built-in functions are available in R software, e.g., the \texttt{BiCopSelect()} function from the \texttt{VineCopula} package~\citep{vinecop_R2023}).   

Relative to the scenarios and measures considered in this work, the main recommendation for the neighborhood measure is thus a copula-based alternative. In particular, parametric variants should be preferred when distributions show a pattern traceable to a parametric family. When parametric measures do not appear appropriate for the data at hand, we would recommend the nonparametric copula-based options, with the exception of scenarios with heavy-tailed marginals. Here, \textbf{$M_1$:$k$NN-Eucl} achieves the best results, and the nonparametric \textbf{$M^{\text{NPCop}}_0$:DE} measure is among the most inferior ones, especially in small samples. On the contrary, the \textbf{$M^{\text{NPCop}}_3$:$\epsilon$-CDF} measure using the $\epsilon$-neighborhood multivariate CDF distance shows promising results. Since the main characteristic of the latter is that it is based on the CDF rather than the PDF, it may suggest that focusing on estimating the former may be advantageous. Similar findings are reported by~\cite{magdon-ismail_density_2002}, who acknowledge that, compared to directly estimating the PDF with e.g., KDE, approximating the CDF is less sensitive to statistical fluctuations and its convergence rate is faster than the convergence rate of KDE methods. We also emphasize that although heuristic considerations for each specific measure are made in terms of their hyperparameters, e.g., the choice of $k$ in the kNN-based approaches, further work may investigate the existence of optimal theoretical values in a similar fashion to the asymptotic works on the optimal bandwidth choice in KDE~\citep[see e.g.,][]{wand_multivariate_1994,chacon_asymptotics_2011}. 

In this work, we focused on estimating HDR for continuous distributions that are dominated by the Lebesgue measure. Furthermore, despite providing a general multidimensional framework for building HDRs, we evaluated the proposed approach in a bivariate context. Future lines of research may examine the problem in the underappreciated setting of discrete probability distributions and in higher dimensions. In the first case, possible connections could be made with the work of~\cite{oneill_smallest_2022}, which established some theory and an algorithm for HDRs for discrete distributions. In the second case, and more specifically with reference to copula-based measures, our aim is to explore the use of \textit{vine copulae}~\citep{nagler_evading_2016} to construct flexible dependence models for an arbitrary number of variables using only bivariate building blocks. We expect, in fact, to see remarkable advantages in using copulae over an increased number of variables, as the extension of the common KDE to high dimensions has proven challenging in terms of both computational efficiency and statistical inference. 

\bibliographystyle{apalike}
\bibliography{main.bib}

\section*{Supplementary Material}

Additional supporting information is provided in this Supplementary material. The implementation of the algorithms is possible with the R package \texttt{HDR2D} provided in the Github repository \href{https://github.com/nina-DL/HDR2D}{https://github.com/nina-DL/HDR2D}.

\appendix

\section{Derivation of the Dirichlet copula} \label{app: proofs}

In this section, we provide the analytical derivation of the Dirichlet copula in the specific case of a $K-1$-dimensional simplex distribution, with $K=3$. Let $X$ be a Dirichlet random vector, denoted by $X \sim \mathcal{D}(\boldsymbol{\alpha})$, where $ \boldsymbol{\alpha} = (\alpha_1, \alpha_2, \alpha_3)$ and $\alpha_j > 0,\ j=1,2,3$. Its joint density is given by
\begin{align*}
    f(x_1, x_2; \boldsymbol{\alpha}) = \frac{1}{B(\boldsymbol{\alpha})}x_1^{\alpha_1-1}x_2^{\alpha_2-1}(1-x_1-x_2)^{\alpha_3-1},\quad x_1, x_2 \in [0,1]; x_1 + x_2 \leq 1,
\end{align*}
where $B(\cdot)$ is the multivariate Beta function defined in terms of the Gamma function $\Gamma(\cdot)$ as:
\begin{align*}
    B(\boldsymbol{\alpha}) = \frac{\prod_{j=1}^K\Gamma(\alpha_j)}{\Gamma\left(\sum_{j=1}^K\alpha_j\right)} = \frac{\Gamma(\alpha_1)\Gamma(\alpha_2)\Gamma(\alpha_3)}{\Gamma (\alpha_1+\alpha_2+\alpha_3)}.
\end{align*}

Using the copula representation given in Eq. (6) in the main paper, we have that
\begin{align*}
    f(x_1, x_2; \boldsymbol{\alpha}) = f_1(x_1; \boldsymbol{\alpha})\ f_2(x_2; \boldsymbol{\alpha})\ c\left(u_1, u_2; \boldsymbol{\alpha}\right),
\end{align*}
where $u_1 = F_1(x_1)$ and $u_2 = F_2(x_2)$.

The marginals of a Dirichlet random vector are known to follow a Beta $X_j \sim \mathcal{B}eta\left(\alpha_j, \sum_{i \neq j} \alpha_i\right)$, for $j = 1,2$. Thus, using the copula representation we have that:
\begin{align} \label{eq: c}
    c\left(u_1, u_2; \boldsymbol{\alpha}\right) &= \frac{f(x_1, x_2; \boldsymbol{\alpha})}{f_1(x_1; \boldsymbol{\alpha})\ f_2(x_2; \boldsymbol{\alpha})} \nonumber\\ &= \frac{\cancel{\Gamma(\alpha_1+\alpha_2+\alpha_3)}\cancel{\Gamma(\alpha_1)}\cancel{\Gamma(\alpha_2)}\Gamma(\alpha_2+\alpha_3)\Gamma(\alpha_1+\alpha_3)\ \cancel{x_1^{\alpha_1-1}}\cancel{x_2^{\alpha_2-1}}(1-x_1-x_2)^{\alpha_3-1}}{\cancel{\Gamma(\alpha_1)}\cancel{\Gamma(\alpha_2)}\Gamma(\alpha_3)\Gamma(\alpha_1+\alpha_2+\alpha_3)\cancel{^2}\ \cancel{x_1^{\alpha_1-1}} (1-x_1)^{\alpha_2+\alpha_3-1} \cancel{x_2^{\alpha_2-1}} (1-x_2)^{\alpha_1+\alpha_3-1}}\nonumber\\
   & =\frac{(1-x_1-x_2)^{\alpha_3-1}}{m \cdot (1-x_1)^{\alpha_2+\alpha_3-1}(1-x_2)^{\alpha_1+\alpha_3-1}}\nonumber\\
   & =\frac{(1-F_1^{-1}(u_1)-F_2^{-1}(u_2))^{\alpha_3-1}}{m \cdot (1-F_1^{-1}(u_1))^{\alpha_2+\alpha_3-1}(1-F_2^{-1}(u_2))^{\alpha_1+\alpha_3-1}},
\end{align}
where $m = \frac{\Gamma(\alpha_3)\Gamma(\alpha_1+\alpha_2+\alpha_3)}{\Gamma(\alpha_2+\alpha_3)\Gamma(\alpha_1+\alpha_3)}$ denotes the normalizing constant and $F_j^{-1}$ is the inverse CDF of $X_j$.

Assume now that $\alpha_1 = 1, \alpha_2 = 1, \alpha_3 = a$, with $a$ being a positive scalar. If $X_j$ is Beta distributed, then $F_j$ can be expressed in terms of the incomplete Beta function, denoted by $I_j$, which gives a closed form solution for both $u_1$ and $u_2$, and therefore $x_1$ and $x_2$:
\begin{align} \label{eq: xj}
    u_j = F_j(x_j) = I_j(x_j; \alpha_1 = 1, \alpha_2 = 1, \alpha_3  = a) &= 1-(1-x_j)^{a+1}, \quad j=1,2 \nonumber \\
    x_j &= 1 - (1-u_j)^{\frac{1}{a+1}},\quad j=1,2.
\end{align}

Taken together, Eq.~\eqref{eq: c} and Eq.~\eqref{eq: xj} give us, up to the normalizing constant $m$,
\begin{align*}
    c\left(u_1, u_2; \alpha_1 = 1, \alpha_2 = 1, \alpha_3  = a\right) \propto \frac{\left[ (1-u_1)^\frac{1}{a+1} + (1-u_2)^\frac{1}{a+1} - 1\right]^{a-1}}{\left[(1-u_1)(1-u_2)\right]^\frac{a}{a+1}}.
\end{align*}
Since $x_1, x_2 \in [0,1]$, and $x_1 + x_2 \leq 1$, we also have the following constraints on $u_1, u_2$, determined by Eq.~\eqref{eq: xj}: 
\begin{align*}
    u_1, u_2 \in [0,1],\quad (1-u_1)^\frac{1}{a+1} + (1-u_2)^\frac{1}{a+1} \geq 1.
\end{align*}

\section{Supporting Details for Hyperparameter Choices} \label{app: hyper_choice}

\begin{figure}[!ht]
    \centering
    \includegraphics[scale=0.65]{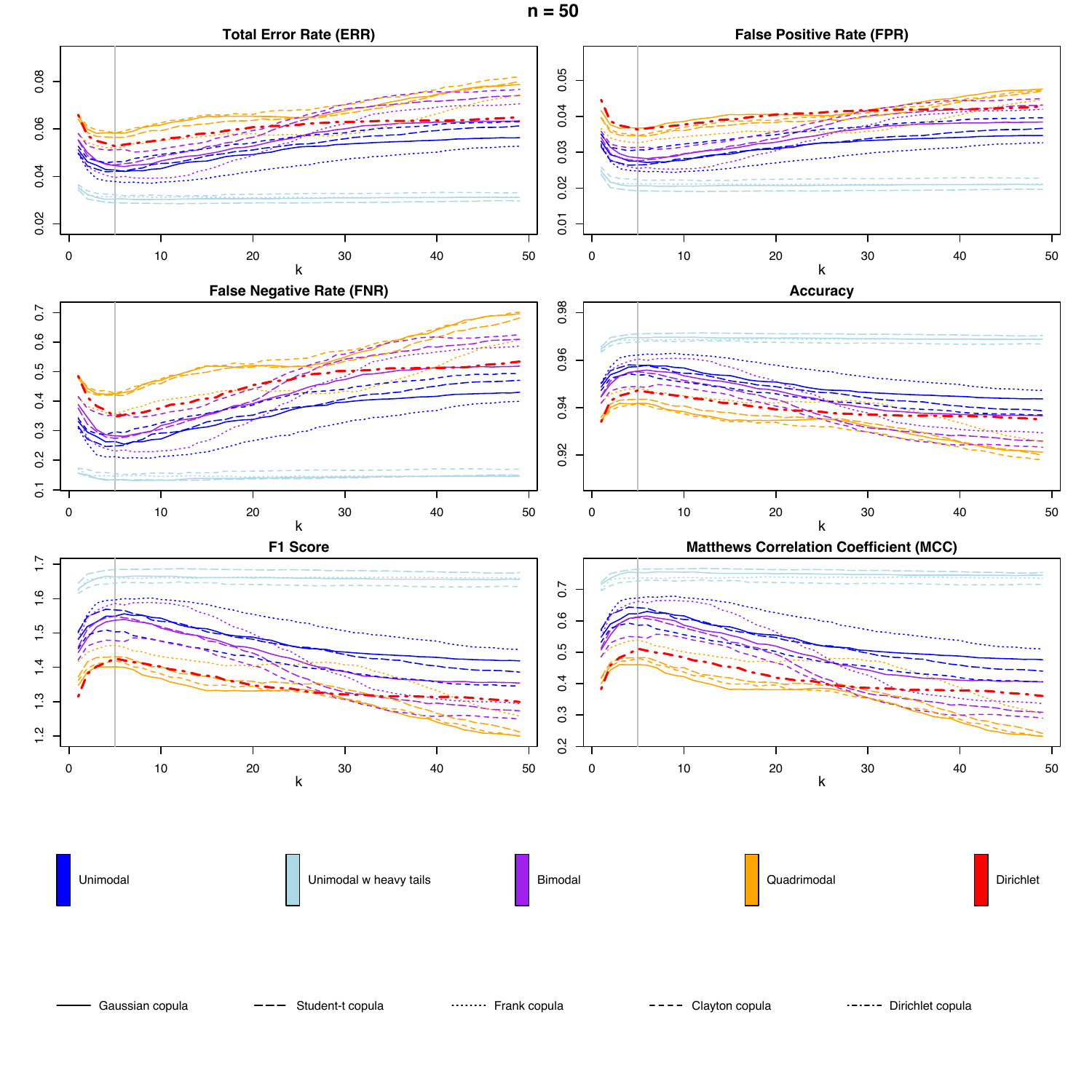}
    \caption{Performance results of \textbf{$M_1$:$k$NN-Eucl} in all evaluated scenarios for different values of the hyperparameter $k$, with $k \in \{1,2,\dots,50\}$. Gray vertical line refers to the rule of thumb for the optimal choice:  $k = \round{\sqrt{n/2}} = 5$, with $n = 50$.}
    \label{fig: choice_k_n50}
\end{figure}

\begin{figure}[!ht]
    \centering
    \includegraphics[scale=0.65]{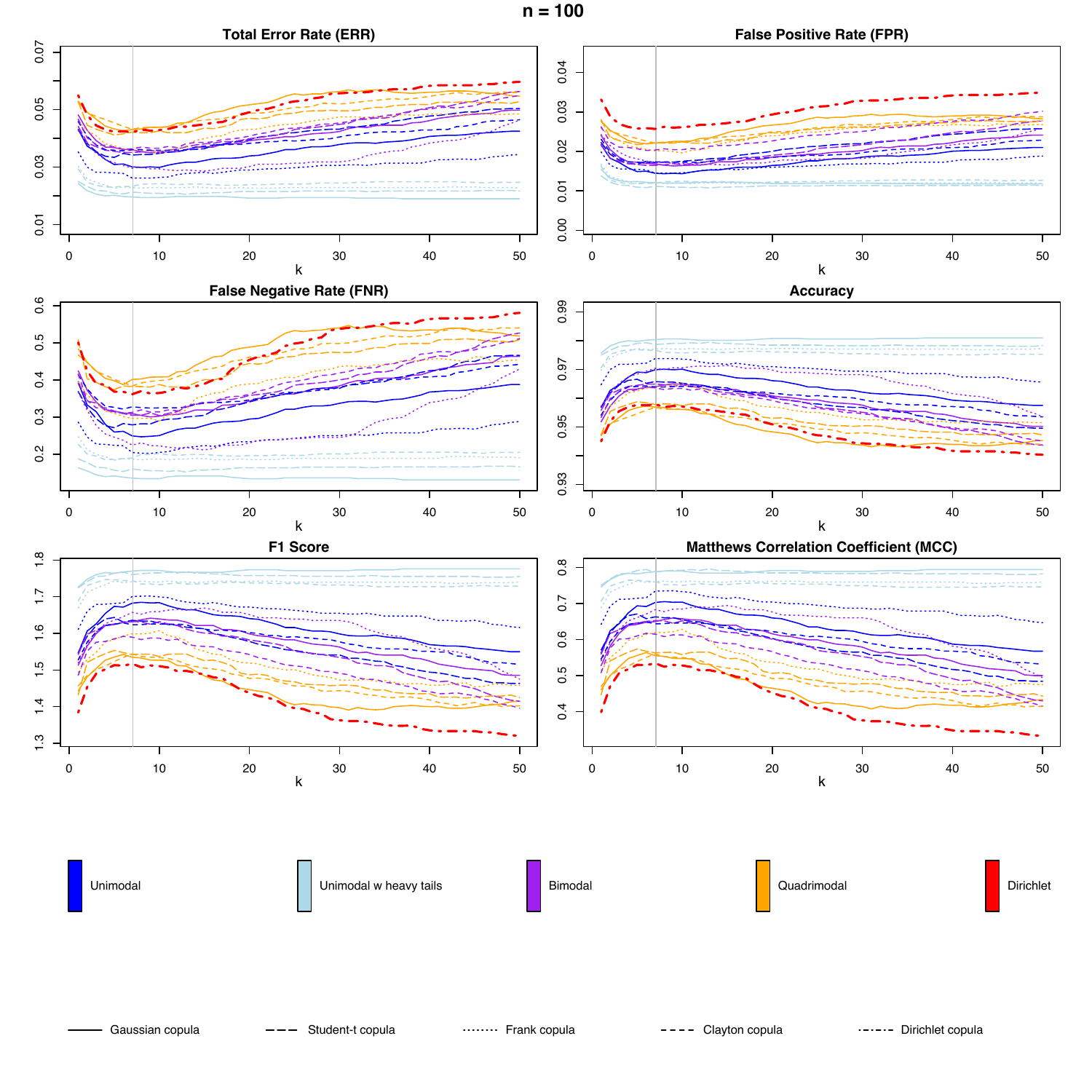}
    \caption{Performance results of \textbf{$M_1$:$k$NN-Eucl} in all evaluated scenarios for different values of the hyperparameter $k$, with $k \in \{1,2,\dots,50\}$. Gray vertical line refers to the rule of thumb for the optimal choice:  $k = \round{\sqrt{n/2}} = 7$, with $n = 100$.}
    \label{fig: choice_k_n100}
\end{figure}

In this section, we illustrate the choices we made for the hyperparameters $k$ and $\epsilon$ of the different measures evaluated in this paper, which we reiterate below. We support our choices with simulation studies in which we compare the performances of our measures for different values of the hyperparameters defined on an appropriate grid. The grid is defined on a sequence of generally equi-spaced points and takes into account the admissible choices of the hyperparameters $k$ and $\epsilon$ (that is, $k \in \{1, 2,\dots, n\}$ and $\epsilon \geq 0$) and the range of variation as well as the trend induced on the different performance metrics for different sample sizes. For example, in Figure~\ref{fig: choice_k_n50}, the grid for $k$ when $n = 50$ is defined on all integers from $1$ to $50$; in Figure~\ref{fig: choice_k_n100} with $n = 100$, we consider the same grid, as no improvements are expected for $k > 50$ given the observed trend. We let the grid vary with $n$ and the measure, but not with the different scenarios S1-S17. The only exception is made for $\epsilon$ in S17, since Dirichlet's support is limited to the simplex, and as such must be $\epsilon$. All the grids of reference are reported in the caption of the corresponding figure. All results for each hyperparameter choice are summarized in terms of their mean across $50$ independent Monte Carlo (MC) replicates.

\begin{figure}[!ht]
    \centering
    \includegraphics[scale=0.65]{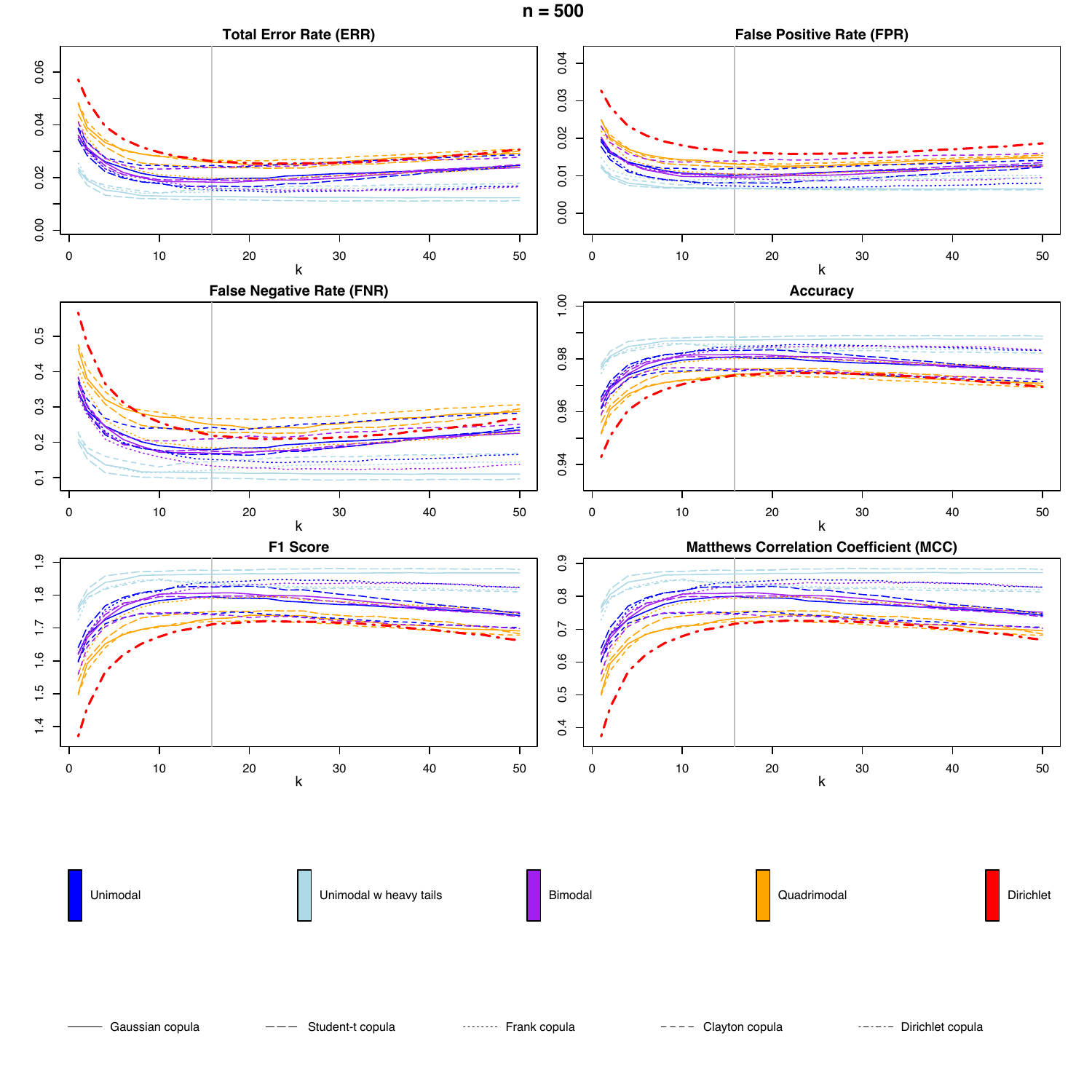}
    \caption{Performance results of \textbf{$M_1$:$k$NN-Eucl} in all evaluated scenarios for different values of the hyperparameter $k$, with $k \in \{1,2,\dots,50\}$. Gray vertical line refers to the rule of thumb for the optimal choice:  $k = \round{\sqrt{n/2}} = 16$, with $n = 500$.}
    \label{fig: choice_k_n500}
\end{figure}

\begin{figure}[!ht]
    \centering
    \includegraphics[scale=0.65]{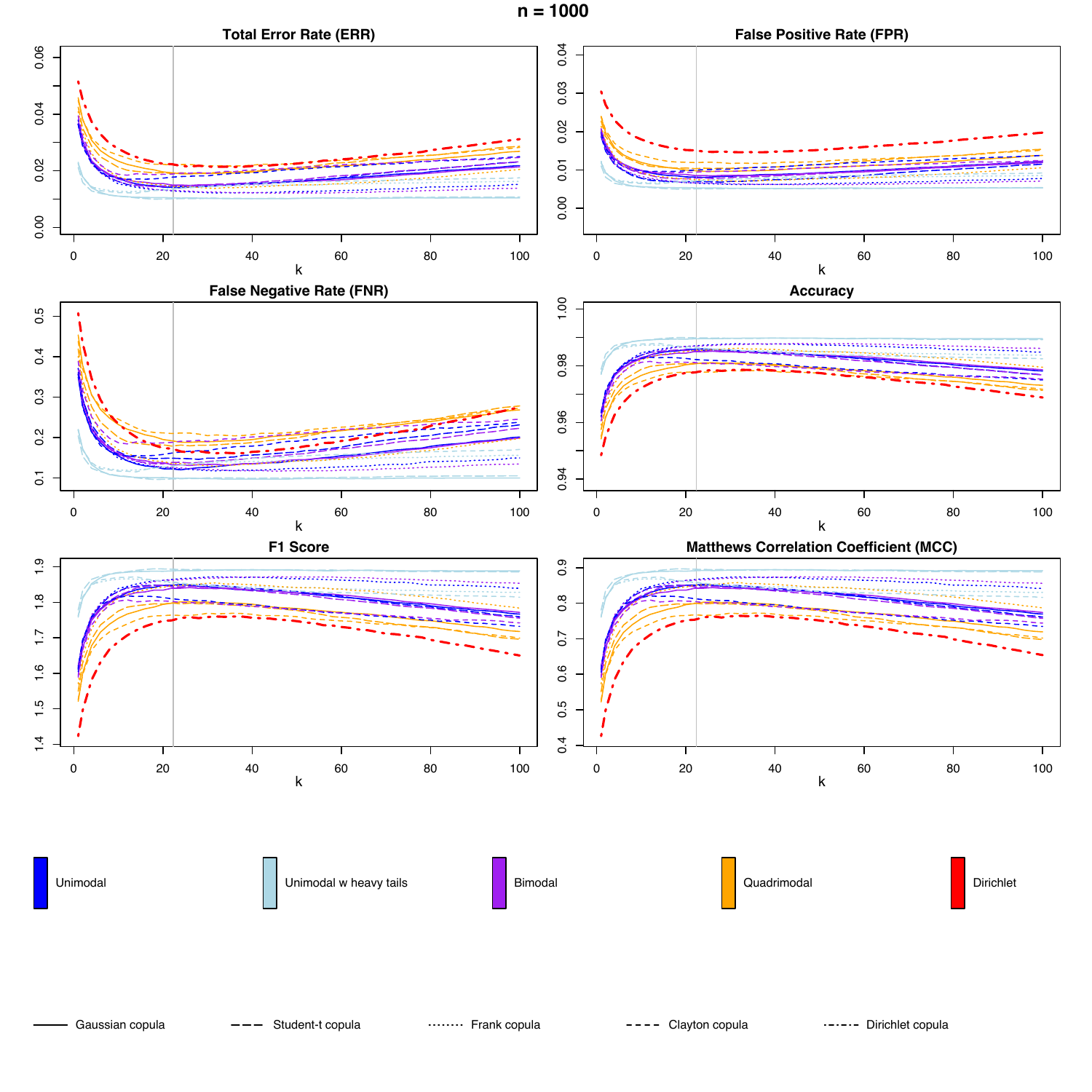}
    \caption{Performance results of \textbf{$M_1$:$k$NN-Eucl} in all evaluated scenarios for different values of the hyperparameter $k$, with $k \in \{1,2,4,6,\dots,98,100\}$. Gray vertical line refers to the rule of thumb for the optimal choice: $k = \round{\sqrt{n/2}} = 22$, with $n = 1000$.}
    \label{fig: choice_k_n1000}
\end{figure}

\begin{figure}[!ht]
    \centering
    \includegraphics[scale=0.65]{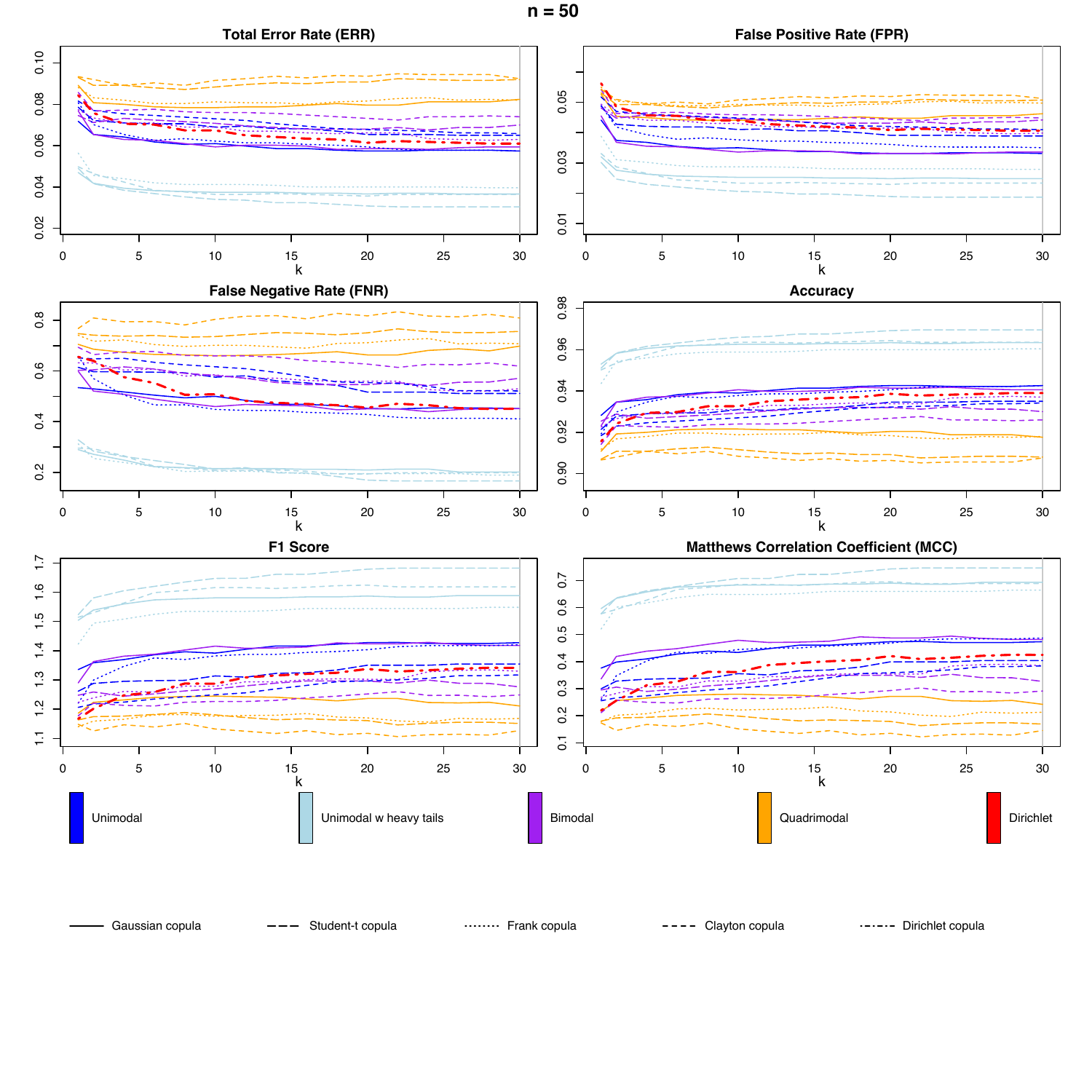}
    \caption{Performance results of \textbf{$M_2$:$k$NN-CDF} in all evaluated scenarios for $n = 50$ and for different values of the hyperparameter $k$, with $k \in \{1,2,4,6\dots,28,30\}$. Gray vertical line refers to the final choice:  $k = 30$.}
    \label{fig: choice_k_CDF_n50}
\end{figure}

\begin{figure}[!ht]
    \centering
    \includegraphics[scale=0.65]{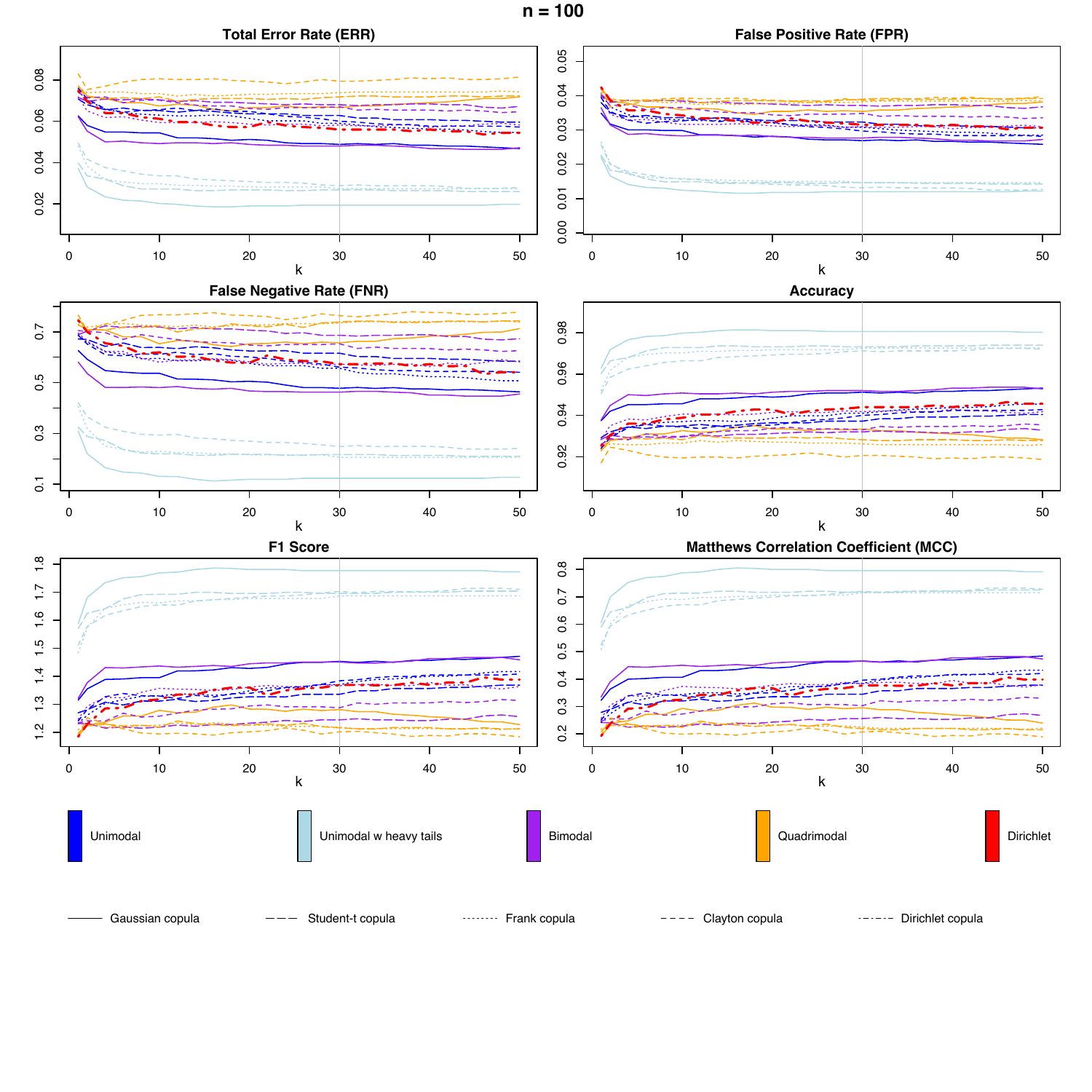}
    \caption{Performance results of \textbf{$M_2$:$k$NN-CDF} in all evaluated scenarios for $n = 100$ and for different values of the hyperparameter $k$, with $k \in \{1,2,4,6\dots,48,50\}$. Gray vertical line refers to the final choice:  $k = 30$.}
    \label{fig: choice_k_CDF_n100}
\end{figure}

\begin{figure}[!ht]
    \centering
    \includegraphics[scale=0.65]{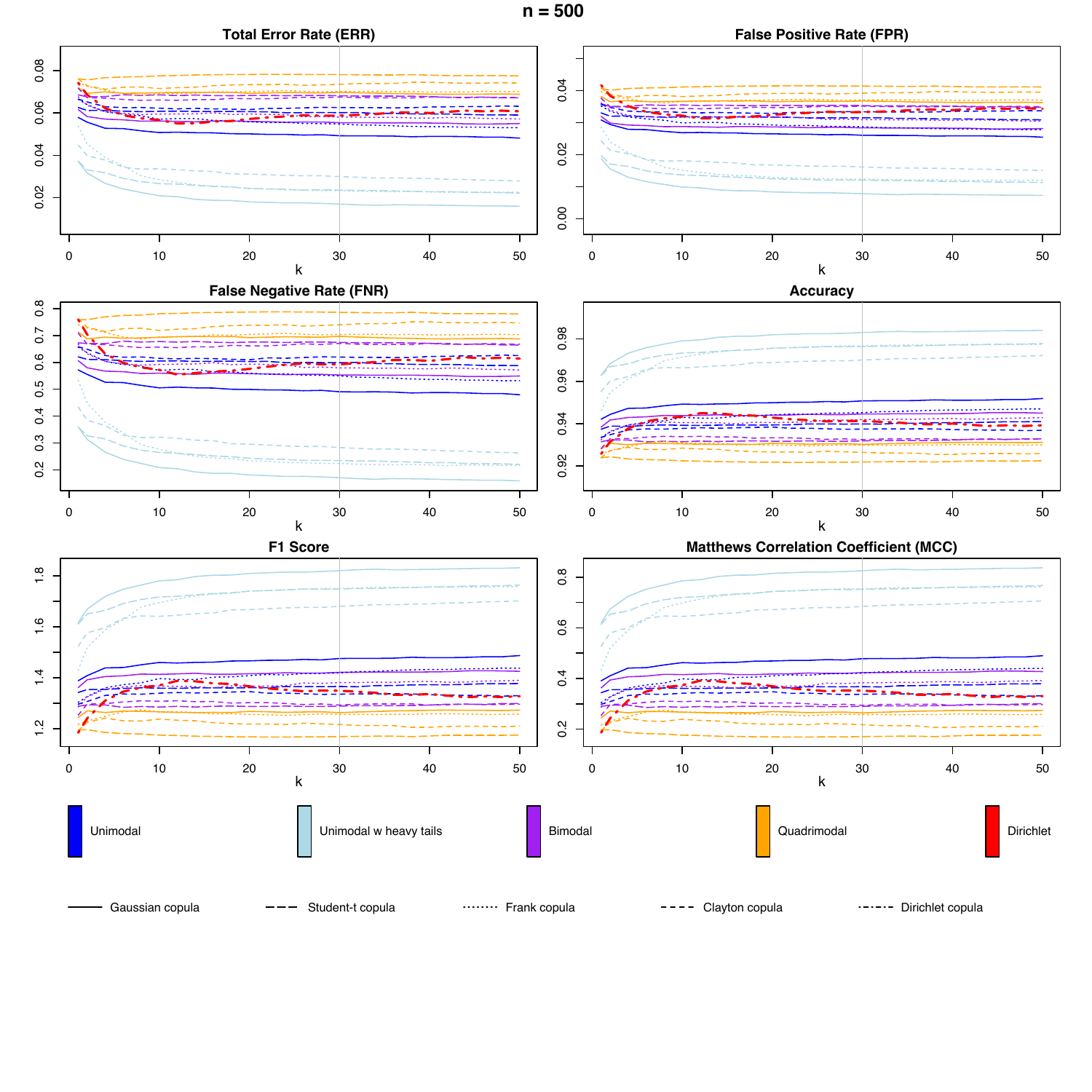}
    \caption{Performance results of \textbf{$M_2$:$k$NN-CDF} in all evaluated scenarios for $n = 500$ and for different values of the hyperparameter $k$, with $k \in \{1,2,4,6\dots,48,50\}$. Gray vertical line refers to the final choice:  $k = 30$.}
    \label{fig: choice_k_CDF_n500}
\end{figure}

\begin{figure}[!ht]
    \centering
    \includegraphics[scale=0.65]{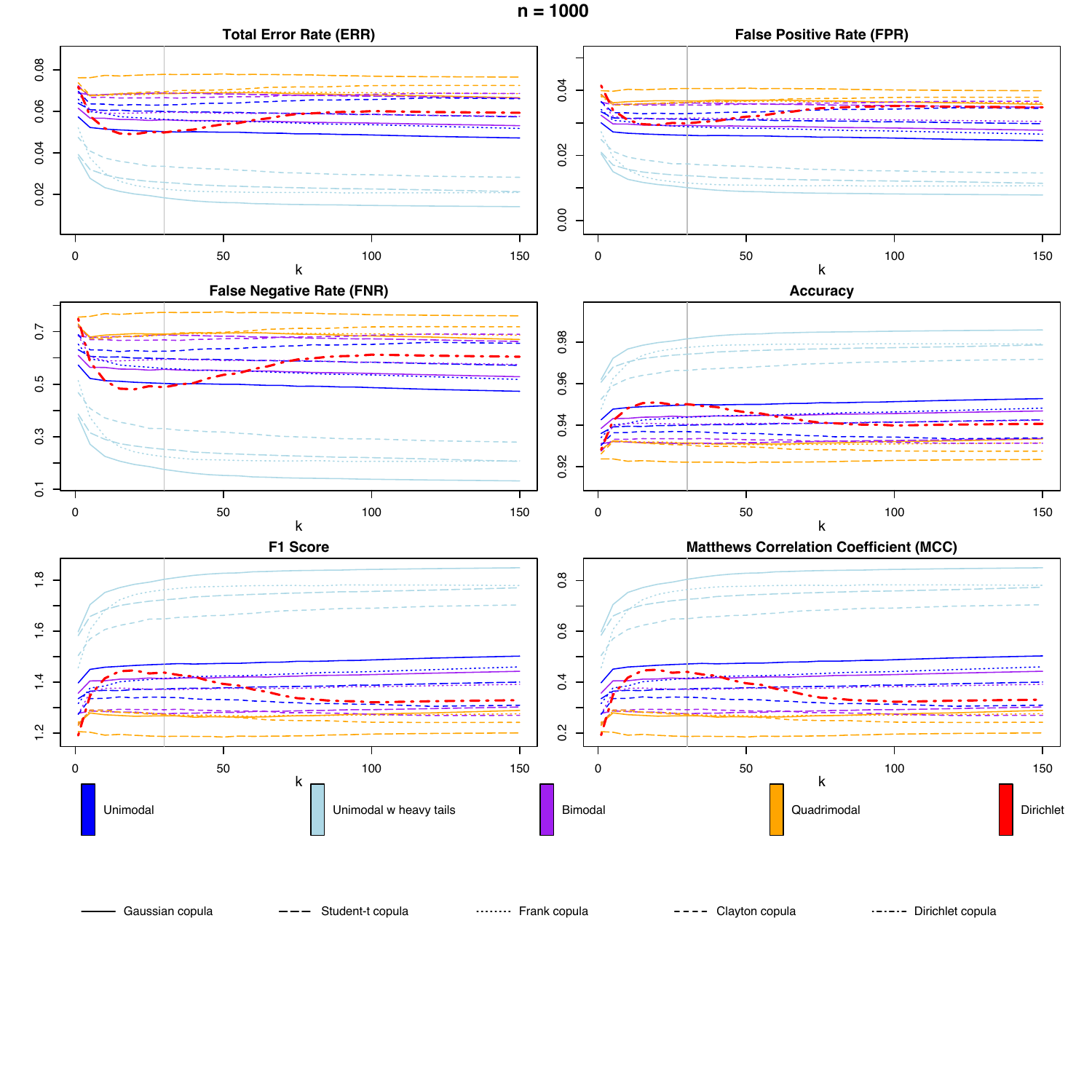}
    \caption{Performance results of \textbf{$M_2$:$k$NN-CDF} in all evaluated scenarios for $n = 1000$ and for different values of the hyperparameter $k$, with $k \in \{1,5,10,15\dots,145,150\}$. Gray vertical line refers to the final choice:  $k = 30$.}
    \label{fig: choice_k_CDF_n1000}
\end{figure}

\begin{figure}[!ht]
    \centering
    \includegraphics[scale=0.65]{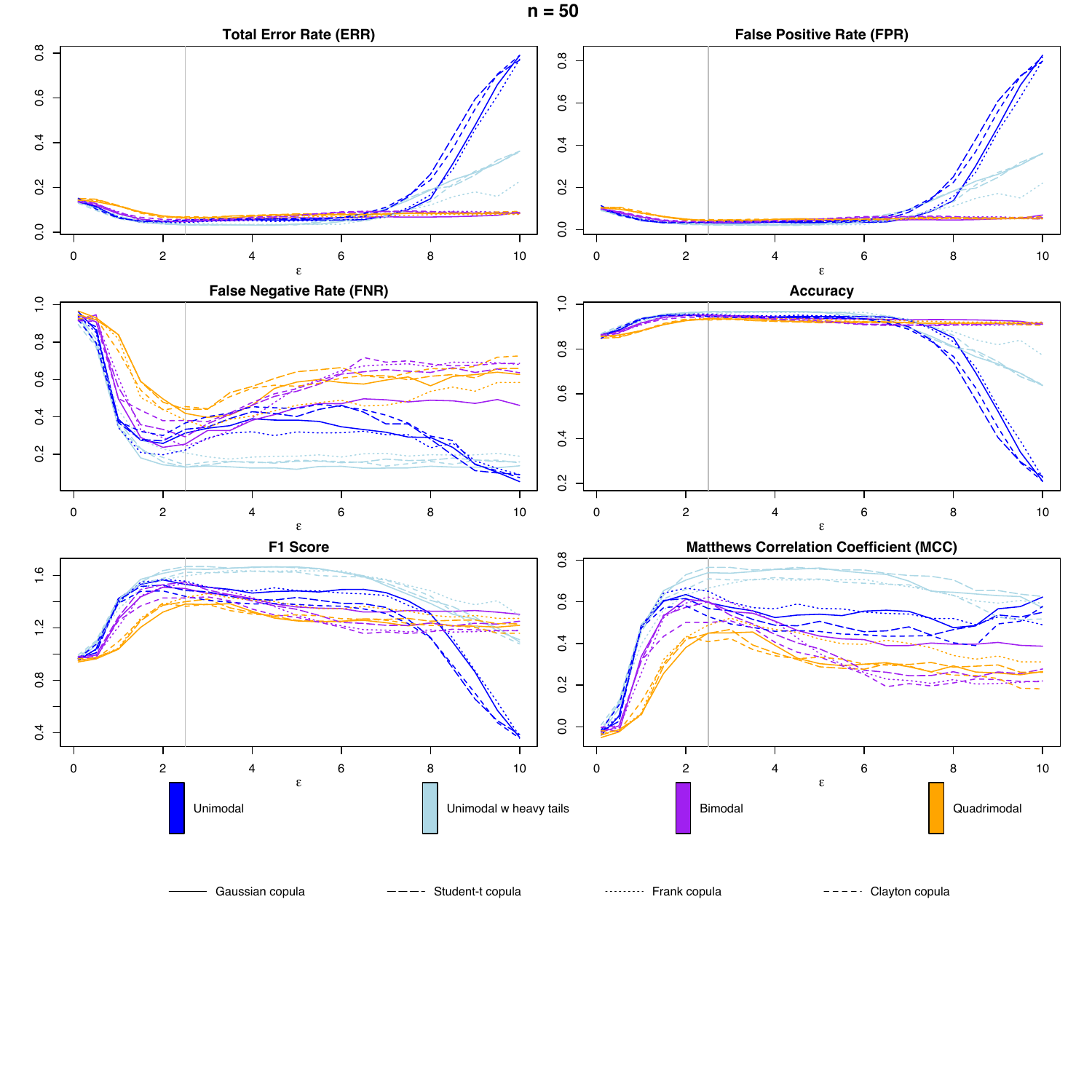}
    \caption{Performance results of \textbf{$M_3$:$\epsilon$-CDF} in all evaluated scenarios for $n = 50$ and for different values of the hyperparameter $\epsilon$, with $\epsilon \in \{0.1,0.5,\dots,9.5,10.0\}$. Gray vertical line reflects the following heuristic: $\epsilon = \exp\left(2.13-0.3\log{n}\right) = 2.60$.}
    \label{fig: choice_eps_CDFmv_n50}
\end{figure}

\begin{figure}[!ht]
    \centering
    \includegraphics[scale=0.65]{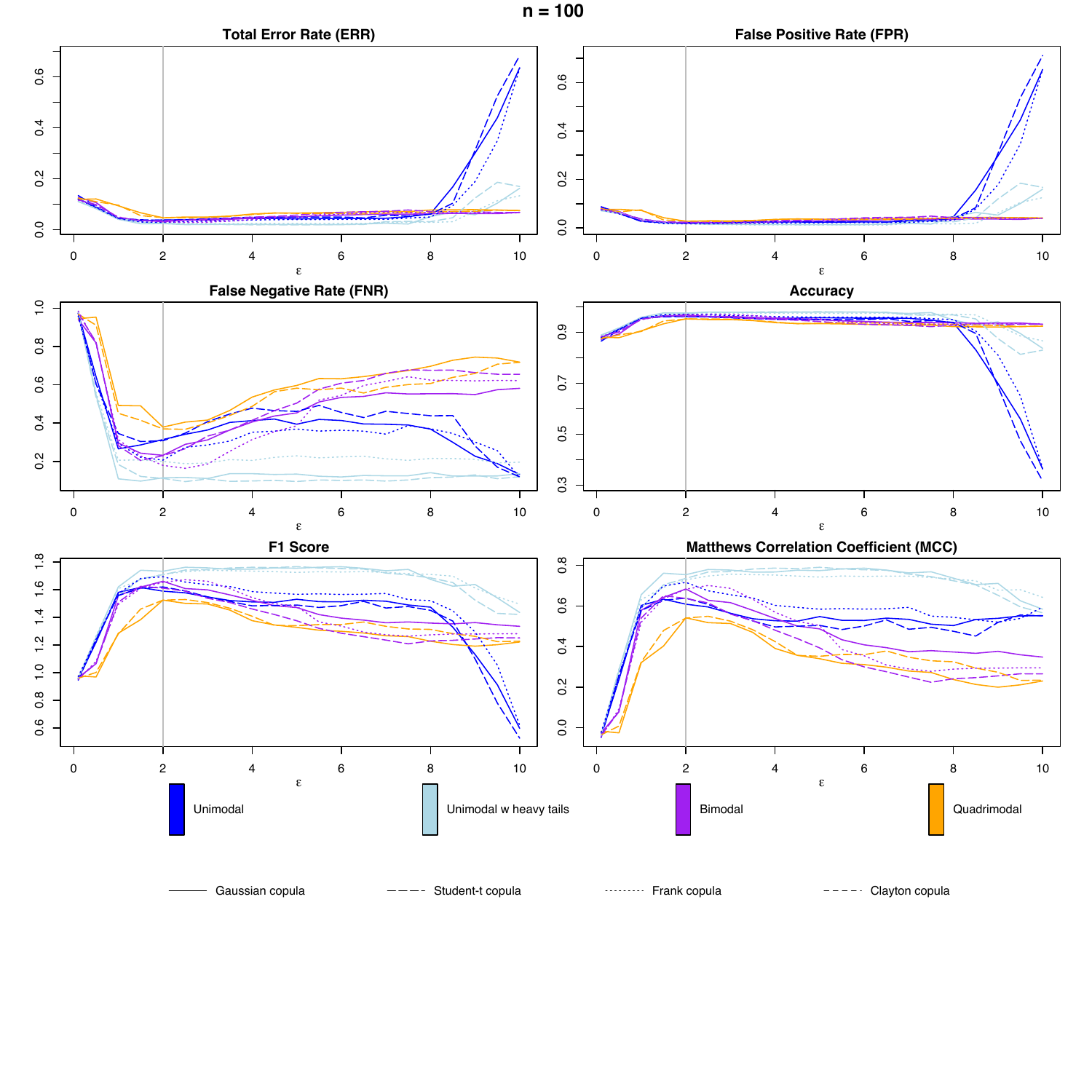}
    \caption{Performance results of \textbf{$M_3$:$\epsilon$-CDF} in all evaluated scenarios for $n = 100$ and for different values of the hyperparameter $\epsilon$, with $\epsilon \in \{0.1,0.5,\dots,9.5,10.0\}$. Gray vertical line reflects the following heuristic: $\epsilon = \exp\left(2.13-0.3\log{n}\right) = 2.11$.}
    \label{fig: choice_eps_CDFmv_n100}
\end{figure}

\begin{figure}[!ht]
    \centering
    \includegraphics[scale=0.65]{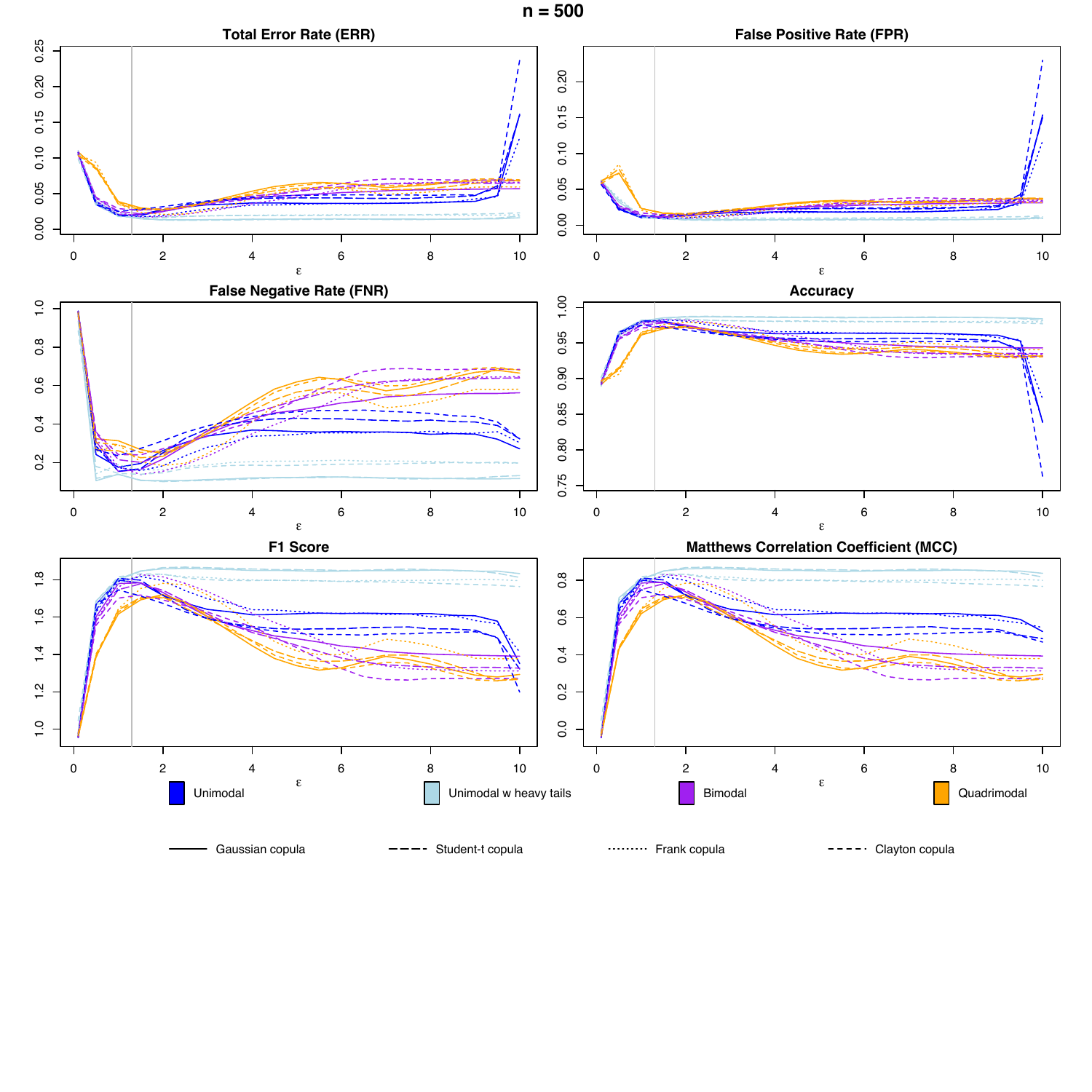}
    \caption{Performance results of \textbf{$M_3$:$\epsilon$-CDF} in all evaluated scenarios for $n = 500$ and for different values of the hyperparameter $\epsilon$, with $\epsilon \in \{0.1,0.5,\dots,9.5,10.0\}$. Gray vertical line reflects the following heuristic: $\epsilon = \exp\left(2.13-0.3\log{n}\right) = 1.30$.}
    \label{fig: choice_eps_CDFmv_n500}
\end{figure}

\begin{figure}[!ht]
    \centering
    \includegraphics[scale=0.65]{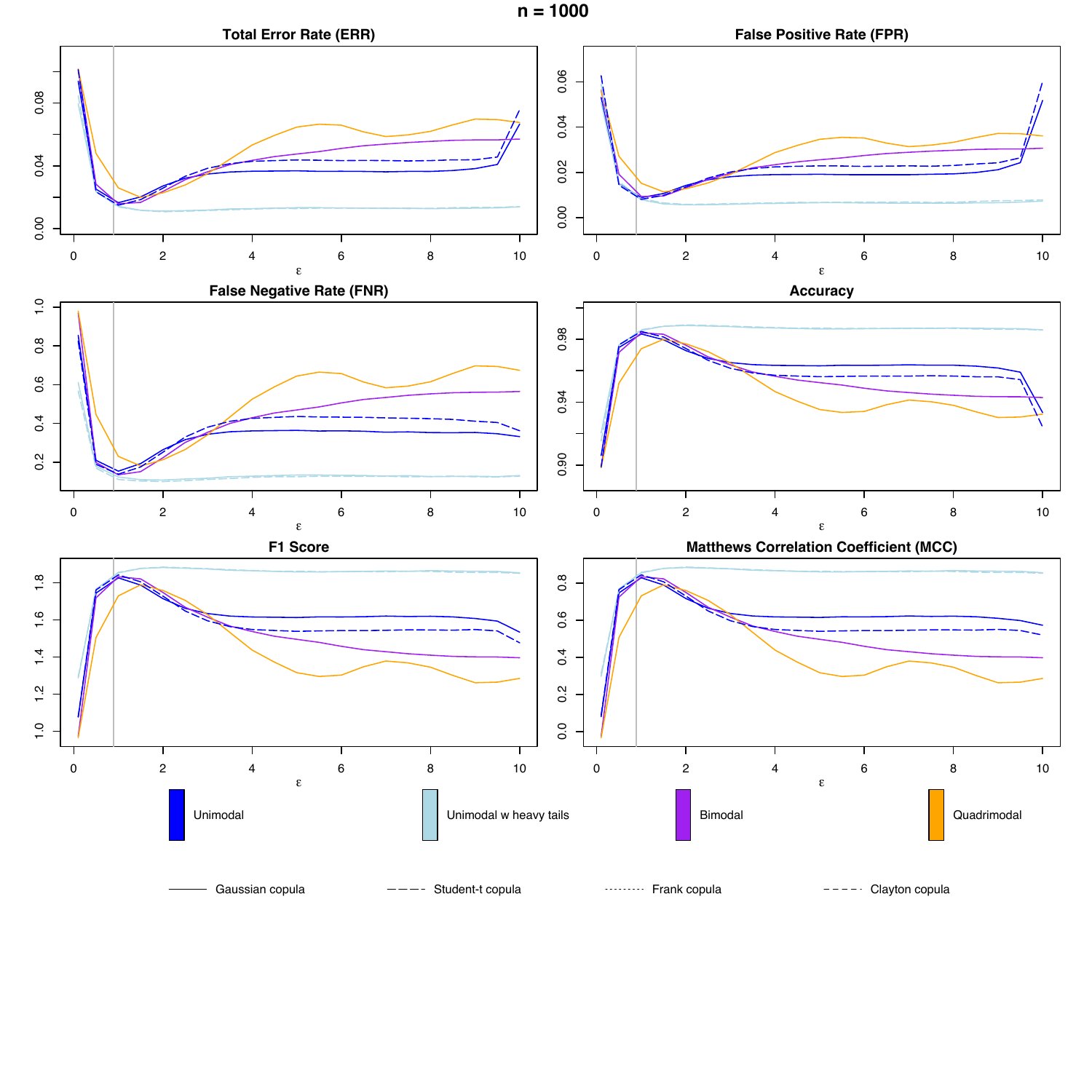}
    \caption{Performance results of \textbf{$M_3$:$\epsilon$-CDF} in all evaluated scenarios for $n = 1000$ and for different values of the hyperparameter $\epsilon$, with $\epsilon \in \{0.1,0.5,\dots,9.5,10.0\}$. Gray vertical line reflects the following heuristic: $\epsilon = \exp\left(2.13-0.3\log{n}\right) = 1.06$.}
    \label{fig: choice_eps_CDFmv_n1000}
\end{figure}

\begin{figure}[!ht]
    \centering
    \includegraphics[scale=0.65]{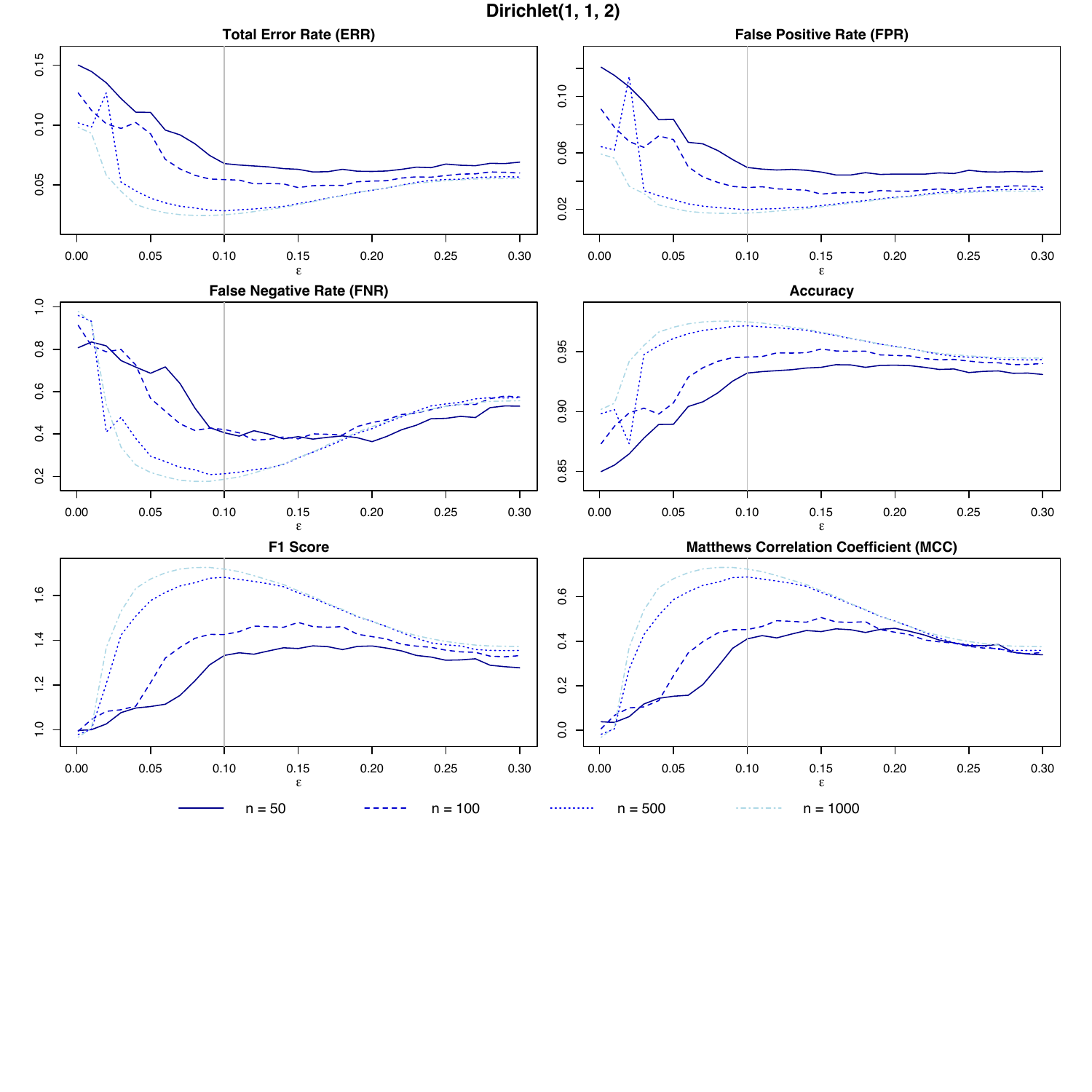}
    \caption{Performance results of \textbf{$M_3$:$\epsilon$-CDF} in the Dirichlet scenario for all sample sizes and for different values of the hyperparameter $\epsilon$, with $\epsilon \in \{0.001,0.010,0.020,\dots,0.290,0.300\}$. Gray vertical line reflects the optimal choice given by $\epsilon = 0.10$.}
    \label{fig: choice_eps_CDFmv_Dir112}
\end{figure}

\begin{figure}[!ht]
    \centering
    \includegraphics[scale=0.65]{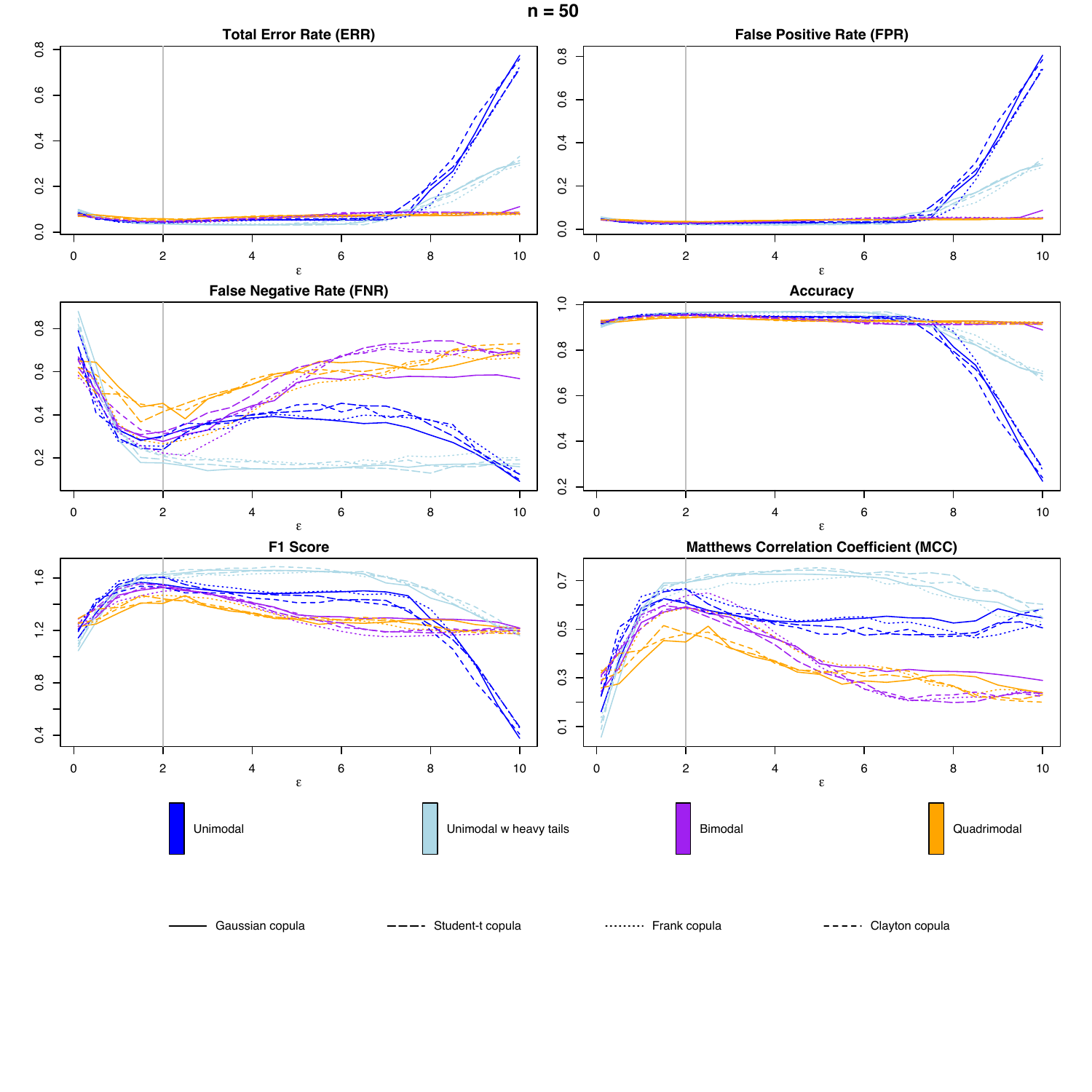}
    \caption{Performance results of \textbf{$M^{\text{NPCop}}_3$:$\epsilon$-CDF} in all evaluated scenarios for $n = 50$ and for different values of the hyperparameter $\epsilon$, with $\epsilon \in \{0.1,0.5,\dots,9.5,10.0\}$. Gray vertical line reflects the following heuristic: $\epsilon = \exp\left(1.74-0.26\log{n}\right) \approx 2$.}
    \label{fig: choice_eps_NPCoP_n50}
\end{figure}

\begin{figure}[!ht]
    \centering
    \includegraphics[scale=0.65]{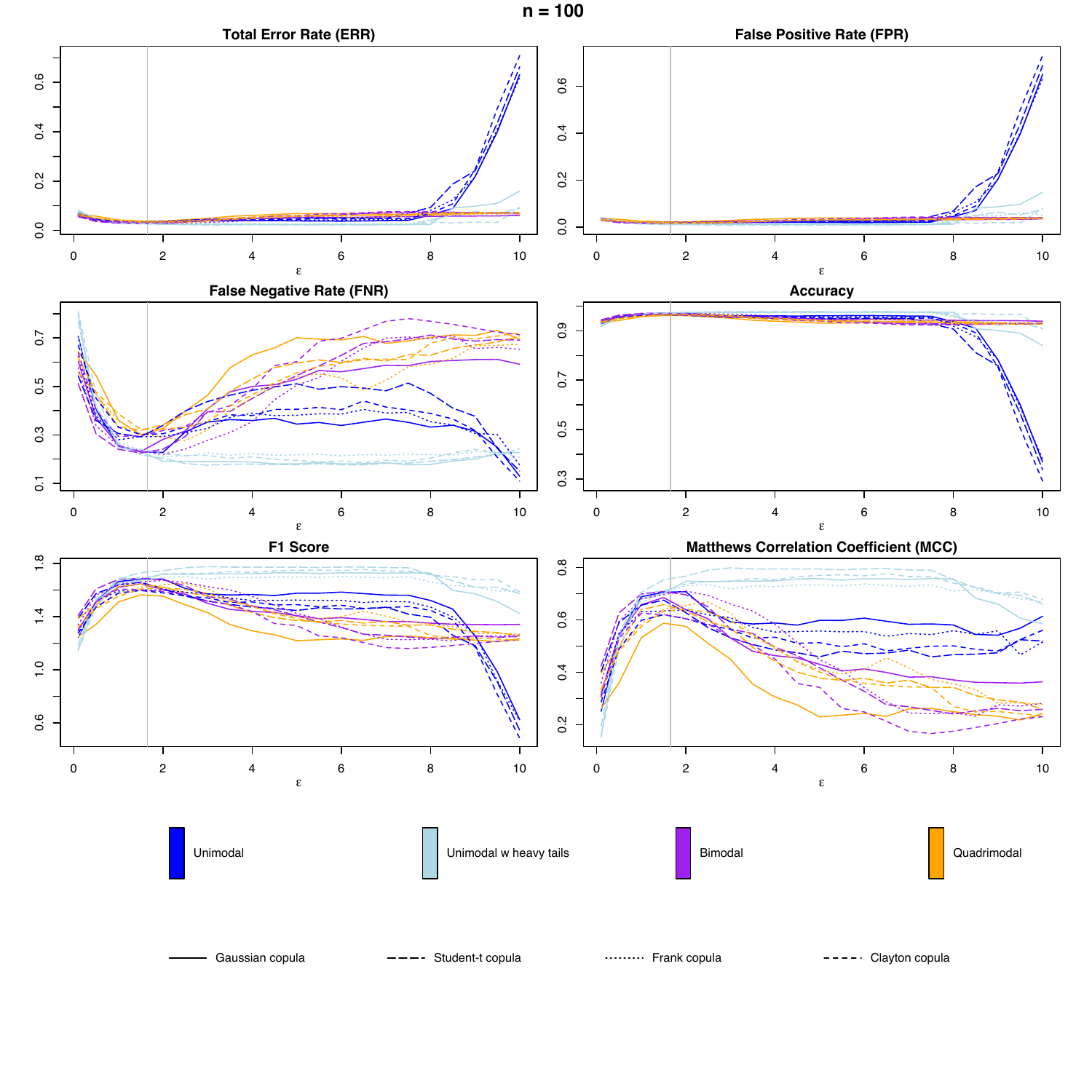}
    \caption{Performance results of \textbf{$M^{\text{NPCop}}_3$:$\epsilon$-CDF} in all evaluated scenarios for $n = 100$ and for different values of the hyperparameter $\epsilon$, with $\epsilon \in \{0.1,0.5,\dots,9.5,10.0\}$. Gray vertical line reflects the following heuristic: $\epsilon = \exp\left(1.74-0.26\log{n}\right) = 1.72 $.}
    \label{fig: choice_eps_NPCoP_n100}
\end{figure}

\begin{figure}[!ht]
    \centering
    \includegraphics[scale=0.65]{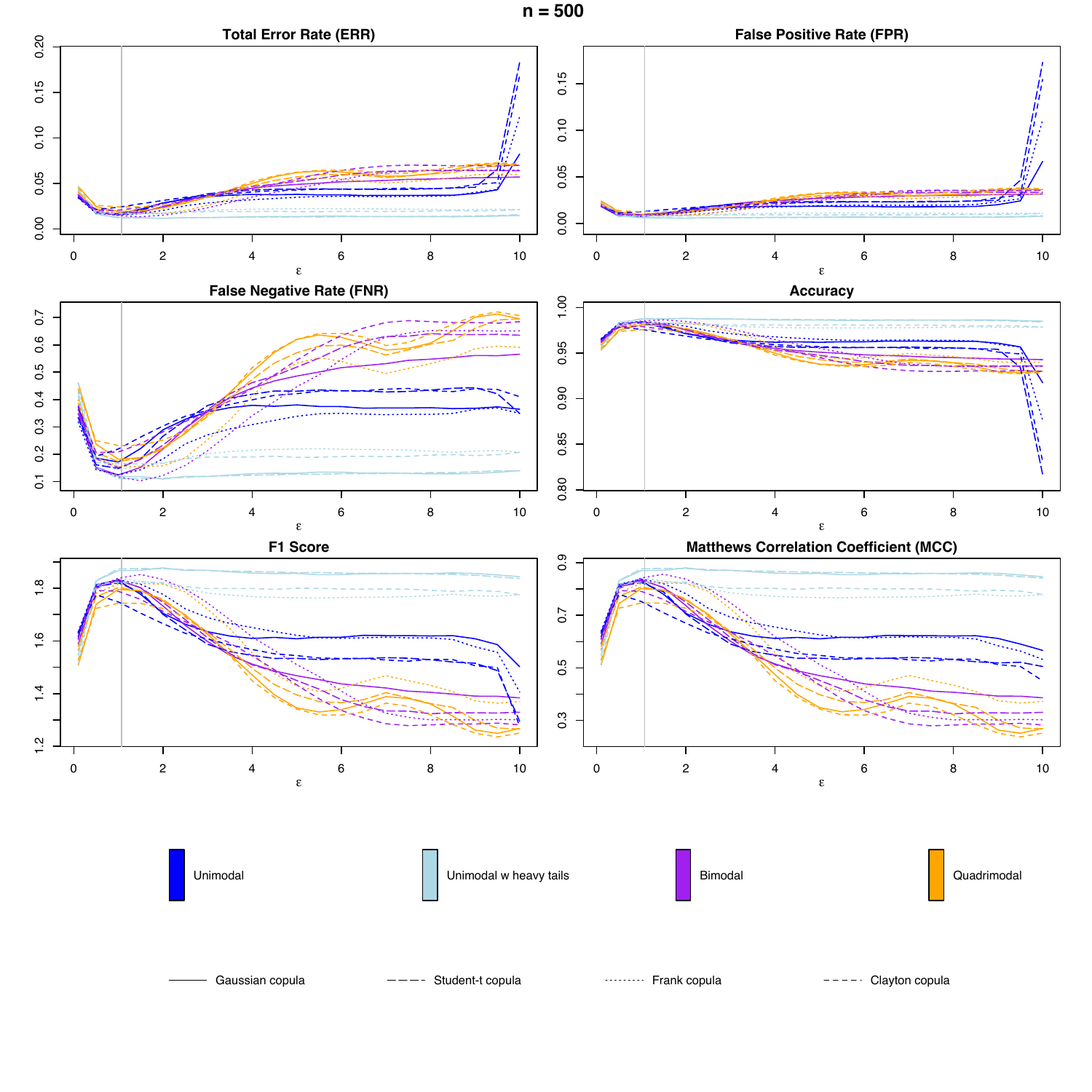}
    \caption{Performance results of \textbf{$M^{\text{NPCop}}_3$:$\epsilon$-CDF} in all evaluated scenarios for $n = 500$ and for different values of the hyperparameter $\epsilon$, with $\epsilon \in \{0.1,0.5,\dots,9.5,10.0\}$. Gray vertical line reflects the following heuristic: $\epsilon = \exp\left(1.74-0.26\log{n}\right) = 1.13$.}
    \label{fig: choice_eps_NPCoP_n500}
\end{figure}

\begin{figure}[!ht]
    \centering
    \includegraphics[scale=0.65]{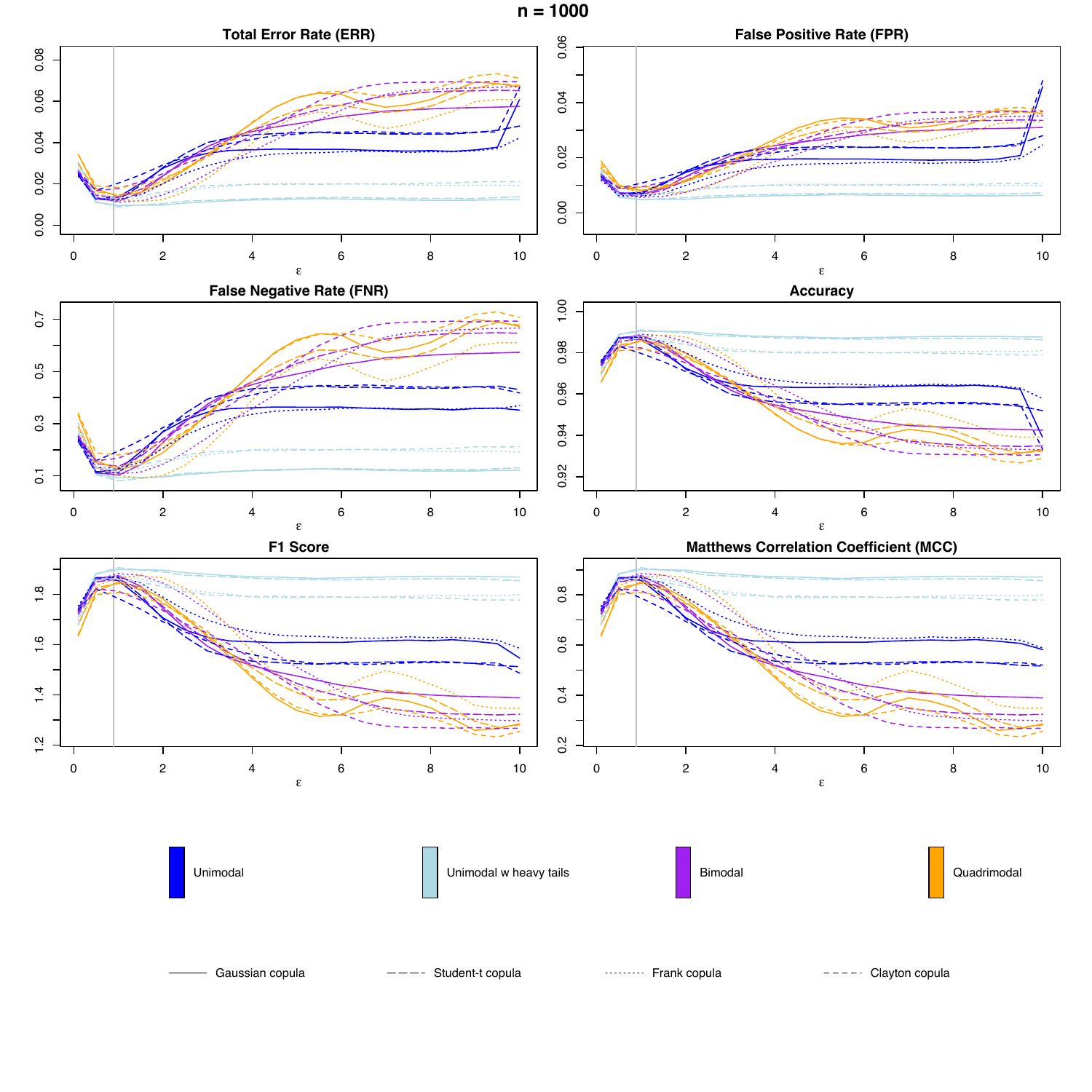}
    \caption{Performance results of \textbf{$M^{\text{NPCop}}_3$:$\epsilon$-CDF} in all evaluated scenarios for $n = 1000$ and for different values of the hyperparameter $\epsilon$, with $\epsilon \in \{0.1,0.5,\dots,9.5,10.0\}$. Gray vertical line reflects the following heuristic: $\epsilon = \exp\left(1.74-0.26\log{n}\right) = 0.95$.}
    \label{fig: choice_eps_NPCoP_n1000}
\end{figure}

\begin{figure}[!ht]
    \centering
    \includegraphics[scale=0.65]{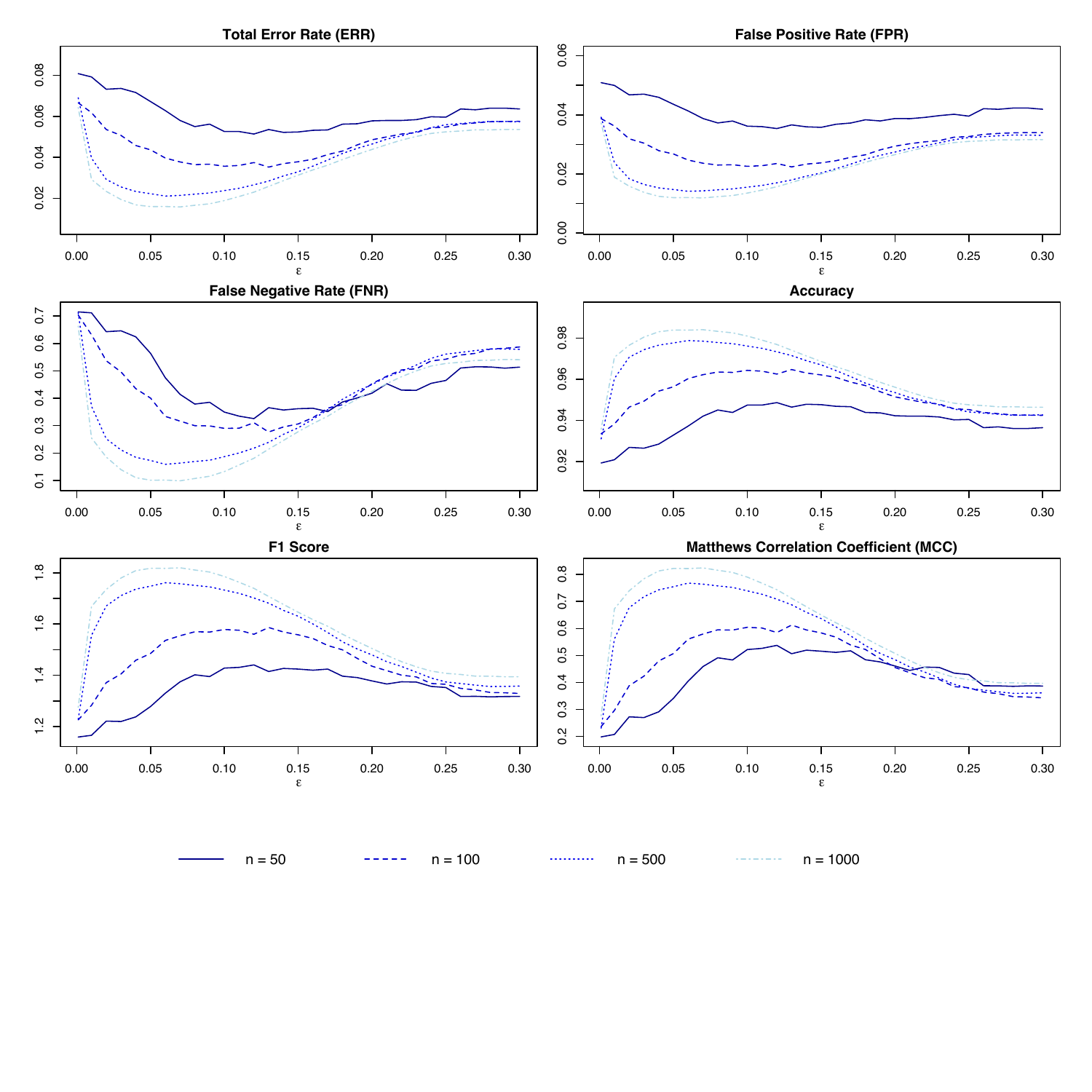}
    \caption{Performance results of \textbf{$M^{\text{NPCop}}_3$:$\epsilon$-CDF} in the Dirichlet scenario for all sample sizes and for different values of the hyperparameter $\epsilon$, with $\epsilon \in \{0.001,0.010,0.020,\dots,0.290,0.300\}$. The optimal choice reflects the following heuristic: $\epsilon = \exp\left(-1.22-0.23\log{n}\right)$.}
    \label{fig: choice_eps_NPCop_Dir112}
\end{figure}

\begin{figure}[!ht]
    \centering
    \includegraphics[scale=0.65]{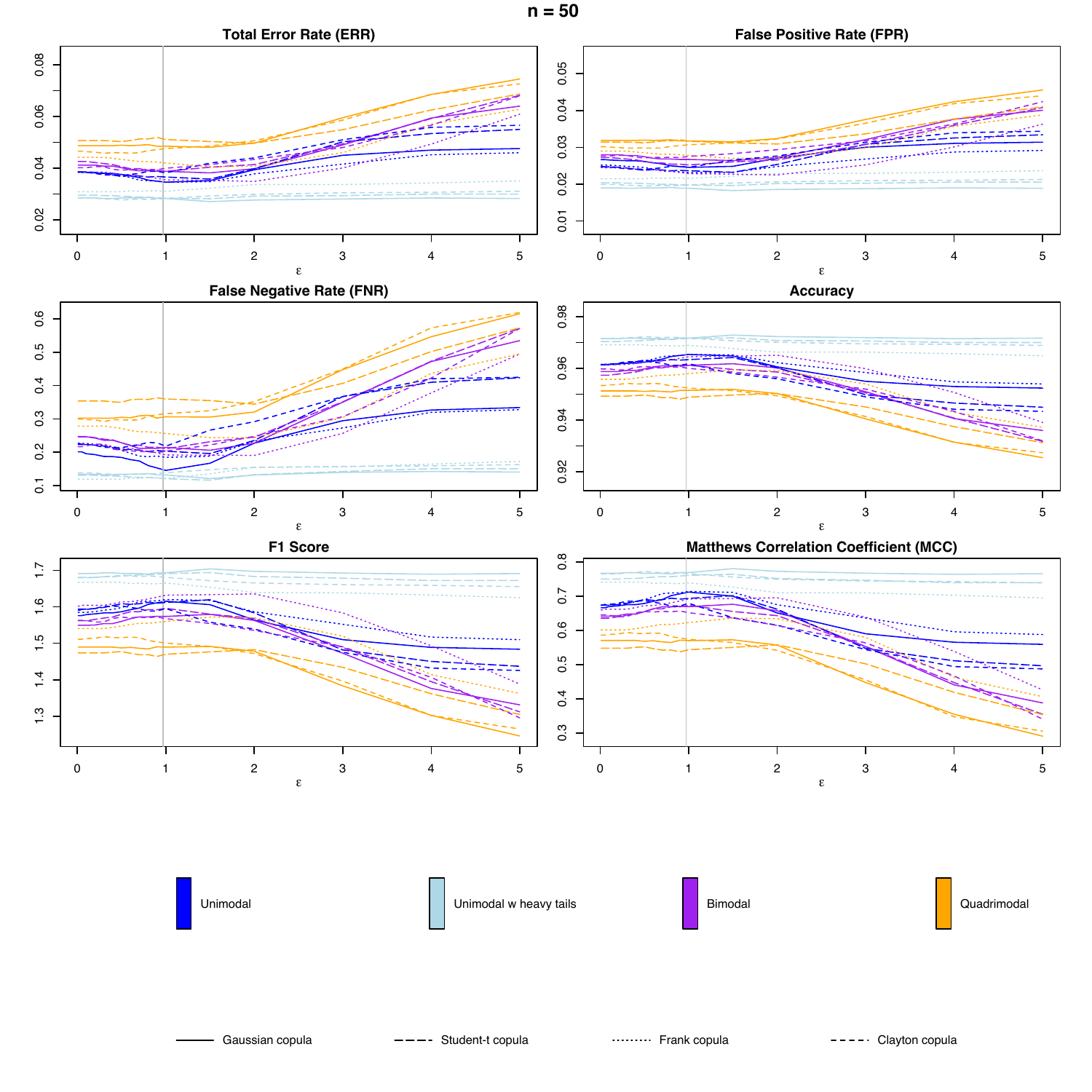}
    \caption{Performance results of \textbf{$M^{\text{PCop}}_3$:$\epsilon$-CDF} in all evaluated scenarios for $n = 50$ and for different values of the hyperparameter $\epsilon$, with $\epsilon \in \{0.01,0.02,0.05,0.10,0.20,0.30,\dots,4.0,5.0\}$. Gray vertical line reflects the following heuristic: $\epsilon = \exp\left(1.60 - 0.41\log{n}\right) \approx 1$.}
    \label{fig: choice_eps_PCoP_n50}
\end{figure}

\begin{figure}[!ht]
    \centering
    \includegraphics[scale=0.65]{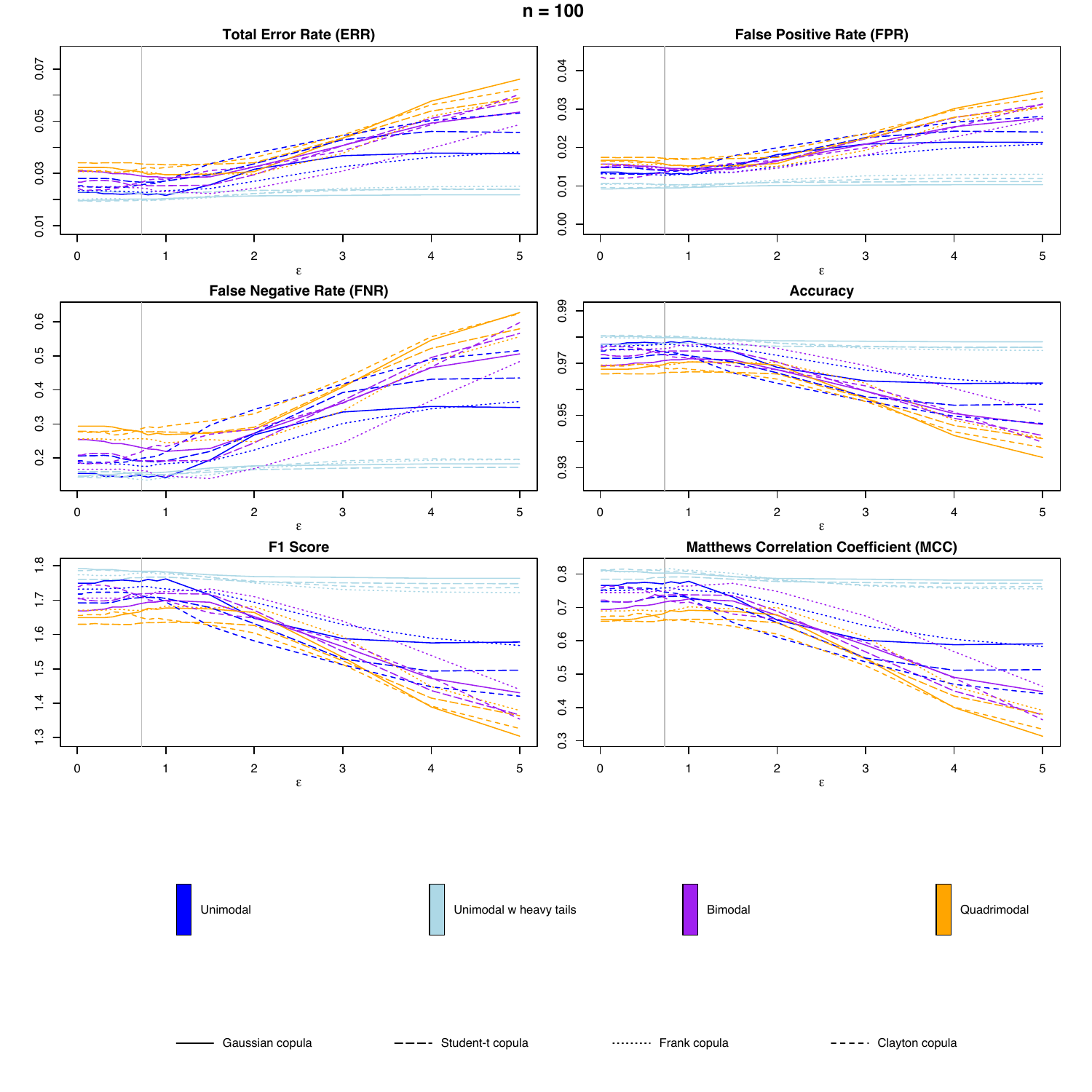}
    \caption{Performance results of \textbf{$M^{\text{PCop}}_3$:$\epsilon$-CDF} in all evaluated scenarios for $n = 100$ and for different values of the hyperparameter $\epsilon$, with $\epsilon \in \{0.01,0.02,0.05,0.10,0.20,0.30,\dots,4.0,5.0\}$. Gray vertical line reflects the following heuristic: $\epsilon = \exp\left(1.60 - 0.41\log{n}\right) = 0.75$.}
    \label{fig: choice_eps_PCoP_n100}
\end{figure}

\begin{figure}[!ht]
    \centering
    \includegraphics[scale=0.65]{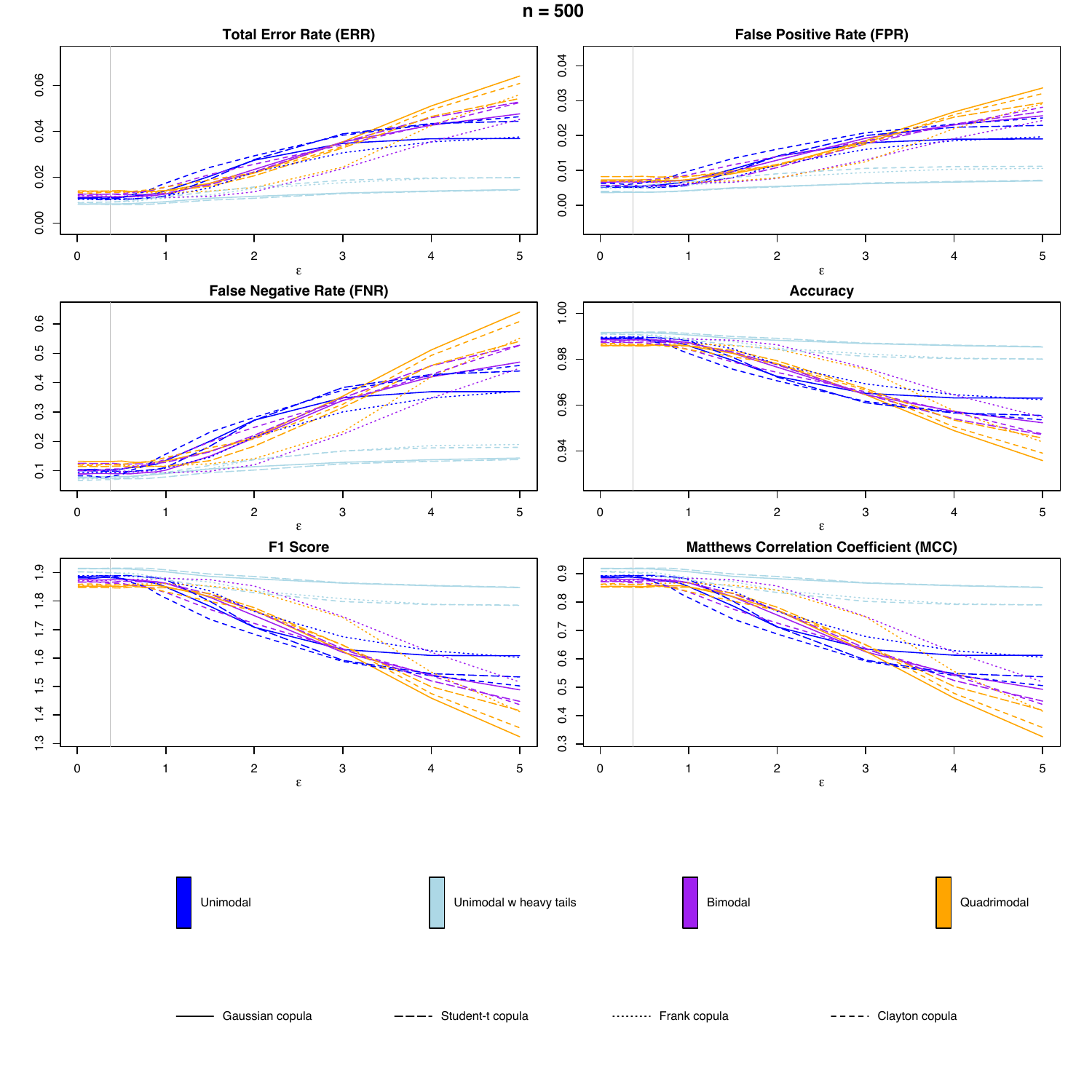}
    \caption{Performance results of \textbf{$M^{\text{PCop}}_3$:$\epsilon$-CDF} in all evaluated scenarios for $n = 500$ and for different values of the hyperparameter $\epsilon$, with $\epsilon \in \{0.01,0.02,0.05,0.10,0.20,0.30,\dots,4.0,5.0\}$. Gray vertical line reflects the following heuristic: $\epsilon = \exp\left(1.60 - 0.41\log{n}\right) = 0.39$.}
    \label{fig: choice_eps_PCoP_n500}
\end{figure}

\begin{figure}[!ht]
    \centering
    \includegraphics[scale=0.65]{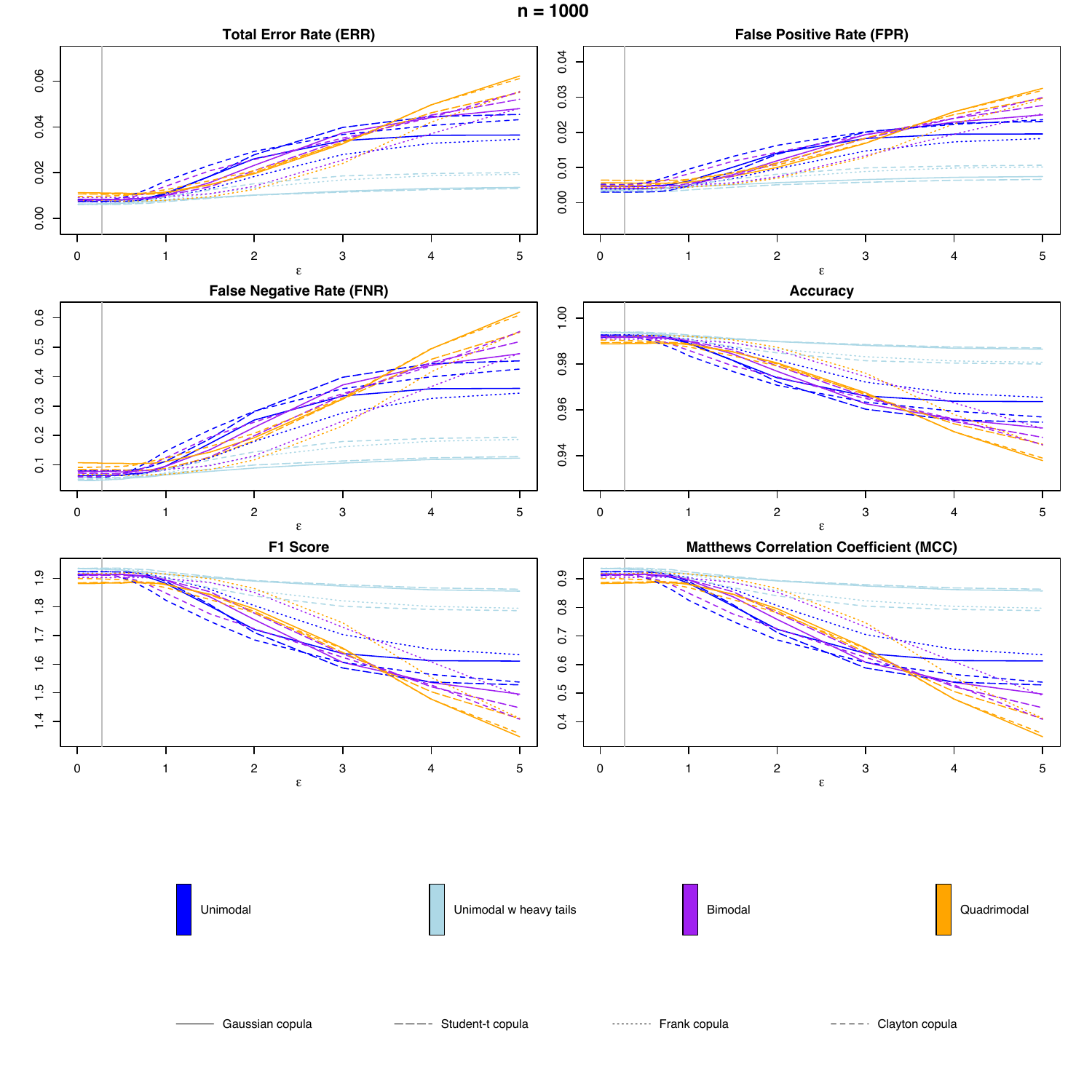}
    \caption{Performance results of \textbf{$M^{\text{PCop}}_3$:$\epsilon$-CDF} in all evaluated scenarios for $n = 1000$ and for different values of the hyperparameter $\epsilon$, with $\epsilon \in \{0.01,0.02,0.05,0.10,0.20,0.30,\dots,4.0,5.0\}$. Gray vertical line reflects the following heuristic: $\epsilon = \exp\left(1.60 - 0.41\log{n}\right) = 0.29$.}
    \label{fig: choice_eps_PCoP_n1000}
\end{figure}

\begin{figure}[!ht]
    \centering
    \includegraphics[scale=0.65]{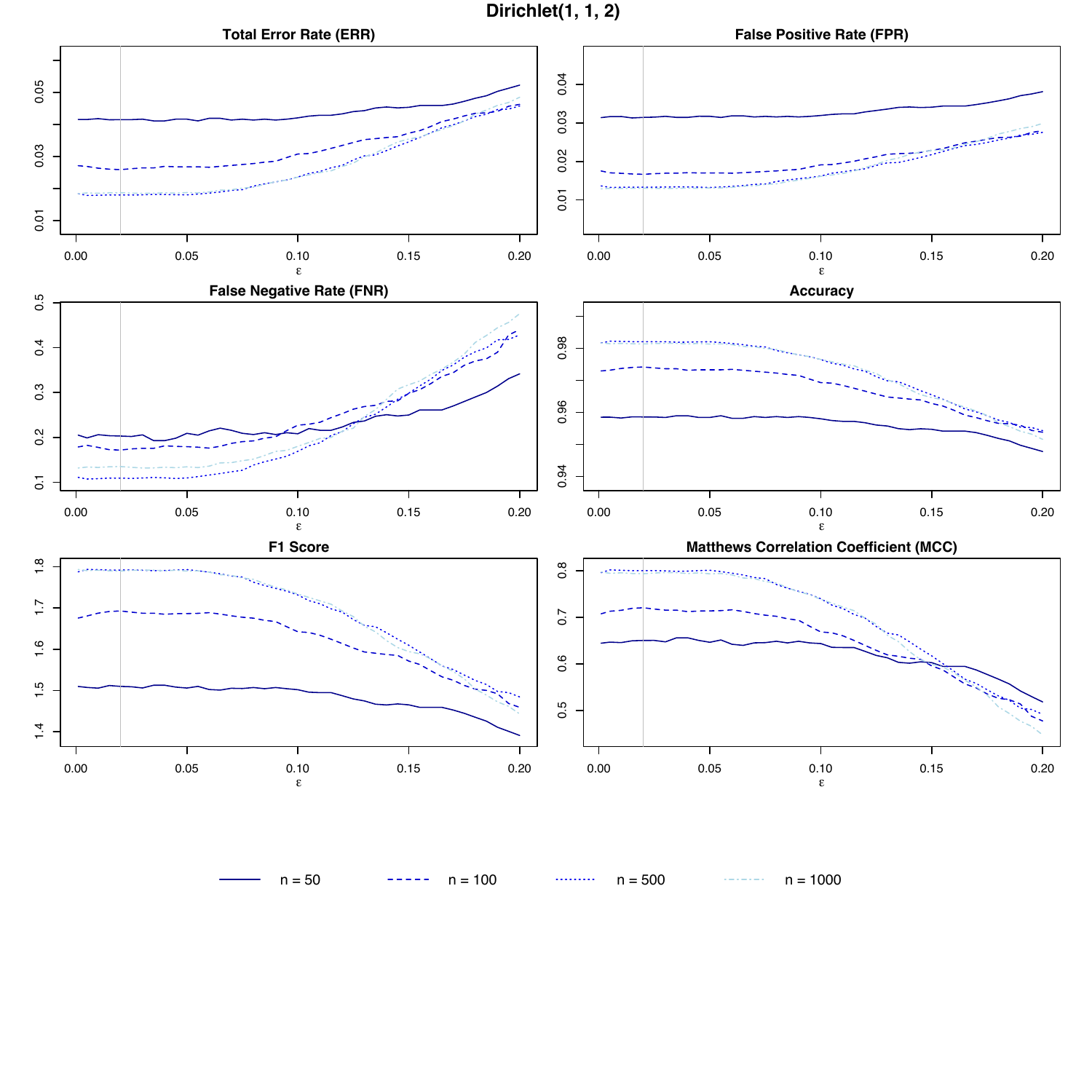}
    \caption{Performance results of \textbf{$M^{\text{PCop}}_3$:$\epsilon$-CDF} in the Dirichlet scenario for all sample sizes and for different values of the hyperparameter $\epsilon$, with $\epsilon \in \{0.001,0.005,0.010,0.015,\dots,0.195,0.200\}$. Gray vertical line reflects the optimal choice given by $\epsilon = 0.02$.}
    \label{fig: choice_eps_PCop_Dir112}
\end{figure}

\begin{description}
    \item[$M_0$:KDE] Direct estimation of the bivariate density. We use KDE, with Gaussian kernel and bandwidth selection based on the asympotically optimal solution proposed in~\cite{chacon_asymptotics_2011}, where its adequacy is shown in general settings, including Gaussian mixture models. Thus, no simulations are performed to empirically choose the hyperparameters. 

    \item[$M^{\text{NPCop}}_0$:DE] Nonparametric indirect density estimation with copula. We use standard KDE with the same optimal bandwidth of~\cite{chacon_asymptotics_2011} for the univariate marginals, and KDE with the transformation local likelihood estimator %of~\cite{geenens2017probit} 
    and nearest-neighbor bandwidth for the copula density~\citep{nagler_evading_2016}. No simulations are performed to empirically choose the hyperparameters. 

    \item[$M^{\text{PCop}}_0$:DE] Parametric indirect density estimation with copula. We adopt a fully parametric approach (with maximum likelihood fitting) to estimate both marginals and the copula. For the copula model, we select the best model using the AIC criterion; no misspecification is introduced for the marginals. No hyperpameters are involved; thus, no simulations are required. 

    \item[$M_1$:$k$NN-Eucl] Cumulative Euclidean distances from the $k$NNs. We sum the Euclidean distances between each point and its $k$ neighbors defined according to the Euclidean metric. The choice of the hyperparameter $k$ is based on an extensive cross-validation procedure, leading to a general rule of thumb aligned with the existing literature: $k = \round{\sqrt{n/2}}$, with $\round{x}$ the integer closest to $x$. The supporting simulation results are given in Figures~\ref{fig: choice_k_n50}-\ref{fig: choice_k_n1000}.

    \item[$M_2$:$k$NN-CDF] Cumulative CDF distances from the $k$NNs. We sum the CDF distances between each point and its $k$ neighbors, which are defined according to the Euclidean metric. The choice of the hyperparameter $k$ is based on an extensive cross-validation procedure, suggesting a uniform choice across different scenarios and sample sizes: $k = 30$. The supporting simulation results are given in Figures~\ref{fig: choice_k_CDF_n50}-\ref{fig: choice_k_CDF_n1000}.

    \item[$M_3$:$\epsilon$-CDF] This represents the $\epsilon$-neighborhood multivariate CDF distance, which uses empirical CDF as the CDF estimator. The optimal choice of the hyperparameter $\epsilon$ is based on an extensive cross-validation procedure, leading to the following heuristic for S1-S16 as a result of an exponential decay model fit with respect to the sample size: $\epsilon = \exp\left(2.13-0.3\log{n}\right)$. For S1-S16, the supporting simulation results are given in Figures~\ref{fig: choice_eps_CDFmv_n50}-\ref{fig: choice_eps_CDFmv_n1000}. For S17 (Dirichlet scenario), performances are robust to sample size, with optimal $\epsilon = 0.10$; see Figure~\ref{fig: choice_eps_CDFmv_Dir112}.

    \item[$M^{\text{NPCop}}_3$:$\epsilon$-CDF] Fully nonparametric indirect estimation of the $\epsilon$-CDF measure with copula. We use the empirical CDF for estimating the univariate marginals, and KDE with the transformation local likelihood estimator and nearest-neighbor bandwidth for the copula density. The optimal choice of the hyperparameter $\epsilon$ follows the same strategy as the previous measure, leading to the following heuristic for S1-S16: $\epsilon = \exp\left(1.74-0.26\log{n}\right)$. For S1-S16, the supporting simulation results are given in Figures~\ref{fig: choice_eps_NPCoP_n50}-\ref{fig: choice_eps_NPCoP_n1000}. For S17 (Dirichlet scenario), performances depend on the sample size, according to a heuristic given by: $\epsilon = \exp\left(-1.22-0.23\log{n}\right)$. This relationship is obtained by fitting a nonlinear regression (exponential decay model), with the empirical optimal $\epsilon$ and the sample size $n$ as dependent and independent variables, respectively; see Figure~\ref{fig: choice_eps_NPCop_Dir112}.
    
    \item[$M^{\text{PCop}}_3$:$\epsilon$-CDF] Parametric indirect CDF estimation with copula. We adopt a fully parametric approach (with maximum likelihood fitting) to estimate both marginals and the copula. For the copula model, we select the best model using the AIC criterion; no misspecification is introduced for the marginals. The optimal choice of the hyperparameter $\epsilon$ follows the heuristic $\epsilon = \exp\left(1.60 - 0.41\log{n}\right)$ for S1-S16 (see Figures~\ref{fig: choice_eps_PCoP_n50}-\ref{fig: choice_eps_PCoP_n1000}). For S17 (Dirichlet scenario), performances are robust to sample size, with optimal $\epsilon = 0.02$; see Figure~\ref{fig: choice_eps_PCop_Dir112}.
\end{description}





\clearpage
\section{Simulation results for all scenarios} \label{app: all_scenarios}

In this section, we report the full set of simulation results (for all scenarios and sample sizes) that were omitted in the main paper for the sake of space. All reported results are summarized in terms of mean (and standard deviation) across $1000$ independent Monte Carlo (MC) replicates.

\begin{table*}[!ht] 
\setlength\tabcolsep{0pt}
\scriptsize
\caption{Performance results in terms of mean (standard deviation) of the compared methods in scenario S1 across the different sample sizes.  \label{tab: performance_results_S1}} 
\begin{tabular*}{\textwidth}{@{\extracolsep\fill}lc@{}c@{}c@{}cc@{}c@{}c@{}c@{}}\toprule
&\multicolumn{4}{@{}@{}@{}c@{}}{\textbf{Measures w/o copula}}
& \multicolumn{4}{@{}@{}@{}c@{}}{\textbf{Measures w copula}} \\
\cmidrule{2-5}\cmidrule{6-9}
\textbf{Metrics} & \textbf{$M_0$:KDE}  & \textbf{$M_1$:$k$NN-Eucl} & \textbf{$M_2$:$k$NN-CDF} & \textbf{$M_3$:$\epsilon$-CDF}
& \multicolumn{1}{@{}c@{}}{\textbf{$M^{\text{NPCop}}_0$:DE}}  & \textbf{$M^{\text{PCop}}_0$:DE} & \textbf{$M^{\text{NPCop}}_3$:$\epsilon$-CDF} & \textbf{$M^{\text{PCop}}_3$:$\epsilon$-CDF}\\
$n = 50$ \\
\toprule
ERR      & 0.035 (0.023) & 0.029 (0.020) & 0.031 (0.021) & 0.034 (0.023) & 0.057 (0.028) & 0.030 (0.020) & 0.034 (0.021) & 0.030 (0.020) \\  
FPR      & 0.024 (0.018) & 0.020 (0.019) & 0.021 (0.019) & 0.025 (0.020) & 0.035 (0.020) & 0.021 (0.019) & 0.023 (0.018) & 0.021 (0.019) \\  
FNR      & 0.175 (0.234) & 0.120 (0.193) & 0.146 (0.221) & 0.133 (0.211) & 0.452 (0.344) & 0.126 (0.196) & 0.158 (0.223) & 0.122 (0.192) \\  
Accuracy & 0.965 (0.023) & 0.971 (0.020) & 0.969 (0.021) & 0.966 (0.023) & 0.943 (0.028) & 0.970 (0.020) & 0.966 (0.021) & 0.970 (0.020) \\  
F1       & 1.628 (0.263) & 1.681 (0.263) & 1.661 (0.272) & 1.649 (0.267) & 1.416 (0.305) & 1.675 (0.262) & 1.643 (0.260) & 1.678 (0.261) \\
\bottomrule
\end{tabular*}
\end{table*}

\begin{table*}[!ht]  \setlength\tabcolsep{0pt} \scriptsize
\caption{Performance results in terms of mean (standard deviation) of the compared methods in scenario S2 across the different sample sizes.  \label{tab: performance_results_S2}} 
\begin{tabular*}{\textwidth}{@{\extracolsep\fill}lc@{}c@{}c@{}cc@{}c@{}c@{}c@{}}\toprule
&\multicolumn{4}{@{}@{}@{}c@{}}{\textbf{Measures w/o copula}}
& \multicolumn{4}{@{}@{}@{}c@{}}{\textbf{Measures w copula}} \\
\cmidrule{2-5}\cmidrule{6-9}
\textbf{Metrics} & \textbf{$M_0$:KDE}  & \textbf{$M_1$:$k$NN-Eucl} & \textbf{$M_2$:$k$NN-CDF} & \textbf{$M_3$:$\epsilon$-CDF}
& \multicolumn{1}{@{}c@{}}{\textbf{$M^{\text{NPCop}}_0$:DE}}  & \textbf{$M^{\text{PCop}}_0$:DE} & \textbf{$M^{\text{NPCop}}_3$:$\epsilon$-CDF} & \textbf{$M^{\text{PCop}}_3$:$\epsilon$-CDF}\\
$n = 50$ \\
\toprule
ERR      & 0.043 (0.025) & 0.043 (0.024) & 0.055 (0.026) & 0.049 (0.026) & 0.042 (0.024) & 0.040 (0.023) & 0.044 (0.024) & 0.037 (0.022) \\  
FPR      & 0.028 (0.018) & 0.028 (0.018) & 0.034 (0.019) & 0.033 (0.022) & 0.027 (0.019) & 0.026 (0.019) & 0.029 (0.019) & 0.025 (0.019) \\  
FNR      & 0.252 (0.280) & 0.251 (0.279) & 0.402 (0.330) & 0.292 (0.297) & 0.247 (0.276) & 0.219 (0.262) & 0.274 (0.281) & 0.195 (0.243) \\  
Accuracy & 0.957 (0.025) & 0.957 (0.024) & 0.945 (0.026) & 0.951 (0.026) & 0.958 (0.024) & 0.960 (0.023) & 0.956 (0.024) & 0.963 (0.022) \\  
F1       & 1.549 (0.274) & 1.553 (0.275) & 1.435 (0.286) & 1.509 (0.284) & 1.557 (0.276) & 1.579 (0.273) & 1.535 (0.274) & 1.600 (0.269) \\
\bottomrule
\end{tabular*}
\end{table*}

\begin{table*}[!ht]  \setlength\tabcolsep{0pt} \scriptsize
\caption{Performance results in terms of mean (standard deviation) of the compared methods in scenario S3 across the different sample sizes.  \label{tab: performance_results_S3}} 
\begin{tabular*}{\textwidth}{@{\extracolsep\fill}lc@{}c@{}c@{}cc@{}c@{}c@{}c@{}}\toprule
&\multicolumn{4}{@{}@{}@{}c@{}}{\textbf{Measures w/o copula}}
& \multicolumn{4}{@{}@{}@{}c@{}}{\textbf{Measures w copula}} \\
\cmidrule{2-5}\cmidrule{6-9}
\textbf{Metrics} & \textbf{$M_0$:KDE}  & \textbf{$M_1$:$k$NN-Eucl} & \textbf{$M_2$:$k$NN-CDF} & \textbf{$M_3$:$\epsilon$-CDF}
& \multicolumn{1}{@{}c@{}}{\textbf{$M^{\text{NPCop}}_0$:DE}}  & \textbf{$M^{\text{PCop}}_0$:DE} & \textbf{$M^{\text{NPCop}}_3$:$\epsilon$-CDF} & \textbf{$M^{\text{PCop}}_3$:$\epsilon$-CDF}\\
$n = 50$ \\
\toprule
ERR      & 0.046 (0.027) & 0.045 (0.025) & 0.062 (0.028) & 0.051 (0.026) & 0.043 (0.024) & 0.043 (0.025) & 0.044 (0.024) & 0.040 (0.023) \\  
FPR      & 0.030 (0.019) & 0.029 (0.018) & 0.038 (0.018) & 0.035 (0.022) & 0.028 (0.019) & 0.028 (0.019) & 0.029 (0.019) & 0.026 (0.019) \\  
FNR      & 0.289 (0.304) & 0.281 (0.294) & 0.492 (0.340) & 0.299 (0.303) & 0.266 (0.283) & 0.261 (0.287) & 0.270 (0.288) & 0.229 (0.267) \\  
Accuracy & 0.954 (0.027) & 0.955 (0.025) & 0.938 (0.028) & 0.949 (0.026) & 0.957 (0.024) & 0.957 (0.025) & 0.956 (0.024) & 0.960 (0.023) \\  
F1       & 1.522 (0.287) & 1.529 (0.278) & 1.363 (0.284) & 1.493 (0.280) & 1.545 (0.279) & 1.549 (0.285) & 1.540 (0.279) & 1.575 (0.278) \\
\bottomrule
\end{tabular*}
\end{table*}

\begin{table*}[!ht]  \setlength\tabcolsep{0pt} \scriptsize
\caption{Performance results in terms of mean (standard deviation) of the compared methods in scenario S4 across the different sample sizes.  \label{tab: performance_results_S4}} 
\begin{tabular*}{\textwidth}{@{\extracolsep\fill}lc@{}c@{}c@{}cc@{}c@{}c@{}c@{}}\toprule
&\multicolumn{4}{@{}@{}@{}c@{}}{\textbf{Measures w/o copula}}
& \multicolumn{4}{@{}@{}@{}c@{}}{\textbf{Measures w copula}} \\
\cmidrule{2-5}\cmidrule{6-9}
\textbf{Metrics} & \textbf{$M_0$:KDE}  & \textbf{$M_1$:$k$NN-Eucl} & \textbf{$M_2$:$k$NN-CDF} & \textbf{$M_3$:$\epsilon$-CDF}
& \multicolumn{1}{@{}c@{}}{\textbf{$M^{\text{NPCop}}_0$:DE}}  & \textbf{$M^{\text{PCop}}_0$:DE} & \textbf{$M^{\text{NPCop}}_3$:$\epsilon$-CDF} & \textbf{$M^{\text{PCop}}_3$:$\epsilon$-CDF}\\
$n = 50$ \\
\toprule
ERR      & 0.061 (0.030) & 0.060 (0.028) & 0.081 (0.030) & 0.068 (0.032) & 0.050 (0.027) & 0.050 (0.027) & 0.055 (0.027) & 0.049 (0.027) \\  
FPR      & 0.038 (0.018) & 0.037 (0.018) & 0.048 (0.016) & 0.046 (0.024) & 0.031 (0.019) & 0.032 (0.019) & 0.034 (0.019) & 0.031 (0.019) \\  
FNR      & 0.464 (0.342) & 0.454 (0.336) & 0.709 (0.321) & 0.444 (0.347) & 0.335 (0.314) & 0.337 (0.314) & 0.400 (0.324) & 0.326 (0.310) \\  
Accuracy & 0.939 (0.030) & 0.940 (0.028) & 0.919 (0.030) & 0.932 (0.032) & 0.950 (0.027) & 0.950 (0.027) & 0.945 (0.027) & 0.951 (0.027) \\  
F1       & 1.379 (0.287) & 1.391 (0.287) & 1.188 (0.260) & 1.363 (0.284) & 1.485 (0.292) & 1.481 (0.290) & 1.434 (0.287) & 1.490 (0.289) \\
\bottomrule
\end{tabular*}
\end{table*}

\begin{table*}[!ht]  \setlength\tabcolsep{0pt} \scriptsize
\caption{Performance results in terms of mean (standard deviation) of the compared methods in scenario S5 across the different sample sizes. \label{tab: performance_results_S5}} 
\begin{tabular*}{\textwidth}{@{\extracolsep\fill}lc@{}c@{}c@{}cc@{}c@{}c@{}c@{}}\toprule
&\multicolumn{4}{@{}@{}@{}c@{}}{\textbf{Measures w/o copula}}
& \multicolumn{4}{@{}@{}@{}c@{}}{\textbf{Measures w copula}} \\
\cmidrule{2-5}\cmidrule{6-9}
\textbf{Metrics} & \textbf{$M_0$:KDE}  & \textbf{$M_1$:$k$NN-Eucl} & \textbf{$M_2$:$k$NN-CDF} & \textbf{$M_3$:$\epsilon$-CDF}
& \multicolumn{1}{@{}c@{}}{\textbf{$M^{\text{NPCop}}_0$:DE}}  & \textbf{$M^{\text{PCop}}_0$:DE} & \textbf{$M^{\text{NPCop}}_3$:$\epsilon$-CDF} & \textbf{$M^{\text{PCop}}_3$:$\epsilon$-CDF}\\
$n = 50$ \\
\toprule
ERR      & 0.033 (0.021) & 0.030 (0.020) & 0.034 (0.022) & 0.034 (0.022) & 0.054 (0.026) & 0.031 (0.020) & 0.034 (0.021) & 0.030 (0.020) \\  
FPR      & 0.023 (0.019) & 0.021 (0.020) & 0.023 (0.020) & 0.025 (0.020) & 0.034 (0.020) & 0.021 (0.020) & 0.023 (0.019) & 0.021 (0.020) \\  
FNR      & 0.150 (0.210) & 0.128 (0.192) & 0.179 (0.238) & 0.132 (0.206) & 0.441 (0.332) & 0.132 (0.193) & 0.166 (0.221) & 0.125 (0.189) \\  
Accuracy & 0.967 (0.021) & 0.970 (0.020) & 0.966 (0.022) & 0.966 (0.022) & 0.946 (0.026) & 0.969 (0.020) & 0.966 (0.021) & 0.970 (0.020) \\  
F1       & 1.639 (0.274) & 1.663 (0.279) & 1.625 (0.291) & 1.638 (0.283) & 1.426 (0.312) & 1.659 (0.277) & 1.626 (0.274) & 1.666 (0.278) \\
\bottomrule
\end{tabular*}
\end{table*}

\begin{table*}[!ht]  \setlength\tabcolsep{0pt} \scriptsize
\caption{Performance results in terms of mean (standard deviation) of the compared methods in scenario S6 across the different sample sizes. \label{tab: performance_results_S6}} 
\begin{tabular*}{\textwidth}{@{\extracolsep\fill}lc@{}c@{}c@{}cc@{}c@{}c@{}c@{}}\toprule
&\multicolumn{4}{@{}@{}@{}c@{}}{\textbf{Measures w/o copula}}
& \multicolumn{4}{@{}@{}@{}c@{}}{\textbf{Measures w copula}} \\
\cmidrule{2-5}\cmidrule{6-9}
\textbf{Metrics} & \textbf{$M_0$:KDE}  & \textbf{$M_1$:$k$NN-Eucl} & \textbf{$M_2$:$k$NN-CDF} & \textbf{$M_3$:$\epsilon$-CDF}
& \multicolumn{1}{@{}c@{}}{\textbf{$M^{\text{NPCop}}_0$:DE}}  & \textbf{$M^{\text{PCop}}_0$:DE} & \textbf{$M^{\text{NPCop}}_3$:$\epsilon$-CDF} & \textbf{$M^{\text{PCop}}_3$:$\epsilon$-CDF}\\
$n = 50$ \\
\toprule
ERR      & 0.042 (0.025) & 0.042 (0.023) & 0.062 (0.027) & 0.049 (0.025) & 0.042 (0.024) & 0.039 (0.023) & 0.043 (0.023) & 0.036 (0.022) \\  
FPR      & 0.027 (0.019) & 0.028 (0.018) & 0.038 (0.018) & 0.033 (0.021) & 0.027 (0.019) & 0.026 (0.019) & 0.028 (0.019) & 0.025 (0.019) \\  
FNR      & 0.242 (0.287) & 0.248 (0.270) & 0.495 (0.338) & 0.296 (0.293) & 0.243 (0.276) & 0.212 (0.255) & 0.258 (0.274) & 0.186 (0.238) \\  
Accuracy & 0.958 (0.025) & 0.958 (0.023) & 0.938 (0.027) & 0.951 (0.025) & 0.958 (0.024) & 0.961 (0.023) & 0.957 (0.023) & 0.964 (0.022) \\  
F1       & 1.560 (0.287) & 1.557 (0.274) & 1.364 (0.286) & 1.502 (0.279) & 1.560 (0.280) & 1.588 (0.277) & 1.548 (0.276) & 1.609 (0.275) \\
\bottomrule
\end{tabular*}
\end{table*}

\begin{table*}[!ht]  \setlength\tabcolsep{0pt} \scriptsize
\caption{Performance results in terms of mean (standard deviation) of the compared methods in scenario S7 across the different sample sizes. \label{tab: performance_results_S7}} 
\begin{tabular*}{\textwidth}{@{\extracolsep\fill}lc@{}c@{}c@{}cc@{}c@{}c@{}c@{}}\toprule
&\multicolumn{4}{@{}@{}@{}c@{}}{\textbf{Measures w/o copula}}
& \multicolumn{4}{@{}@{}@{}c@{}}{\textbf{Measures w copula}} \\
\cmidrule{2-5}\cmidrule{6-9}
\textbf{Metrics} & \textbf{$M_0$:KDE}  & \textbf{$M_1$:$k$NN-Eucl} & \textbf{$M_2$:$k$NN-CDF} & \textbf{$M_3$:$\epsilon$-CDF}
& \multicolumn{1}{@{}c@{}}{\textbf{$M^{\text{NPCop}}_0$:DE}}  & \textbf{$M^{\text{PCop}}_0$:DE} & \textbf{$M^{\text{NPCop}}_3$:$\epsilon$-CDF} & \textbf{$M^{\text{PCop}}_3$:$\epsilon$-CDF}\\
$n = 50$ \\
\toprule
ERR      & 0.046 (0.026) & 0.043 (0.025) & 0.073 (0.030) & 0.050 (0.027) & 0.043 (0.025) & 0.042 (0.024) & 0.043 (0.025) & 0.040 (0.024) \\  
FPR      & 0.030 (0.019) & 0.028 (0.018) & 0.044 (0.017) & 0.034 (0.023) & 0.028 (0.019) & 0.027 (0.019) & 0.028 (0.019) & 0.026 (0.019) \\  
FNR      & 0.296 (0.301) & 0.258 (0.279) & 0.620 (0.334) & 0.287 (0.294) & 0.257 (0.286) & 0.259 (0.277) & 0.255 (0.277) & 0.226 (0.265) \\  
Accuracy & 0.954 (0.026) & 0.957 (0.025) & 0.927 (0.030) & 0.950 (0.027) & 0.958 (0.025) & 0.958 (0.024) & 0.957 (0.025) & 0.960 (0.024) \\  
F1       & 1.522 (0.283) & 1.551 (0.274) & 1.267 (0.279) & 1.511 (0.280) & 1.555 (0.283) & 1.555 (0.276) & 1.556 (0.276) & 1.582 (0.277) \\
\bottomrule
\end{tabular*}
\end{table*}

\begin{table*}[!ht]  \setlength\tabcolsep{0pt} \scriptsize
\caption{Performance results in terms of mean (standard deviation) of the compared methods in scenario S8 across the different sample sizes. \label{tab: performance_results_S8}} 
\begin{tabular*}{\textwidth}{@{\extracolsep\fill}lc@{}c@{}c@{}cc@{}c@{}c@{}c@{}}\toprule
&\multicolumn{4}{@{}@{}@{}c@{}}{\textbf{Measures w/o copula}}
& \multicolumn{4}{@{}@{}@{}c@{}}{\textbf{Measures w copula}} \\
\cmidrule{2-5}\cmidrule{6-9}
\textbf{Metrics} & \textbf{$M_0$:KDE}  & \textbf{$M_1$:$k$NN-Eucl} & \textbf{$M_2$:$k$NN-CDF} & \textbf{$M_3$:$\epsilon$-CDF}
& \multicolumn{1}{@{}c@{}}{\textbf{$M^{\text{NPCop}}_0$:DE}}  & \textbf{$M^{\text{PCop}}_0$:DE} & \textbf{$M^{\text{NPCop}}_3$:$\epsilon$-CDF} & \textbf{$M^{\text{PCop}}_3$:$\epsilon$-CDF}\\
$n = 50$ \\
\toprule
ERR      & 0.058 (0.029) & 0.057 (0.027) & 0.087 (0.031) & 0.066 (0.031) & 0.046 (0.025) & 0.048 (0.025) & 0.052 (0.027) & 0.048 (0.025) \\  
FPR      & 0.036 (0.018) & 0.035 (0.018) & 0.051 (0.015) & 0.045 (0.025) & 0.029 (0.019) & 0.031 (0.018) & 0.033 (0.019) & 0.030 (0.018) \\  
FNR      & 0.425 (0.339) & 0.428 (0.335) & 0.770 (0.295) & 0.430 (0.344) & 0.291 (0.299) & 0.321 (0.306) & 0.372 (0.327) & 0.311 (0.299) \\  
Accuracy & 0.942 (0.029) & 0.943 (0.027) & 0.913 (0.031) & 0.934 (0.031) & 0.954 (0.025) & 0.952 (0.025) & 0.948 (0.027) & 0.952 (0.025) \\  
F1       & 1.413 (0.290) & 1.414 (0.288) & 1.140 (0.241) & 1.378 (0.285) & 1.522 (0.284) & 1.497 (0.281) & 1.460 (0.294) & 1.504 (0.278) \\
\bottomrule
\end{tabular*}
\end{table*}

\begin{table*}[!ht]  \setlength\tabcolsep{0pt} \scriptsize
\caption{Performance results in terms of mean (standard deviation) of the compared methods in scenario S9 across the different sample sizes. \label{tab: performance_results_S9}} 
\begin{tabular*}{\textwidth}{@{\extracolsep\fill}lc@{}c@{}c@{}cc@{}c@{}c@{}c@{}}\toprule
&\multicolumn{4}{@{}@{}@{}c@{}}{\textbf{Measures w/o copula}}
& \multicolumn{4}{@{}@{}@{}c@{}}{\textbf{Measures w copula}} \\
\cmidrule{2-5}\cmidrule{6-9}
\textbf{Metrics} & \textbf{$M_0$:KDE}  & \textbf{$M_1$:$k$NN-Eucl} & \textbf{$M_2$:$k$NN-CDF} & \textbf{$M_3$:$\epsilon$-CDF}
& \multicolumn{1}{@{}c@{}}{\textbf{$M^{\text{NPCop}}_0$:DE}}  & \textbf{$M^{\text{PCop}}_0$:DE} & \textbf{$M^{\text{NPCop}}_3$:$\epsilon$-CDF} & \textbf{$M^{\text{PCop}}_3$:$\epsilon$-CDF}\\
$n = 50$ \\
\toprule
ERR      & 0.037 (0.023) & 0.033 (0.020) & 0.036 (0.021) & 0.038 (0.024) & 0.060 (0.027) & 0.030 (0.020) & 0.038 (0.023) & 0.031 (0.021) \\  
FPR      & 0.025 (0.019) & 0.022 (0.019) & 0.024 (0.020) & 0.028 (0.020) & 0.037 (0.019) & 0.021 (0.020) & 0.026 (0.019) & 0.022 (0.020) \\  
FNR      & 0.188 (0.239) & 0.147 (0.202) & 0.192 (0.243) & 0.165 (0.228) & 0.478 (0.340) & 0.125 (0.192) & 0.197 (0.240) & 0.130 (0.199) \\  
Accuracy & 0.963 (0.023) & 0.967 (0.020) & 0.964 (0.021) & 0.962 (0.024) & 0.940 (0.027) & 0.970 (0.020) & 0.962 (0.023) & 0.969 (0.021) \\  
F1       & 1.600 (0.275) & 1.641 (0.271) & 1.607 (0.283) & 1.602 (0.276) & 1.382 (0.296) & 1.660 (0.276) & 1.593 (0.271) & 1.656 (0.279) \\
\bottomrule
\end{tabular*}
\end{table*}

\begin{table*}[!ht]  \setlength\tabcolsep{0pt} \scriptsize
\caption{Performance results in terms of mean (standard deviation) of the compared methods in scenario S10 across the different sample sizes. \label{tab: performance_results_S10}} 
\begin{tabular*}{\textwidth}{@{\extracolsep\fill}lc@{}c@{}c@{}cc@{}c@{}c@{}c@{}}\toprule
&\multicolumn{4}{@{}@{}@{}c@{}}{\textbf{Measures w/o copula}}
& \multicolumn{4}{@{}@{}@{}c@{}}{\textbf{Measures w copula}} \\
\cmidrule{2-5}\cmidrule{6-9}
\textbf{Metrics} & \textbf{$M_0$:KDE}  & \textbf{$M_1$:$k$NN-Eucl} & \textbf{$M_2$:$k$NN-CDF} & \textbf{$M_3$:$\epsilon$-CDF}
& \multicolumn{1}{@{}c@{}}{\textbf{$M^{\text{NPCop}}_0$:DE}}  & \textbf{$M^{\text{PCop}}_0$:DE} & \textbf{$M^{\text{NPCop}}_3$:$\epsilon$-CDF} & \textbf{$M^{\text{PCop}}_3$:$\epsilon$-CDF}\\
$n = 50$ \\
\toprule
ERR      & 0.044 (0.026) & 0.039 (0.022) & 0.058 (0.027) & 0.044 (0.025) & 0.041 (0.023) & 0.037 (0.023) & 0.041 (0.023) & 0.036 (0.022) \\  
FPR      & 0.029 (0.019) & 0.026 (0.019) & 0.037 (0.019) & 0.031 (0.022) & 0.027 (0.019) & 0.025 (0.019) & 0.027 (0.019) & 0.024 (0.019) \\  
FNR      & 0.252 (0.288) & 0.213 (0.260) & 0.459 (0.341) & 0.248 (0.280) & 0.229 (0.266) & 0.194 (0.251) & 0.238 (0.272) & 0.174 (0.235) \\  
Accuracy & 0.956 (0.026) & 0.961 (0.022) & 0.942 (0.027) & 0.956 (0.025) & 0.959 (0.023) & 0.963 (0.023) & 0.959 (0.023) & 0.964 (0.022) \\  
F1       & 1.537 (0.290) & 1.577 (0.286) & 1.389 (0.298) & 1.537 (0.294) & 1.560 (0.283) & 1.591 (0.288) & 1.557 (0.287) & 1.608 (0.286) \\
\bottomrule
\end{tabular*}
\end{table*}

\begin{table*}[!ht]  \setlength\tabcolsep{0pt} \scriptsize
\caption{Performance results in terms of mean (standard deviation) of the compared methods in scenario S11 across the different sample sizes. \label{tab: performance_results_S11}} 
\begin{tabular*}{\textwidth}{@{\extracolsep\fill}lc@{}c@{}c@{}cc@{}c@{}c@{}c@{}}\toprule
&\multicolumn{4}{@{}@{}@{}c@{}}{\textbf{Measures w/o copula}}
& \multicolumn{4}{@{}@{}@{}c@{}}{\textbf{Measures w copula}} \\
\cmidrule{2-5}\cmidrule{6-9}
\textbf{Metrics} & \textbf{$M_0$:KDE}  & \textbf{$M_1$:$k$NN-Eucl} & \textbf{$M_2$:$k$NN-CDF} & \textbf{$M_3$:$\epsilon$-CDF}
& \multicolumn{1}{@{}c@{}}{\textbf{$M^{\text{NPCop}}_0$:DE}}  & \textbf{$M^{\text{PCop}}_0$:DE} & \textbf{$M^{\text{NPCop}}_3$:$\epsilon$-CDF} & \textbf{$M^{\text{PCop}}_3$:$\epsilon$-CDF}\\
$n = 50$ \\
\toprule
ERR      & 0.047 (0.027) & 0.041 (0.024) & 0.072 (0.029) & 0.044 (0.026) & 0.043 (0.024) & 0.040 (0.024) & 0.040 (0.023) & 0.038 (0.024) \\  
FPR      & 0.030 (0.019) & 0.027 (0.019) & 0.043 (0.017) & 0.031 (0.024) & 0.028 (0.019) & 0.027 (0.020) & 0.027 (0.019) & 0.026 (0.020) \\  
FNR      & 0.287 (0.306) & 0.234 (0.277) & 0.609 (0.339) & 0.231 (0.268) & 0.253 (0.280) & 0.226 (0.269) & 0.226 (0.262) & 0.203 (0.255) \\  
Accuracy & 0.953 (0.027) & 0.959 (0.024) & 0.928 (0.029) & 0.956 (0.026) & 0.957 (0.024) & 0.960 (0.024) & 0.960 (0.023) & 0.962 (0.024) \\  
F1       & 1.512 (0.296) & 1.561 (0.294) & 1.271 (0.284) & 1.548 (0.293) & 1.544 (0.289) & 1.566 (0.293) & 1.567 (0.287) & 1.585 (0.290) \\
\bottomrule
\end{tabular*}
\end{table*}

\begin{table*}[!ht]  \setlength\tabcolsep{0pt} \scriptsize
\caption{Performance results in terms of mean (standard deviation) of the compared methods in scenario S12 across the different sample sizes. \label{tab: performance_results_S12}} 
\begin{tabular*}{\textwidth}{@{\extracolsep\fill}lc@{}c@{}c@{}cc@{}c@{}c@{}c@{}}\toprule
&\multicolumn{4}{@{}@{}@{}c@{}}{\textbf{Measures w/o copula}}
& \multicolumn{4}{@{}@{}@{}c@{}}{\textbf{Measures w copula}} \\
\cmidrule{2-5}\cmidrule{6-9}
\textbf{Metrics} & \textbf{$M_0$:KDE}  & \textbf{$M_1$:$k$NN-Eucl} & \textbf{$M_2$:$k$NN-CDF} & \textbf{$M_3$:$\epsilon$-CDF}
& \multicolumn{1}{@{}c@{}}{\textbf{$M^{\text{NPCop}}_0$:DE}}  & \textbf{$M^{\text{PCop}}_0$:DE} & \textbf{$M^{\text{NPCop}}_3$:$\epsilon$-CDF} & \textbf{$M^{\text{PCop}}_3$:$\epsilon$-CDF}\\
$n = 50$ \\
\toprule
ERR      & 0.055 (0.028) & 0.052 (0.027) & 0.084 (0.031) & 0.059 (0.029) & 0.045 (0.024) & 0.046 (0.026) & 0.047 (0.025) & 0.045 (0.025) \\  
FPR      & 0.035 (0.018) & 0.033 (0.018) & 0.050 (0.015) & 0.041 (0.024) & 0.029 (0.019) & 0.030 (0.019) & 0.030 (0.019) & 0.029 (0.019) \\  
FNR      & 0.385 (0.332) & 0.357 (0.322) & 0.732 (0.321) & 0.351 (0.318) & 0.275 (0.287) & 0.293 (0.306) & 0.302 (0.301) & 0.271 (0.294) \\  
Accuracy & 0.945 (0.028) & 0.948 (0.027) & 0.916 (0.031) & 0.941 (0.029) & 0.955 (0.024) & 0.954 (0.026) & 0.953 (0.025) & 0.955 (0.025) \\  
F1       & 1.429 (0.286) & 1.455 (0.289) & 1.162 (0.251) & 1.430 (0.280) & 1.521 (0.283) & 1.508 (0.294) & 1.503 (0.291) & 1.525 (0.291) \\
\bottomrule
\end{tabular*}
\end{table*}

\begin{table*}[!ht]  \setlength\tabcolsep{0pt} \scriptsize
\caption{Performance results in terms of mean (standard deviation) of the compared methods in scenario S13 across the different sample sizes. \label{tab: performance_results_S13}} 
\begin{tabular*}{\textwidth}{@{\extracolsep\fill}lc@{}c@{}c@{}cc@{}c@{}c@{}c@{}}\toprule
&\multicolumn{4}{@{}@{}@{}c@{}}{\textbf{Measures w/o copula}}
& \multicolumn{4}{@{}@{}@{}c@{}}{\textbf{Measures w copula}} \\
\cmidrule{2-5}\cmidrule{6-9}
\textbf{Metrics} & \textbf{$M_0$:KDE}  & \textbf{$M_1$:$k$NN-Eucl} & \textbf{$M_2$:$k$NN-CDF} & \textbf{$M_3$:$\epsilon$-CDF}
& \multicolumn{1}{@{}c@{}}{\textbf{$M^{\text{NPCop}}_0$:DE}}  & \textbf{$M^{\text{PCop}}_0$:DE} & \textbf{$M^{\text{NPCop}}_3$:$\epsilon$-CDF} & \textbf{$M^{\text{PCop}}_3$:$\epsilon$-CDF}\\
$n = 50$ \\
\toprule
ERR      & 0.037 (0.021) & 0.033 (0.021) & 0.036 (0.021) & 0.036 (0.022) & 0.055 (0.026) & 0.033 (0.021) & 0.037 (0.022) & 0.033 (0.021) \\  
FPR      & 0.025 (0.019) & 0.023 (0.020) & 0.024 (0.020) & 0.026 (0.021) & 0.034 (0.020) & 0.022 (0.020) & 0.025 (0.019) & 0.022 (0.020) \\  
FNR      & 0.185 (0.233) & 0.161 (0.221) & 0.198 (0.246) & 0.167 (0.235) & 0.433 (0.337) & 0.150 (0.218) & 0.192 (0.242) & 0.151 (0.216) \\  
Accuracy & 0.963 (0.021) & 0.967 (0.021) & 0.964 (0.021) & 0.964 (0.022) & 0.945 (0.026) & 0.967 (0.021) & 0.963 (0.022) & 0.967 (0.021) \\  
F1       & 1.606 (0.270) & 1.632 (0.278) & 1.604 (0.284) & 1.613 (0.283) & 1.421 (0.301) & 1.640 (0.280) & 1.600 (0.275) & 1.639 (0.279) \\
\bottomrule
\end{tabular*}
\end{table*}

\begin{table*}[!ht]  \setlength\tabcolsep{0pt} \scriptsize
\caption{Performance results in terms of mean (standard deviation) of the compared methods in scenario S14 across the different sample sizes. \label{tab: performance_results_S14}} 
\begin{tabular*}{\textwidth}{@{\extracolsep\fill}lc@{}c@{}c@{}cc@{}c@{}c@{}c@{}}\toprule
&\multicolumn{4}{@{}@{}@{}c@{}}{\textbf{Measures w/o copula}}
& \multicolumn{4}{@{}@{}@{}c@{}}{\textbf{Measures w copula}} \\
\cmidrule{2-5}\cmidrule{6-9}
\textbf{Metrics} & \textbf{$M_0$:KDE}  & \textbf{$M_1$:$k$NN-Eucl} & \textbf{$M_2$:$k$NN-CDF} & \textbf{$M_3$:$\epsilon$-CDF}
& \multicolumn{1}{@{}c@{}}{\textbf{$M^{\text{NPCop}}_0$:DE}}  & \textbf{$M^{\text{PCop}}_0$:DE} & \textbf{$M^{\text{NPCop}}_3$:$\epsilon$-CDF} & \textbf{$M^{\text{PCop}}_3$:$\epsilon$-CDF}\\
$n = 50$ \\
\toprule
ERR      & 0.046 (0.025) & 0.047 (0.025) & 0.065 (0.028) & 0.051 (0.026) & 0.045 (0.024) & 0.038 (0.023) & 0.048 (0.025) & 0.039 (0.024) \\  
FPR      & 0.029 (0.019) & 0.030 (0.019) & 0.039 (0.017) & 0.034 (0.021) & 0.029 (0.019) & 0.025 (0.019) & 0.031 (0.019) & 0.025 (0.020) \\  
FNR      & 0.301 (0.298) & 0.304 (0.297) & 0.517 (0.338) & 0.342 (0.310) & 0.288 (0.292) & 0.208 (0.258) & 0.323 (0.307) & 0.223 (0.271) \\  
Accuracy & 0.954 (0.025) & 0.953 (0.025) & 0.935 (0.028) & 0.949 (0.026) & 0.955 (0.024) & 0.962 (0.023) & 0.952 (0.025) & 0.961 (0.024) \\  
F1       & 1.518 (0.284) & 1.514 (0.281) & 1.344 (0.278) & 1.480 (0.288) & 1.529 (0.280) & 1.596 (0.283) & 1.501 (0.287) & 1.586 (0.292) \\
\bottomrule
\end{tabular*}
\end{table*}

\begin{table*}[!ht]  \setlength\tabcolsep{0pt} \scriptsize
\caption{Performance results in terms of mean (standard deviation) of the compared methods in scenario S15 across the different sample sizes. \label{tab: performance_results_S15}} 
\begin{tabular*}{\textwidth}{@{\extracolsep\fill}lc@{}c@{}c@{}cc@{}c@{}c@{}c@{}}\toprule
&\multicolumn{4}{@{}@{}@{}c@{}}{\textbf{Measures w/o copula}}
& \multicolumn{4}{@{}@{}@{}c@{}}{\textbf{Measures w copula}} \\
\cmidrule{2-5}\cmidrule{6-9}
\textbf{Metrics} & \textbf{$M_0$:KDE}  & \textbf{$M_1$:$k$NN-Eucl} & \textbf{$M_2$:$k$NN-CDF} & \textbf{$M_3$:$\epsilon$-CDF}
& \multicolumn{1}{@{}c@{}}{\textbf{$M^{\text{NPCop}}_0$:DE}}  & \textbf{$M^{\text{PCop}}_0$:DE} & \textbf{$M^{\text{NPCop}}_3$:$\epsilon$-CDF} & \textbf{$M^{\text{PCop}}_3$:$\epsilon$-CDF}\\
$n = 50$ \\
\toprule
ERR      & 0.049 (0.027) & 0.048 (0.026) & 0.075 (0.031) & 0.054 (0.027) & 0.046 (0.025) & 0.042 (0.024) & 0.049 (0.025) & 0.042 (0.024) \\  
FPR      & 0.031 (0.019) & 0.031 (0.019) & 0.045 (0.017) & 0.036 (0.023) & 0.030 (0.018) & 0.027 (0.019) & 0.031 (0.019) & 0.027 (0.019) \\  
FNR      & 0.333 (0.315) & 0.324 (0.308) & 0.629 (0.343) & 0.337 (0.315) & 0.290 (0.294) & 0.249 (0.277) & 0.335 (0.307) & 0.250 (0.278) \\  
Accuracy & 0.951 (0.027) & 0.952 (0.026) & 0.925 (0.031) & 0.947 (0.027) & 0.954 (0.025) & 0.958 (0.024) & 0.951 (0.025) & 0.958 (0.024) \\  
F1       & 1.487 (0.292) & 1.496 (0.288) & 1.249 (0.276) & 1.468 (0.291) & 1.520 (0.282) & 1.556 (0.286) & 1.488 (0.287) & 1.557 (0.289) \\
\bottomrule
\end{tabular*}
\end{table*}

\begin{table*}[!ht]  \setlength\tabcolsep{0pt} \scriptsize
\caption{Performance results in terms of mean (standard deviation) of the compared methods in scenario S16 across the different sample sizes. \label{tab: performance_results_S16}} 
\begin{tabular*}{\textwidth}{@{\extracolsep\fill}lc@{}c@{}c@{}cc@{}c@{}c@{}c@{}}\toprule
&\multicolumn{4}{@{}@{}@{}c@{}}{\textbf{Measures w/o copula}}
& \multicolumn{4}{@{}@{}@{}c@{}}{\textbf{Measures w copula}} \\
\cmidrule{2-5}\cmidrule{6-9}
\textbf{Metrics} & \textbf{$M_0$:KDE}  & \textbf{$M_1$:$k$NN-Eucl} & \textbf{$M_2$:$k$NN-CDF} & \textbf{$M_3$:$\epsilon$-CDF}
& \multicolumn{1}{@{}c@{}}{\textbf{$M^{\text{NPCop}}_0$:DE}}  & \textbf{$M^{\text{PCop}}_0$:DE} & \textbf{$M^{\text{NPCop}}_3$:$\epsilon$-CDF} & \textbf{$M^{\text{PCop}}_3$:$\epsilon$-CDF}\\
$n = 50$ \\
\toprule
ERR      & 0.062 (0.030) & 0.058 (0.029) & 0.086 (0.032) & 0.069 (0.032) & 0.050 (0.026) & 0.048 (0.027) & 0.055 (0.028) & 0.048 (0.027) \\  
FPR      & 0.038 (0.018) & 0.036 (0.018) & 0.051 (0.015) & 0.046 (0.025) & 0.032 (0.018) & 0.031 (0.019) & 0.035 (0.019) & 0.030 (0.019) \\  
FNR      & 0.471 (0.343) & 0.428 (0.341) & 0.754 (0.310) & 0.452 (0.342) & 0.341 (0.305) & 0.318 (0.310) & 0.400 (0.331) & 0.316 (0.311) \\  
Accuracy & 0.938 (0.030) & 0.942 (0.029) & 0.914 (0.032) & 0.931 (0.032) & 0.950 (0.026) & 0.952 (0.027) & 0.945 (0.028) & 0.952 (0.027) \\  
F1       & 1.371 (0.288) & 1.406 (0.291) & 1.148 (0.246) & 1.359 (0.284) & 1.478 (0.281) & 1.497 (0.294) & 1.432 (0.293) & 1.500 (0.297) \\
\bottomrule
\end{tabular*}
\end{table*}

\begin{table*}[!ht]  \setlength\tabcolsep{0pt} \scriptsize
\caption{Performance results in terms of mean (standard deviation) of the compared methods in scenario S17 across the different sample sizes. \label{tab: performance_results_S17}} 
\begin{tabular*}{\textwidth}{@{\extracolsep\fill}lc@{}c@{}c@{}cc@{}c@{}c@{}c@{}}\toprule
&\multicolumn{4}{@{}@{}@{}c@{}}{\textbf{Measures w/o copula}}
& \multicolumn{4}{@{}@{}@{}c@{}}{\textbf{Measures w copula}} \\
\cmidrule{2-5}\cmidrule{6-9}
\textbf{Metrics} & \textbf{$M_0$:KDE}  & \textbf{$M_1$:$k$NN-Eucl} & \textbf{$M_2$:$k$NN-CDF} & \textbf{$M_3$:$\epsilon$-CDF}
& \multicolumn{1}{@{}c@{}}{\textbf{$M^{\text{NPCop}}_0$:DE}}  & \textbf{$M^{\text{PCop}}_0$:DE} & \textbf{$M^{\text{NPCop}}_3$:$\epsilon$-CDF} & \textbf{$M^{\text{PCop}}_3$:$\epsilon$-CDF}\\
$n = 50$ \\
\toprule
ERR      & 0.072 (0.027) & 0.053 (0.027) & 0.060 (0.026) & 0.070 (0.033) & 0.049 (0.025) & 0.041 (0.024) & 0.053 (0.026) & 0.041 (0.024) \\  
FPR      & 0.047 (0.015) & 0.037 (0.017) & 0.041 (0.017) & 0.052 (0.025) & 0.035 (0.017) & 0.031 (0.019) & 0.037 (0.017) & 0.030 (0.019) \\  
FNR      & 0.620 (0.358) & 0.343 (0.346) & 0.466 (0.360) & 0.420 (0.368) & 0.295 (0.322) & 0.200 (0.276) & 0.342 (0.338) & 0.200 (0.277) \\  
Accuracy & 0.928 (0.027) & 0.947 (0.027) & 0.940 (0.026) & 0.930 (0.033) & 0.951 (0.025) & 0.959 (0.024) & 0.947 (0.026) & 0.959 (0.024) \\  
F1       & 1.229 (0.267) & 1.417 (0.291) & 1.342 (0.290) & 1.321 (0.272) & 1.454 (0.289) & 1.531 (0.298) & 1.422 (0.290) & 1.538 (0.300) \\
\bottomrule
\end{tabular*}
\end{table*}

\begin{figure}[!ht]
    \centering
    \includegraphics[scale=0.56]{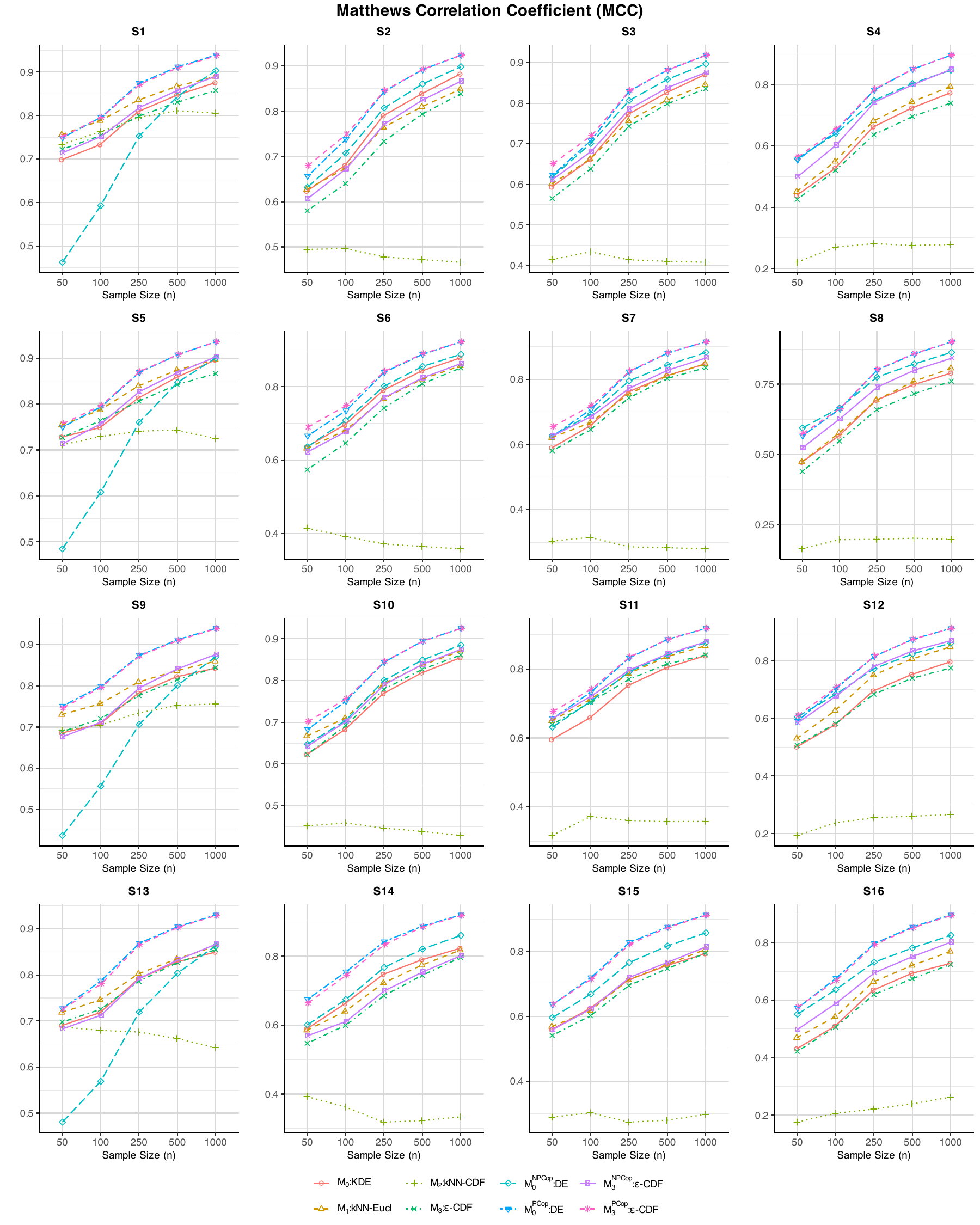}
    \caption{Global comparison between the evaluated measures in terms of the Matthews correlation coefficient (MCC) metric, for scenarios S1-S16 and sample sizes $n \in \{50, 100, 250, 500, 1000\}$.}
    \label{fig: global_MCC}
\end{figure}

\end{document}